\documentclass[pra,reprint,superscriptaddress,twocolumn,preprintnumbers,amsmath,amssymb]{revtex4-1}
\usepackage{amssymb,latexsym}
\usepackage{amsbsy} 
\usepackage{amsmath}
\usepackage{xfrac}
\usepackage{tabularx}
\usepackage{hyperref}
\usepackage{graphicx,color}
\usepackage{verbatim}

\DeclareMathOperator*{\argmin}{argmin}

\newcommand\bra[1] {\langle {#1} |}
\newcommand\ket[1] {| {#1} \rangle}

\newcommand\orbdown[2] {\text{#1}_{\text{#2}}}
\newcommand\orbdouble[2] {\text{#1}_{\text{#2}}^2}
\newcommand\orbNoSpin[2] {\text{#1}_{\text{#2}}}

\newcommand\IIorbdown[3] {{#1}_{\text{#3}}^{{#2}}}
\newcommand\IIorbdouble[3] {{#1}_{\text{#3}}^{{#2}2}}
\newcommand\IIorbNoSpin[3] {{#1}_{\text{#3}}^{{#2}}}

\newcommand\IITorbdown[3] {\text{#1}_{\text{#3}}}
\newcommand\IITorbdouble[3] {\text{#1}_{\text{#3}}^{2}}
\newcommand\IITorbNoSpin[3] {\text{#1}_{\text{#3}}}

\newcommand\subidx {\alpha}
\newcommand\subidxB {\beta}

\begin{document}

\title{Bonding in the helium dimer in strong magnetic fields: the role of spin and angular momentum} 

\author{Jon Austad}
\affiliation{Hylleraas Centre for Quantum Molecular Sciences, Department of Chemistry,  University of Oslo, P.O. Box 1033 Blindern, N-0315 Oslo, Norway}

\author{Alex Borgoo}
\affiliation{Hylleraas Centre for Quantum Molecular Sciences, Department of Chemistry,  University of Oslo, P.O. Box 1033 Blindern, N-0315 Oslo, Norway}

\author{Erik I. Tellgren}
\email{erik.tellgren@kjemi.uio.no}
\affiliation{Hylleraas Centre for Quantum Molecular Sciences, Department of Chemistry,  University of Oslo, P.O. Box 1033 Blindern, N-0315 Oslo, Norway}

\author{Trygve Helgaker}
\email{t.u.helgaker@kjemi.uio.no}
\affiliation{Hylleraas Centre for Quantum Molecular Sciences, Department of Chemistry,  University of Oslo, P.O. Box 1033 Blindern, N-0315 Oslo, Norway}

\begin{abstract}
We investigate the helium dimer in strong magnetic fields, focusing on the spectrum of low-lying electronic states and their dissociation curves, 
at the full configuration-interaction level of theory. To address the loss of cylindrical symmetry and angular momentum as a good quantum number 
for nontrivial angles between the bond axis and magnetic field, we introduce the almost quantized angular momentum (AQAM) and show that it provides 
useful information about states in arbitrary orientations. In general, strong magnetic fields dramatically rearrange the spectrum, with the orbital 
Zeeman effect bringing down states of higher angular momentum below the states with pure $\sigma$ character as the field strength increases. 
In addition, the spin Zeeman effect pushes triplet states below the lowest singlet; in particular, a field of one atomic unit is strong enough to 
push a quintet state below the triplets. In general, the angle between the bond axis and the magnetic field also continuously modulates 
the degree of $\sigma$, $\pi$, and $\delta$ character of bonds and the previously identified perpendicular paramagnetic bonding mechanism 
is found to be common among excited states. Electronic states with preferred skew field orientations are identified and rationalized
in terms of permanent and induced electronic  currents.
\end{abstract}

\maketitle 

\section{Introduction}

It has long been known that strong magnetic fields dramatically affect the physics and chemistry of molecules~\cite{GARSTANG_RPP40_105,LAI_RMP73_629}. 
In the atmospheres of neutron stars, intense magnetic
fields, orders of magnitudes stronger than one atomic unit $B_0 = 235$\,kT, dominate the electrostatic forces,
resulting in highly prolate, or even needle-like, charge distributions
around atoms. In such ultrastrong magnetic fields, matter is expected to consist of long chains of atoms,
oriented parallel to the magnetic field vector. The strong field regime $0.1B_0 < B < B_0$ is interesting as the direct magnetic effects and
electrostatic forces in small molecules are on the same order of
magnitude, leading to novel and complicated bonding mechanisms. This
regime corresponds to the upper range of magnetic field strengths
encountered in magnetic white dwarf (MWD) stars.

In the strong and ultrastrong field regimes, atomic spectra and chemical
bonding become modified. Calculated helium spectra have assisted the
interpretation of observed spectra from the atmosphere of MWDs~\cite{JORDAN_AA376_614,JORDAN_AA336_33}, 
supplementing the well-established use of hydrogen lines to analyse MWDs. The magnetic
field dependence of energy levels in
hydrogen~\cite{KRAVCHENKO_PRA54_287}, hydrogen
anions~\cite{ALHUJAJ_PRA61_063413},
helium~\cite{JONES_PRA59_2875,BECKEN_PRA65_033416,THIRUMALAI_PRA79_012514},
and other small atoms~\cite{ALHUJAJ_PRA70_023411,IVANOV_PRA61_022505,IVANOV_PRA60_3558,THIRUMALAI_PRA90_052501} have
been subject to several studies. Even one-electron molecular ions
exhibit a rich phenomenology to
explore~\cite{TURBINER_PR424_309}. Many otherwise unstable
few-electron ions, such as He$^{-}$, HeH$^+$, and He$_2^{2+}$, become
stabilized in external magnetic
fields~\cite{AVRON_PRL39_1068,TURBINER_JPB40_3249,TURBINER_PRA74_063419}. Several
studies have focused on potential-energy surfaces and the modification of
bonding in H$_2^+$ and H$_2$ subject to strong
fields~\cite{OZAKI_CPL203_184,SCHMELCHER_PRA41_4936,KAPPES_PRA50_3775,DETMER_PRA57_1767,DETMER_JCP109_9694}. Most
studies have been restricted to the parallel orientation as this is
by far the easiest to study. However, a few studies of varying
accuracy have found that the H$_2$ triplet state becomes stabilized in
a perpendicular magnetic
field~\cite{ZAUCER_PRA18_1320,LOZOVIK_PLA66_282,BASILE_INC9_457,KOROLEV_PRA45_1762,KUBO_JPCA111_5572},
subsequently explained based on high-quality quantum-chemical
calculations as an orientation-dependent stabilization of the
antibonding $\sigma$-orbital~\cite{LANGE_S337_327}. This effect,
termed perpendicular paramagnetic bonding, is also seen in singlet
helium clusters and other diatomic
molecules~\cite{TELLGREN_PCCP14_9492,STOPKOWICZ_JCP143_074110}.

While the highest field strengths
available in the laboratory are two to three orders of
magnitude below $B_0$~\cite{MOTOKAWA_RPP67_1995,NAKAMURA_RSI84_044702,NAKAMURA_RSI89_095106,BYKOV_PB294_574},
quasiparticles in semiconductors can have effective masses much below
that of a bare electron and exhibit analogous effects at lower field
strengths. Notably, quasiparticle analogues to perpendicular
paramagnetic bonding have already been reported~\cite{MURDIN_NC4_1469,LITVINENKO_PRB90_115204}.  Rydberg states, which are sensitive to magnetic fields due
to their diffuseness and high angular momenta~\cite{KIMURA_RSI82_013108,MONTENEIRO_JPB23_427}, are another promising candidate for
analogous effects.

In what follows, we report a computational study of the chemical
bonding of the helium dimer. Potential-energy surfaces are mapped for
low-lying states of singlet, triplet, and quintet total spin, subject to strong magnetic
fields of arbitrary orientation. We use a finite-field approach,
where the magnetic-field effects are incorporated directly without
perturbative approximations. Although higher-order perturbation theory
is sometimes an alternative to probe high-field
effects~\cite{PAGOLA_CPL400_133,PAGOLA_JCP120_9556,PAGOLA_PRA72_033401,PAGOLA_JCTC5_3049,VAARA_CPL372_750,MANNINEN_PRA69_022503},
a nonperturbative approach is needed to study reliably 
potential-energy surfaces and level crossings in a strong field. To
handle the gauge-origin problem and ensure faster basis-set
convergence, we employ London atomic
orbitals~\cite{LONDON_JPR8_397,HAMEKA_MP1_203,DITCHFIELD_JCP65_3123,HELGAKER_JCP95_2595}. Without
a solution the gauge-origin problem, potential-energy surfaces suffer
from a spurious parabolic distance dependence and become qualitatively
wrong in a magnetic field. Unlike perturbative approaches, the present
non-perturbative approach necessitates an unconventional integral
evaluation scheme, such as the one reported for the {\sc London}
program package~\cite{TELLGREN_JCP129_154114,LondonProgram} or the
subsequent approaches in the {\sc Bagel}~\cite{REYNOLDS_PCCP17_14280},
{\sc Quest}~\cite{IRONS_JCTC13_3636}, and {\sc ChronusQ}~\cite{WILLIAMSYOUNG_WIRECMS,SUN_JCTC15_348} packages. For the smallest systems, the
extremely accurate free-complement method is also an
option~\cite{ISHIKAWA_CP401_62,NAKASHIMA_AJ725_528}.

The outline of this article is as follows. First, in
Sec.~\ref{secTHEORY}, we specify the electronic Hamiltonian and the
quantum-chemical model. We also introduce a new way to classify
electronic states and discuss perpendicular paramagnetic bonding
involving higher-angular-momentum states. Moreover, we discuss a simple
analytical model that gives insight into bonding in strong fields. In
Sec.~\ref{secRESULTS}, we present results for singlet,
triplet, and quintet states of the helium dimer in a strong magnetic
field. Finally, we summarize the conclusions in Sec.~\ref{secCONCL}.

\section{Theory}
\label{secTHEORY}

In the presence of a uniform magnetic field $\mathbf{B}$, the standard nonrelativistic Hamiltonian for $N$ electrons is 
in SI-based atomic units given by
\begin{equation}
  \hat{H} = \frac{1}{2} \sum_{j=1}^N \hat{\pi}_j^2 + \sum_{j=1}^N \mathbf{B}\cdot\hat{\mathbf{S}}_j + \sum_{j=1}^N v(\mathbf{r}_j) + \sum_{j<l} \frac{1}{r_{jl}}.
\end{equation}
where
$\hat{\mathbf{S}}_j$ is the spin operator for the $j$th electron, $v(\mathbf{r}_j)$ is the electrostatic potential from the nuclei at the position of the $j$th electron, $\hat{\boldsymbol{\pi}}_j = -\mathrm i\nabla_j + \mathbf{A}(\mathbf{r}_j)$ is the mechanical momentum operator, to be distinguished from the canonical momentum operator $\hat{\mathbf{p}}_j=-i\nabla_j$, and $\mathbf{A}(\mathbf{r}_j)$ is the magnetic vector potential at $\mathbf r_j$. Restriction of the vector potential to the linear form $\mathbf{A}(\mathbf{r}) = \tfrac{1}{2} \mathbf{B} \times (\mathbf{r}-\mathbf{G})$ reduces the gauge freedom to the position of the gauge origin $\mathbf{G}$. 

An efficient way to handle this gauge-origin freedom is to use London atomic orbitals~\cite{LONDON_JPR8_397,HAMEKA_MP1_203,DITCHFIELD_JCP65_3123,HELGAKER_JCP95_2595}, 
leading to gauge-origin invariant results and faster basis-set convergence; see Ref.~\cite{TELLGREN_JCP139_164118} for a more general perspective.
Given a Gaussian-type orbital $\chi(\mathbf{r})$ centred at $\mathbf{C}$, 
the corresponding London atomic orbital is $\omega(\mathbf{r}) = \mathrm e^{-\mathrm i\mathbf{A}(\mathbf{C})\cdot\mathbf{r}} \chi(\mathbf r)$. 
Hence, $\omega$ is product of a Gaussian and a plane wave with wave vector $\mathbf{q}=\mathbf{A}(\mathbf{C})$. The resulting nonstandard integrals, 
including the two-electron four-centre Coulomb integrals, are evaluated using the {\sc London} program~\cite{TELLGREN_JCP129_154114,LondonProgram}. 
This program package also contains a number of electronic structure models~\cite{LANGE_S337_327,TELLGREN_JCP140_034101,FURNESS_JCTC11_4169,STOPKOWICZ_JCP143_074110,SEN_JCTC15_3974}. 
We here use the full configuration-interaction (FCI) model~\cite{LANGE_S337_327} to be able to handle exact degeneracies and quasidegeneracies that inevitably arise when parameters such as bond distances and external magnetic fields are varied over large intervals.

\subsection{Classification of states using an approximately quantized
  angular momentum}

In the present section, we shall not be concerned with the spin contribution to angular momentum. For a given state $\Psi$, the gauge-invariant, 
physical angular momentum relative to a point $\mathbf{D}$ may then be defined as $\mathbf{J}_{\mathbf{D}} = \bra{\Psi} \sum_j (\mathbf{r}_j-\mathbf{D})\times\hat{\boldsymbol{\pi}}_j \ket{\Psi}$.
  In fact, since $\bra{\Psi} \sum_j \hat{\boldsymbol{\pi}}_j \ket{\Psi}$ vanishes in the complete basis-set limit for any variationally optimized state, the physical angular momentum is independent of the reference point. 
The gauge-dependent, canonical angular momentum is likewise given by the expectation value $\mathbf{L}_{\mathbf{D}} = \bra{\Psi} \sum_j (\mathbf{r}_j-\mathbf{D})\times\hat{\mathbf{p}} \ket{\Psi}$.
Introducing the density and paramagnetic current density,
\begin{align}
  \rho(\mathbf{r}) & = \sum_{j=1}^N \bra{\Psi} \delta(\mathbf{r}-\mathbf{r}_j) \ket{\Psi}, \\
    \mathbf{j}_{\mathrm{p}}(\mathbf{r}) & = \frac{1}{2} \sum_{j=1}^N \bra{\Psi} \delta(\mathbf{r}-\mathbf{r}_j) \hat{\mathbf{p}}_j + \hat{\mathbf{p}}_j \delta(\mathbf{r}-\mathbf{r}_j) \ket{\Psi},
\end{align}
the canonical momentum can also be calculated as 
$\mathbf{L}_{\mathbf{D}} = \int (\mathbf{r}-\mathbf{D})\times\mathbf{j}_{\mathrm{p}} \mathrm d\mathbf{r}$. 
Under a gauge transformation with gauge function $f$, we have $\mathbf{A} \mapsto \mathbf{A} + \nabla f$, $\mathbf{j}_{\mathrm{p}} \mapsto \mathbf{j}_{\mathrm{p}} - \rho \nabla f$, and $\mathbf{L}_{\mathbf{D}} \mapsto \mathbf{L}_{\mathbf{D}} - \int (\mathbf{r}-\mathbf{D})\times \rho \nabla f \mathrm d\mathbf{r}$.
Despite its gauge dependence, the canonical angular momentum is sometimes useful for classifying states. 

When both the electrostatic potential and the magnetic vector potential are cylindrically symmetric, the component of $\mathbf{L}_{\mathbf{G}}$ parallel to the symmetry axis is a good quantum number. In
general, for a diatomic molecule in a non-parallel magnetic field,
canonical momentum ceases to be a good quantum number---also the dissociation limit, since the total system is not
cylindrically symmetric even though cylindrical symmetry is restored
for the individual subsystems (dissociated atoms). Unlike the physical
angular momentum, the canonical momentum depends on a global reference
position. To restore quantization in the dissociation limit, the angular
momentum of a subsystem instead needs to be evaluated with respect to the
symmetry centre of that subsystem, and the wave function must be gauge transformed to correspond to what is obtained in a calculation with gauge origin adapted to the subsystem.

We now consider the idealized case where each isolated subsystem $\subidx$ is
cylindrically symmetric about its electronic centre of mass $\mathbf{C}_{\subidx}$. In a calculation
of the isolated system, with the gauge origin placed at
$\mathbf{C}_{\subidx}$, the resulting density $\rho_{\subidx}$ and paramagnetic current density
$\mathbf{j}'_{\mathrm{p};\subidx}$ are cylindrically symmetric too. Moreover,
the canonical angular momentum relative to $\mathbf{C}_{\subidx}$ is
\begin{equation}
  \mathbf{L}'_{\subidx} = \int (\mathbf{r}-\mathbf{C}_{\subidx}) \times
  \mathbf{j}'_{\mathrm{p};\subidx}(\mathbf{r}) \, \mathrm d\mathbf{r},
\end{equation}
and the component parallel to $\mathbf{B}$ is quantized. In the limit
of a complete basis, the mechanical linear momentum must vanish for any energy eigenstate. Using the fact that $\mathbf{C}_{\subidx}$ is the subsystem centre of mass, we see that the paramagnetic and diamagnetic contributions must vanish separately,
\begin{align}
  \label{eqBasLimPvanish}
  \boldsymbol{\pi}_{\subidx} &= \int \! \big( \mathbf{j}'_{\mathrm{p};\subidx}(\mathbf{r}) +
  \frac{1}{2} \rho_{\subidx}(\mathbf{r}) \, \mathbf{B} \times (\mathbf{r}-\mathbf{C}_{\subidx})
  \big) \mathrm d\mathbf{r} \nonumber \\ &= \int \! \mathbf{j}'_{\mathrm{p};\subidx}(\mathbf{r}) \, \mathrm d\mathbf{r} = \mathbf{0}.
\end{align}
Next, consider the total system. The gauge origin $\mathbf{G}$ cannot
coincide with all subsystem centres $\mathbf{C}_{\subidx}$. Hence, the subsystem
paramagnetic current densities obtained from a calculation on the total systems are gauge
transformed according to
\begin{equation}
  \mathbf{j}_{\mathrm{p};\subidx}(\mathbf{r}) =
  \mathbf{j}'_{\mathrm{p};\subidx}(\mathbf{r}) + \frac{1}{2}
  \rho_{\subidx}(\mathbf{r}) \, \mathbf{B}\times(\mathbf{G}-\mathbf{C}_{\subidx}).
\end{equation}
The subsystem contribution to the total angular momentum about a
global reference point $\mathbf{D}$ thus becomes
\begin{equation}
  \mathbf{L}_{\mathbf{D},\subidx} = \int \! (\mathbf{r}-\mathbf{D}) \times
  \mathbf{j}_{\mathrm{p};\subidx}(\mathbf{r}) \, \mathrm d\mathbf{r},
\end{equation}
or, using the relations established above,
\begin{align}
  \mathbf{L}_{\mathbf{D},\subidx} = \int \!&\Big(  (\mathbf{r}-\mathbf{D}) \times
  \mathbf{j}'_{\mathrm{p};\subidx}(\mathbf{r})   \Big. \nonumber \\ &+ \Bigr. \frac{1}{2}
  \rho_{\subidx}(\mathbf{r}) \, (\mathbf{r}-\mathbf{D})
  \times (\mathbf{B}\times(\mathbf{G}-\mathbf{C}_{\subidx})) \Big) \, \mathrm d\mathbf{r}.
\end{align}
Writing $N_{\subidx} = \int \! \rho_{\subidx}(\mathbf{r}) \mathrm d\mathbf{r}$ for the number of
electrons in a subsystem and using Eq.\,\eqref{eqBasLimPvanish}, we obtain in the basis-set limit
\begin{equation}
  \mathbf{L}_{\mathbf{D},\subidx} = \mathbf{L}'_{\subidx} +  \frac{N_{\subidx}}{2} (\mathbf{C}_{\subidx}-\mathbf{D})
  \times (\mathbf{B}\times(\mathbf{G}-\mathbf{C}_{\subidx})).
\end{equation}
Whereas the total canonical angular momentum
\begin{equation}
  \mathbf{L}_{\mathbf{D}} = \sum_{\subidx} \mathbf{L}_{\mathbf{D},\subidx}
\end{equation}
exhibits a gauge-dependent quadratic growth with the
distances $|\mathbf{C}_{\subidx}-\mathbf{C}_{\subidxB}|^2$ between different
subsystems, we can now subtract the quadratic terms  to obtain
\begin{align}
  \boldsymbol{\Lambda} &= \sum_{\subidx} \Big( \mathbf{L}_{\mathbf{D},\subidx} - \frac{N_{\subidx}}{2} (\mathbf{C}_{\subidx}-\mathbf{D})
  \times (\mathbf{B}\times(\mathbf{G}-\mathbf{C}_{\subidx})) \Big) \nonumber \\ &= \sum_{\subidx} \mathbf{L}'_{\subidx}.
\end{align}
We term $\boldsymbol{\Lambda}$ the {\it approximately quantized angular
  momentum} (AQAM) since, for a diatomic molecule, its projection
$\Lambda_{\mathbf{B}} = \mathbf{e}_{\mathbf{B}} \cdot \boldsymbol{\Lambda}$ onto the field direction $\mathbf{e}_{\mathbf{B}} = \mathbf{B}/|\mathbf{B}|$ exhibits exact
quantization for all parallel orientations as well as in the dissociation
limit. In other cases,
$\Lambda_{\mathbf{B}}$ is often approximately quantized, despite the
presence of interactions between subsystems. This quantity therefore
provides a useful generalization of the atomic quantum number $m_l$
for classifying the states of a diatomic molecule. A closely related quantity was considered for a different purpose (and with different notation) in a formal density-functional context in Sec.~IV.C of Ref.~\cite{TELLGREN_JCP148_024101}.

Finally, we remark that some care is required when interpreting $\frac{1}{2} \mathbf{B}\cdot\boldsymbol{\Lambda}$ as an energy. The physical angular momentum is a sum of two terms: the canonical angular momentum and the diamagnetic contribution. However, the gauge invariant kinetic energy is a sum of three terms: the canonical kinetic energy, the orbital Zeeman term, and the diamagnetic term. Only one of these terms (and the sum of the other two) can be modified to have a well-defined dissociation limit, not all three simultaneously.

\subsection{Symmetry properties of molecular orbitals}

In a magnetic field, the point-group symmetry of He$_2$ is lower than the symmetry D$_{\infty \text h}$ of the molecule
in the absence of a field. In all field orientations, inversion symmetry exists and the molecule therefore belongs to the
C$_\text i$ point group with the irreps A$_\text g$ and A$_\text u$. In the parallel and perpendicular field orientations 
additional symmetry operations exist. In the parallel orientation, rotation about the molecular axis give rise to the C$_{\infty \text h}$ point
group with the one-dimensional irreps $\Sigma_\text{g}$ and $\Sigma_\text u$ and the two-dimensional irreps $\Pi_\text{g}$, $\Pi_\text u$, $\Delta_\text{g}$, 
$\Delta_\text u$,\dots. The C$_{\infty \text h}$ symmetry group (which does not occur for molecules in the absence of a magnetic field)
differs from D$_{\infty \text h}$ by the absence of vertical mirror planes and two-fold perpendicular axes. Finally, in
the perpendicular field orientation, we have in addition to inversion symmetry a two-fold symmetry axis along the field direction, giving rise to 
the C$_{2 \text h}$ symmetry group with the A$_\text{g}$, A$_\text u$, B$_\text{g}$ and B$_\text u$ irreps.

In Table~\ref{tabsym}, we compare
the symmetries of the molecular orbitals in the field orientations and also with the atomic orbitals in the united atoms limit.
\begin{table}
\caption{\label{tabsym} Symmetries and bonding properties of molecular
  orbitals of homonuclear diatomic molecule in a magnetic field. The
  symbol $\angle$ here indicates an intermediate angle.}
\begin{center}
\begin{tabular}{ccccccc}
\hline\hline
D$_{\infty \text h}$&  C$_{\infty \text h}$ & C$_{2 \text h}$& C$_\text i$ &  united-atom& preferred & chemical \\
$B=0$&  $B_\parallel$ & $B_\perp$ & $B_\angle$ &  limit& orientation& bonding \\
\hline
$\sigma^+_\text g$ &$\sigma_\text g$      &  a$_{\text{g}}$      & a$_{\text{g}}$       & s           & $\parallel$ & covalent  \\
$\sigma^+_\text u$ &$\sigma_\text u$ &  b$_{\text{u}}$ & a$_{\text{u}}$  & p$_0$       & $\perp$     & magnetic \\
$\pi_\text u$      & $\pi_\text u$        &  $\text{a}_{\text{u}} + \text{b}_{\text{u}}$  & a$_{\text{u}}$       & p$_{\pm 1}$ & $\parallel$ & covalent \\
$\pi_\text g$      &$\pi_\text g$    &  $\text{a}_{\text{g}} + \text{b}_{\text{g}}$ & a$_{\text{g}}$  & d$_{\pm 1}$ & $\angle$    & magnetic \\
$\delta_\text g$   & $\delta_\text g$     &  $\text{a}_{\text{g}} + \text{b}_{\text{g}}$      & a$_{\text{g}}$       & d$_{\pm 2}$ & $\parallel$ & covalent \\
$\delta_\text u$  &$\delta_\text u$  &  $\text{a}_{\text{u}} + \text{b}_{\text{u}}$ & a$_{\text{u}}$  & f$_{\pm 2}$ & $\angle$    & magnetic \\
\hline
\hline
\end{tabular}
\end{center}
\end{table}

\subsection{Perpendicular paramagnetic bonding}
\label{secParaBond}

Strong magnetic fields can lead to new exotic bonding
mechanisms. Previous work has established that the normally unbound
lowest triplet state of the H$_2$ becomes bound in a perpendicular magnetic field of strength on
the order of $B \sim B_0$. Also the lowest singlet state of the
He$_2$ molecule becomes substantially stabilized and the equilibrium
bond length substantially compressed in a perpendicular field. The
underlying bonding mechanism, termed {\it perpendicular paramagnetic
  bonding}, is that the antibonding $\sigma_{\text{u}}^*$ orbital develops an
angular momentum, which leads to an energetic stabilization by the
orbital Zeeman effect~\cite{LANGE_S337_327}. This is true even in a
minimal basis of only s orbitals, provided that they are equipped with
London gauge factors. The magnitude of the angular momentum of
$\sigma^*_\text u$ vanishes in the parallel orientation and the net
energetic effect is largest in
the perpendicular orientation at intermediate bond lengths. By
contrast, the bonding $\sigma_{\text{g}}$ orbital does not develop an angular
momentum and is not stabilized by this mechanism.

The above considerations generalize and apply in a somewhat stronger
form to higher angular-momentum states. In the parallel orientation,
a linear combination of atomic orbitals with atomic quantum numbers
$|m_l| \leq M$ can never lead to an angular momentum exceeding $M$. By
contrast, this becomes possible in nonparallel orientations. For
example, the cc-pVDZ basis has two s orbitals and three p orbitals for
each helium atom. In a dimer with the helium atoms placed on the $x$ axis
at $(\pm 1,0,0)$~bohr, one finds by diagonalizing the canonical
angular-momentum operator (relative to the mid-bond position) that the
largest perpendicular components are $L_{\mathbf{0};z} = \pm 1.83\hbar$. If we omit the 1s orbitals, the 2s orbitals, and  both the 1s and 2s orbitals on the two atoms, we obtain 
$L_{\mathbf{0};z} = \pm 1.35\hbar$, 
$L_{\mathbf{0};z} = \pm 1.06\hbar$ and $L_{\mathbf{0};z} = \pm 1.03\hbar$, respectively. With London gauge factors and a
perpendicular field $\mathbf{B} = 0.5 \hbar \mathbf{e}_z$, the most negative
eigenvalue becomes $L_{\mathbf{0};z} = -2.08\hbar$. Hence,
some combination of s  and p orbitals acquires a d-orbital
character when the orientation is changed from parallel to
perpendicular, leading to a lower orbital Zeeman energy,
which competes with the diamagnetic energy.

In light of the visual similarity of antisymmetric combinations of
{\it real-valued} p orbitals to real valued d orbitals, it may be
surprising that s functions play such a large role in the above
example---for example, an antisymmetric linear combination of two p$_y$
orbitals centred at different points on the $x$ axis 
resembles a $\mathrm d_{xy}$ orbital. We remark, however, that the canonical angular momentum
vanishes for all real valued orbitals and the s functions are needed
to produce complex-valued orbitals of the right form to represent an angular
momentum of about $\pm 2 \hbar$.

A simple analytical model provides further insight into magnetic-field
effects on bonding and
antibonding orbitals in homonuclear diatomic molecules.
Let $G_{\ell m}(\mathbf r, \alpha, \mathbf K,\mathbf B)$ denote a solid-harmonic Gaussian orbital of exponent $\alpha$ centred at $\mathbf K$ and equipped with
a London phase factor for the magnetic field $\mathbf B$:
\begin{align}
G_{\ell m}\left(\mathbf r, \alpha, \mathbf K, \mathbf B\right) = & c_{\ell,m}(\alpha)\exp\left(- \mathrm i (\tfrac{1}{2} \mathbf B \times \mathbf K) \cdot \mathbf r \right)  \nonumber \\ 
& \times S_{\ell m}(\mathbf r_K) \exp\left(- \alpha \vert \mathbf r \vert^2_K\right).
\end{align}
Here $c_{\ell,m}(\alpha)$ is a normalization constant and
$S_{\ell m}(\mathbf r_K)$ with $\mathbf r_K =  \mathbf r - \mathbf K$ 
is a solid-harmonic function centred at $\mathbf K$
of angular-momentum quantum numbers $\ell$ and $m$ about the z axis. 

Consider now the normalized bonding and antibonding orbitals along the z axis:
\begin{align}
&g_{\ell m}^\pm(\mathbf r, \alpha,\delta, \mathbf B) = C_{\ell m}(\delta, \alpha)  \times \nonumber \\ &\times \left( 
G_{\ell m}(\mathbf r, \alpha, (0,0,+\delta), \mathbf B)  \pm G_{\ell m}(\mathbf r, \alpha, (0,0,-\delta), \mathbf B) \right),
\end{align}
where $C_{\ell m}(\delta,\alpha)$ is a normalization constant.
We are interested in the united-atom limits of these orbitals,
\begin{align}
&G_{\ell m}^\pm(\mathbf r, \alpha, \mathbf B) = \lim_{\delta \to 0^+}  g_{\ell m}^\pm(\mathbf r, \alpha, \delta, \mathbf B).
\end{align}
Clearly, for the bonding orbitals, we have the field-free standard Gaussian orbital positioned at the origin,
\begin{align}
G_{\ell m}^+(\mathbf r, \alpha, \mathbf B) 
= G_{\ell m}(\mathbf r, \alpha),
\end{align}
in the notation $G_{\ell m}(\mathbf r, \alpha) = G_{\ell m}(\mathbf r, \alpha,\mathbf 0,\mathbf 0)$.
For the antibonding orbitals, the limit is less trivial. 
To illustrate, we consider the special case when the magnetic
field is oriented perpendicular to the bonding and antibonding orbitals $\mathbf B_x = (B,0,0)$. We furthermore set the Gaussian exponent
equal to the optimal exponent of a free electron in uniform magnetic field,
$\alpha_B = B/4$.  For $\ell \leq 1$, we then find
\begin{align}
G^-_{0,0}(\mathbf r, \alpha_B,\mathbf B_x) &= \sqrt{\tfrac{B}{2}}     (y - \mathrm i z) G_{0,0}(\mathbf r, \alpha_B), \label{gm00}\\
G^-_{1,\pm 1}(\mathbf r, \alpha_B,\mathbf B_x) &= \sqrt{\tfrac{B}{3}} (y - \mathrm i z) G_{1,\pm 1}(\mathbf r, \alpha_B), \label{gm11}\\
G^-_{1,0}(\mathbf r, \alpha_B,\mathbf B_x) &= \sqrt{\tfrac{B}{4}}     (y - \mathrm i z) G_{1,0}(\mathbf r, \alpha_B)  \nonumber \\ &\qquad\qquad\qquad \phantom{c}- G_{0,0}(\mathbf r, \alpha_B). \label{gm10}
\end{align}
Noting that
$[L_x, y - \mathrm i z] = - \hbar (y - \mathrm i z)$
we conclude that a magnetic field perpendicular to the antibonding orbital induces a component of the angular momentum about the field axis and 
perpendicular to the angular momentum about the bond axis. To first order, this will reduce the energy of the antibonding orbital in the united-atom limit
relative to the dissociation limit. It is a reasonable assumption that this stabilization of antibonding orbitals occurs at all atomic separations but is
stronger the closer the two atoms are to each other---that is, to the united-atom limit.

\begin{figure*}
  \includegraphics[width=\linewidth]{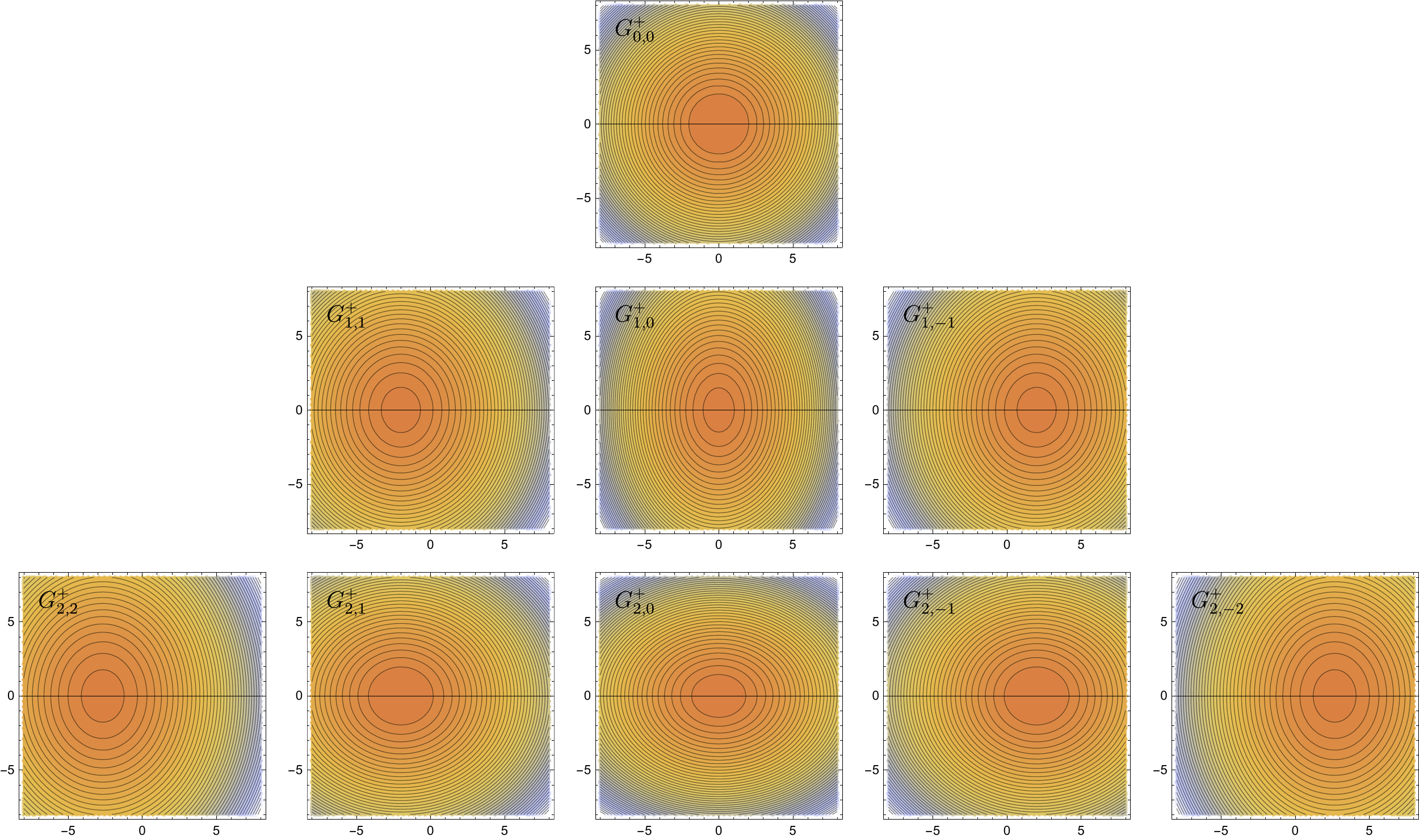}
  \includegraphics[width=\linewidth]{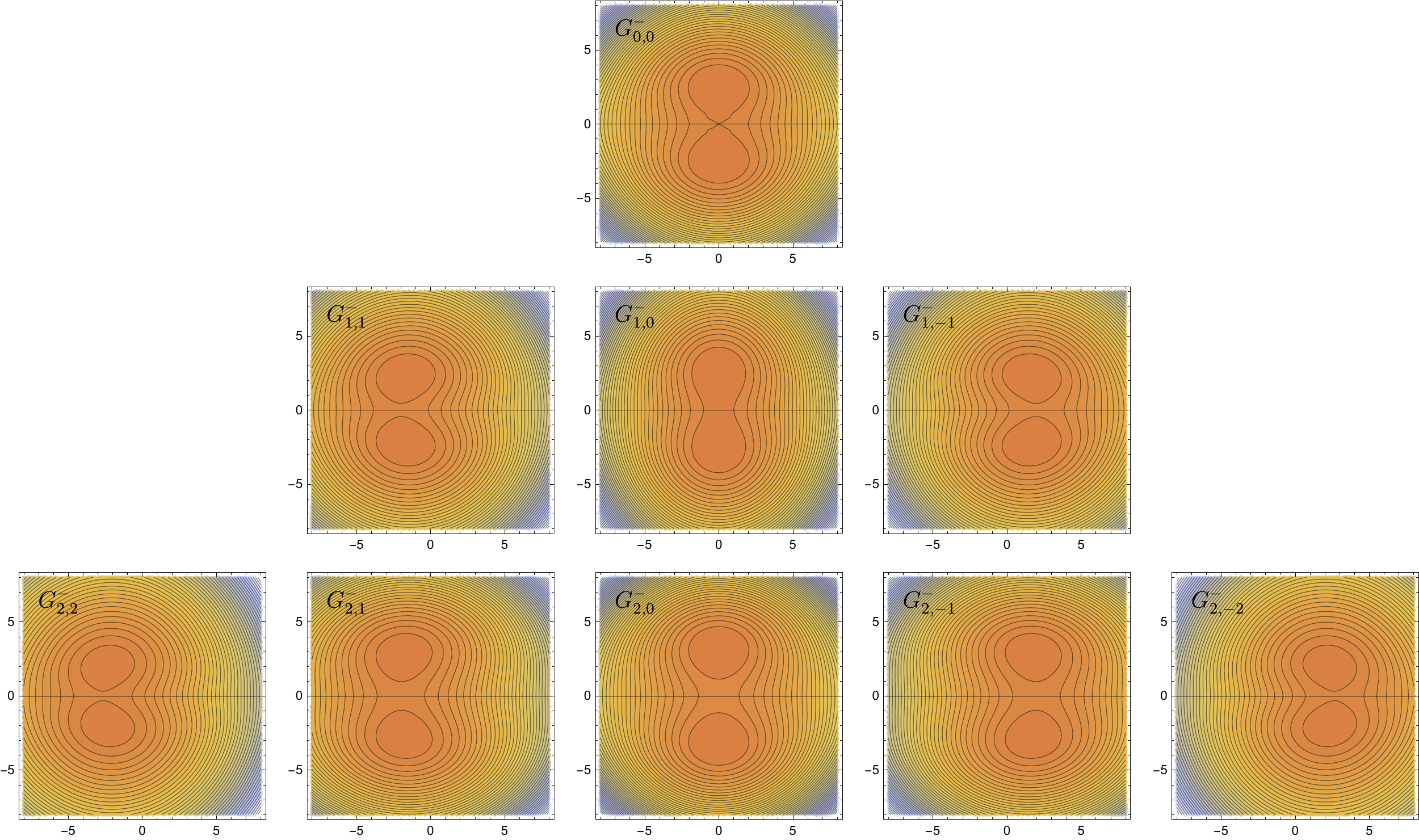}
  \caption{Contour plots of the kinetic energy of bonding and antibonding orbitals in the united-atom limit as a function of magnetic field strength $\mathbf B$ in the zx plane; with the
z axis marked by a horizontal line. For bonding orbitals, the minimum is located on the z axis; for the antibonding orbitals, the minima are located away from the z axis, symmetrically on
each side.}
  \label{fig:spd0}
\end{figure*}

The total kinetic energy and angular momentum of bonding and antibonding atomic orbitals in the united-atom limit with $\alpha = 1$ are given by
\begin{align}
T_{\ell m}(\mathbf B) &=  \int  \!\! G_{\ell m}^\pm(\mathbf r,1,\mathbf B)^\ast \tfrac{1}{2} \pi^2_\mathbf B  \, G^\pm_{\ell m}(\mathbf r,1,\mathbf B) \, \mathrm d \mathbf r,  \\
\mathbf L_{\ell m}(\mathbf B) &= \int  \!\! G_{\ell m}^\pm(\mathbf r,1,\mathbf B)^\ast \,  {\mathbf L}  \, G^\pm_{\ell m}(\mathbf r,1,\mathbf B) \, \mathrm d \mathbf r,
\end{align}
where $\frac{1}{2} \pi^2_\mathbf B$ is the kinetic-energy operator in the magnetic field and $\mathbf L$ the canonical angular-momentum operator. For a fixed Gaussian exponent $\alpha = 1$,
we have in Table~\ref{tabsorb} calculated the kinetic energy and angular momentum of the bonding and antibonding orbitals in the united-atom limit for zero field
$\mathbf B = \mathbf 0$ and for the magnetic field $\mathbf B = \mathbf B_\text{min}$ that minimizes the kinetic energy for $\alpha = 1$:
\begin{equation}
\mathbf B_{\ell m} = \argmin_{\mathbf{B}}  T_{\ell m}(\mathbf B).
\end{equation}
In the table, we have also listed $B_{\ell m} = \vert \mathbf B_{\ell m} \vert$ and the angle $\theta_{\ell m}$ of $\mathbf B_{\ell m}$ with the z axis (bond axis).
We note that, for a free electron in a magnetic field, the optimal Gaussian exponent is $B/4$;
for an electron in an atom or molecule up to field strengths of about $B_0$, the electronic wave function responds less directly to the magnetic field strength. 
In Fig.\,\ref{fig:spd0}, we have plotted the kinetic energy of the bonding and antibonding orbitals in the united-atom limit in the xz plane. 

\begin{table}
\renewcommand{\arraystretch}{1.3}
\addtolength{\tabcolsep}{2pt}
\caption{Kinetic energy and angular momentum of bonding and antibonding atomic orbitals in the united-atom limit $G^\pm_{\ell,m}$ in a zero magnetic field $\mathbf B = \mathbf 0$ and
in the minimizing magnetic field $\mathbf B = \mathbf B_{\ell m}$. Here
$\Delta T_{\ell m}(\mathbf B_{\ell m}) = T_{\ell m}(\mathbf B_{\ell m}) - T_{\ell m}(\mathbf 0)$ and the components of the angular momentum not listed are zero. Units
are $E_\text h$ for energy, $\hbar$ for angular momentum, and $B_0$ for magnetic field strength.
\label{tabsorb}}
\begin{center}
\scalebox{0.85}{
\begin{tabular}{c|cc|cccc}
\hline\hline
& $T_{\ell m}(\mathbf 0)$ & $L_{\ell m}^z(\mathbf 0)$ & $\Delta T_{\ell m}(\mathbf B_{\ell m})$ & $L_{\ell m}^x(\mathbf B_{\ell m})$ & $B_{\ell m}$ & $\theta_{\ell m}$ \\
\hline
$G^+_{0,0\phantom{\pm}}$  & 3/2 & $\phantom{\pm}0$ & $ 0$   & $0$ & 0   & $\checkmark$ \\
$G^+_{1,\pm 1}$             & 5/2 & $\pm 1$          & $-\tfrac{1}{2}$ & $0$ & 2   & $(90 \pm 90)^\circ$\\
$G^+_{1,0\phantom{\pm}}$             & 5/2 & $\phantom{\pm}0$ & $ 0$   & $0$ & 0   & $\checkmark$ \\
$G^+_{2,\pm 2}$& 7/2 & $\pm 2$          & $-\tfrac{4}{3}$ & $0$ & $\tfrac{8}{3}$ & $(90 \pm 90)^\circ$ \\
$G^+_{2,\pm 1}$& 7/2 & $\pm 1$          & $-\tfrac{1}{2}$ & $0$ & 2   & $(90 \pm 90)^\circ$\\
$G^+_{2,0\phantom{\pm}}$& 7/2 & $\phantom{\pm}0$ & $0$    & $0$ & 0   & $\checkmark$\\
\hline
$G^-_{0,0\phantom{\pm}}$& 5/2 & $\phantom{\pm}0$ & $-0.3$ & $-0.9$ & $2.6$ & $90^\circ$ \\
$G^-_{1,\pm 1}$& 7/2 & $\pm 1$          & $-1.2$ & $-1.0$ & $2.9$ & $(90 \pm 35)^\circ$\\
$G^-_{1,0\phantom{\pm}}$& 7/2 & $\phantom{\pm}0$ & $-0.2$ & $-0.8$ & $2.7$ & $90^\circ$\\
$G^-_{2,\pm 2}$& 9/2 & $\pm 2$          & $-2.1$ & $-1.1$ & $3.1$ & $(90 \pm 48)^\circ$ \\
$G^-_{2,\pm 1}$& 9/2 & $\pm 1$          & $-1.2$ & $-1.2$ & $3.4$ & $(90 \pm 32)^\circ$\\
$G^-_{2,0\phantom{\pm}}$& 9/2 & $\phantom{\pm}0$ & $-0.7$ & $-1.4$ & $3.2$ & $90^\circ$\\
\hline
\hline
\end{tabular}}
\end{center}
\end{table}

The first two columns in Table~\ref{tabsorb} contains information about the energies and angular momentum in the absence of a magnetic field. The kinetic energy
is $(\ell + 3/2)E_\text h$ and $(\ell + 5/2)E_\text h$ for bonding and antibonding orbitals, respectively, the higher energy of the antibonding arising from
the presence of an additional nodal plane in the orbital. 

Turning our attention to
the orbitals in the minimizing magnetic field $\mathbf B_{\ell m}$, we note that 
\begin{equation}
\Delta T_{\ell m}(\mathbf B_{\ell m}) = T_{\ell m}(\mathbf B_{\ell m}) - T_{\ell m}(\mathbf 0)
\end{equation}
is zero or negative. Furthermore, the only orbitals whose global energy minimum occurs at
zero field are the bonding orbitals with $m = 0$. For bonding orbitals with $m <0 $, the energy is lowered 
by applying a field parallel with the quantization axis (bond axis);
if $m > 0$, the same minimum energy is obtained by applying a magnetic field of the same magnitude but in the opposite direction. As expected, the energy minimum becomes deeper and
the minimizing field stronger with increasing value of $\vert m \vert$. We note that the energy minimization of the bonding orbitals is the same in the united-atom and dissociation
limits, being associated with the permanent angular momentum in the system. 
We also note that, in a sufficiently strong field, the energy of the orbitals will increase diamagnetically, 
for all field orientations 

For antibonding orbitals in the united-atom limit, the energy is in all cases reduced by the magnetic field and in all cases significantly more than for the corresponding bonding orbitals.
At the same time, an angular momentum is induced in the direction of the magnetic field, as predicted from~Eqs.\,\eqref{gm00}--\eqref{gm10}. The resulting total angular momentum is then
no longer parallel to the bond axis and the minimizing magnetic field is no longer parallel or antiparallel to the z axis. Indeed, for orbitals with $m = 0$, the preferred field
orientation is perpendicular to the bond axis, while for orbitals with $m \neq 0$, the preferred field orientation is skewed relative to the bond axis. We note that the energy lowering
arising from the induced angular momentum vanishes in the dissociation limit, unlike the energy lowering arising from the permanent angular momentum.

\section{Results}
\label{secRESULTS}

The spectrum of He$_2$ depends on the bond length $R$, the strength of
the magnetic field $B$, and the angle $\theta$ between the field and
the bond axis. We have employed the {\sc London}
program~\cite{TELLGREN_JCP129_154114,LondonProgram}  to map out the
spectrum as a function of these parameters. Basis sets are denoted by
standard notation amended by prefixes `L' and `u' to indicate London
gauge factors and uncontracted functions, respectively. The calculations have been 
carried out at the
FCI/Lu-aug-cc-pVTZ level unless otherwise indicated. All bond
distances are reported in units of $a_0 = 1$~bohr.

\subsection{Dissociation limit: helium atom}

\begin{figure}[h]
  \includegraphics[width=\linewidth]{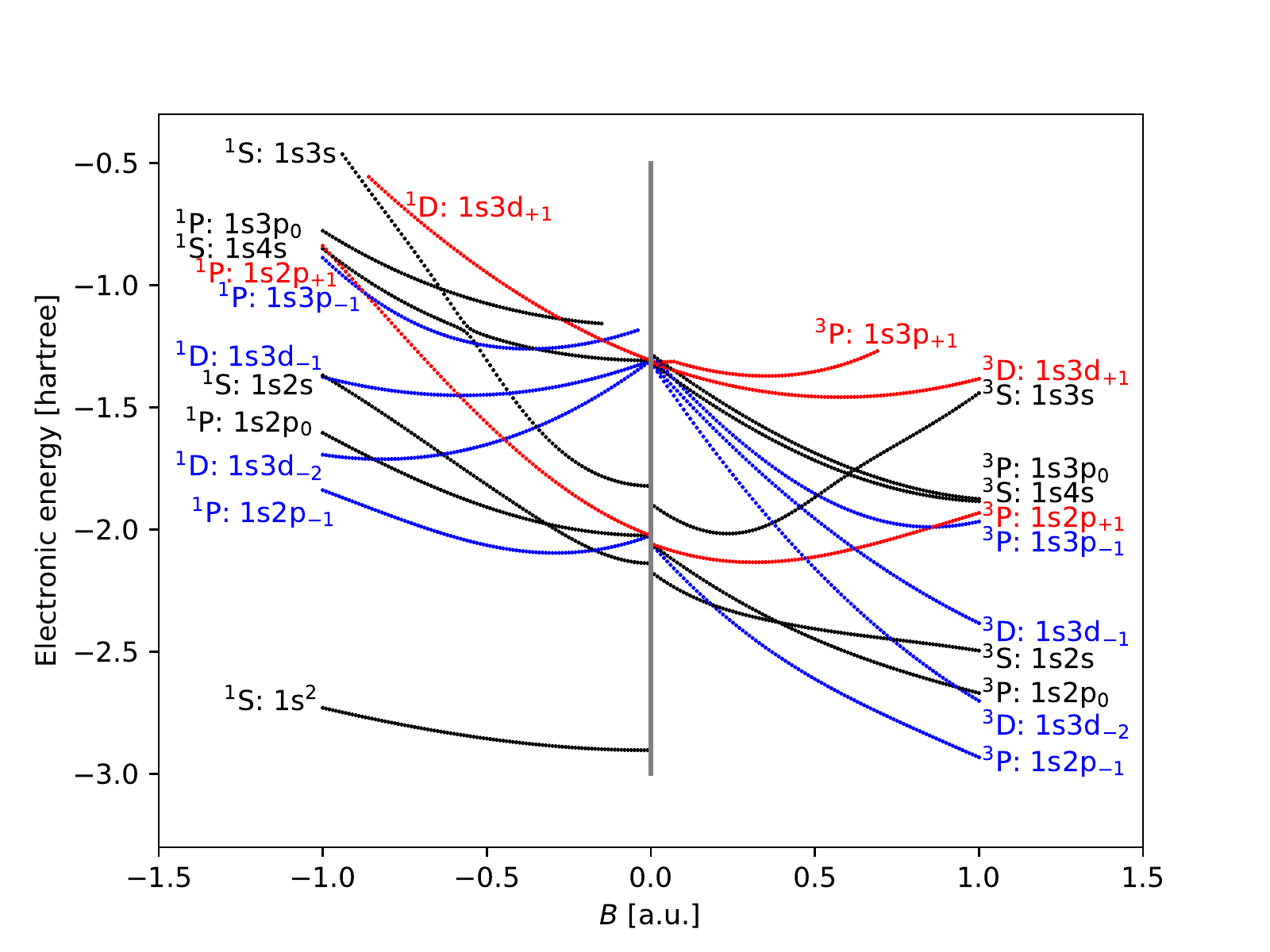}
  \caption{Spectrum of the helium atom as a function of magnetic field strength. Singlet states are shown along the negative horizontal axis and triplet states 
along the positive axis.}
  \label{figHeAtom}
\end{figure}

In the limit of an infinite bond distance, the helium dimer becomes two isolated helium atoms. 
The atomic spectrum, calculated at the FCI/Lu-aug-cc-pVQZ level, is shown in Fig.~\ref{figHeAtom}, with singlet- and triplet-state energies plotted along the negative and
positive axes, respectively.

While the diamagnetic $^{1}\Sigma_\text g(1\text s^2)$ singlet state is the lowest singlet in the plotted field interval (and also the
ground state up to about 0.8$B_0$), the remaining singlets in the plot undergo several level crossings. 
In particular, due to the orbital Zeeman interaction, the $^{1}\Pi_\text u(1\text s2\text p_{-1}$) state 
crosses the $^{1}\Sigma_\text g(1\text s2\text s)$ state to become the first excited singlet state at about 0.1$B_0$.  At a magnetic field 
strength of about $1B_0$, the 
paramagnetic ${}^1\Delta_\text g(1\text s3\text d_{-2})$ state has been sufficiently stabilized to become the second excited singlet state, 
having crossed in turn the four diamagnetic states
$^{1}\Sigma_\text g(1\text s3\text s)$, $^{1}\Pi_\text u(1\text s2\text p_{+1})$, $^{1}\Sigma_\text g(1\text s2\text s)$, and \smash{$^{1}\Sigma_\text u(1\text s2\text p_{0})$} with increasing field strength.

Most singlet states have analogues in the triplet spectrum. However, because of the
the spin Zeeman interaction, the triplet states are split, 
the $m_\text s=-1$ components (with two spin-down electrons) being stabilized 
more than the corresponding singlet states.  Additional stabilization may be provided by the orbital Zeeman interaction.
Thus, while the lowest triplet state is $^{3}\Sigma_{\text{g}}(1\text{s}\,2\text{s})$ in weak magnetic fields,
the $^{3}\Pi_{\text{u}}(1\text{s} \,2\text{p}_{-1})$ state becomes the lowest triplet at
$0.2B_0$ and the ground state at about $0.8B_0$.  
In even stronger fields, the ground state becomes $^{3}\Delta_{\text{g}}(1\text{s} \,3\text{d}_{-2})$, and so on.

\subsection{United-atom limit: beryllium atom}

\begin{figure}[h]
  \includegraphics[width=\linewidth]{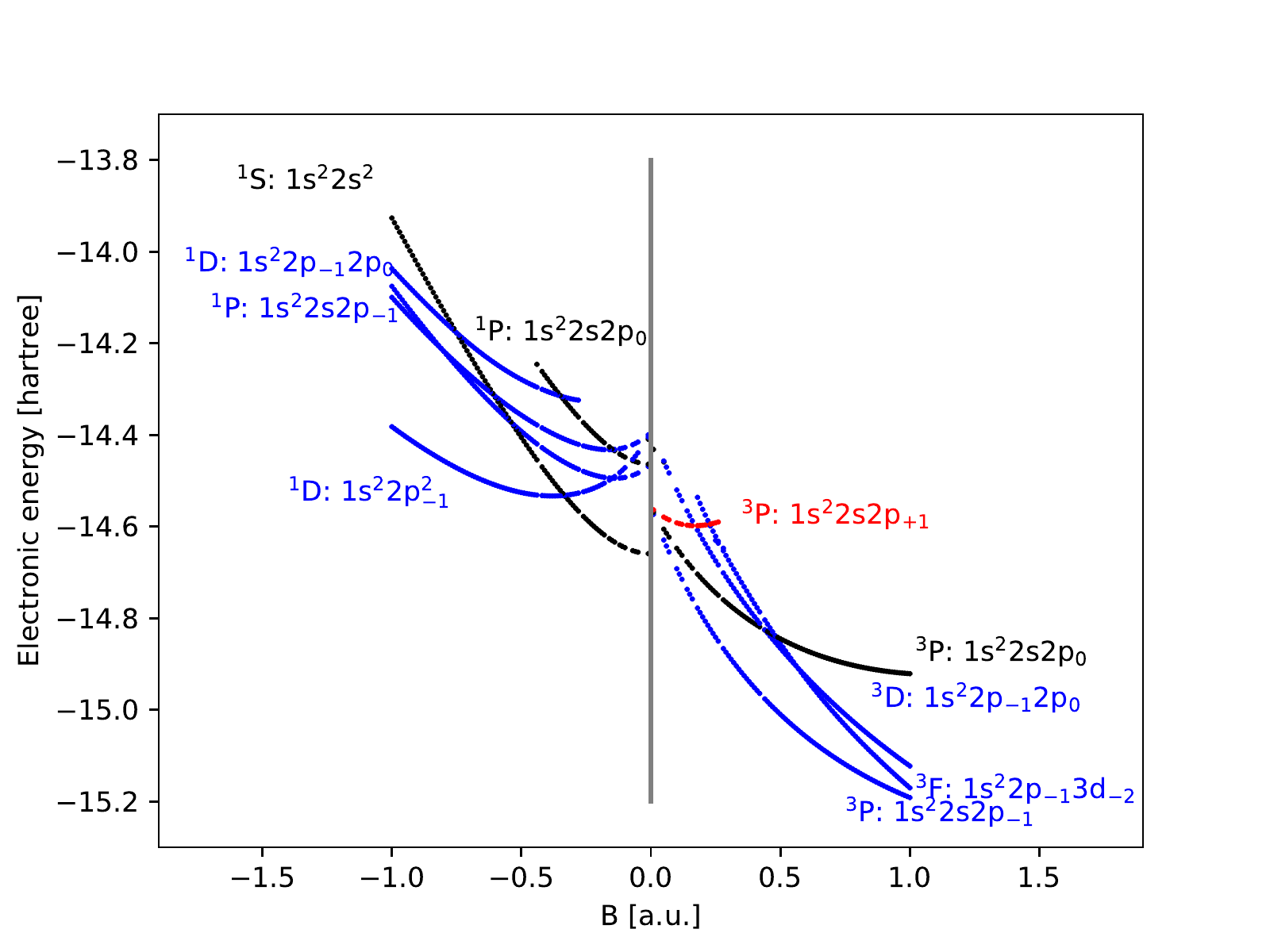}
  \caption{Spectrum of the beryllium atom as a function of magnetic field strength. Singlet and triplet states are shown along the negative 
and positive horizontal axes, respectively.}
  \label{figBeAtom}
\end{figure}

It is also instructive to consider the united-atom limit, in which the helium dimer becomes the beryllium atom. 
The corresponding spectrum, obtained at the FCI/Lu-cc-pVDZ level of theory, is shown in Fig.~\ref{figBeAtom}. Again, the Zeeman interactions 
result in a reordering of the spectrum. 
As the zero-field singlet ground state $^{1}\Sigma_\text g(1\text s ^22\text s^2)$ is increasingly destabilized by the magnetic field, 
the ${}^1\Delta_\text g(1\text s^22\text p_{-1}^2)$ state becomes 
the lowest singlet in the strongest fields plotted. 

However, because of the spin Zeeman interaction, the $m_s=-1$ triplet components 
are stabilized even faster. Indeed, already at about
$0.05B_0$, the ground state is ${}^3\Pi_{\text{u}}(1\text{s}^2 2\text{s}\,2\text{p}_{-1})$. 
For the strongest field strengths shown, the first excited state is ${}^3\Phi_{\text{u}}(1\text{s}^2 2\text{p}_{-1}3\text{d}_{-2})$,
which appears to become the ground state at a field strength slightly stronger than one atomic unit. 

\begin{figure}[h]
  \includegraphics[width=\linewidth]{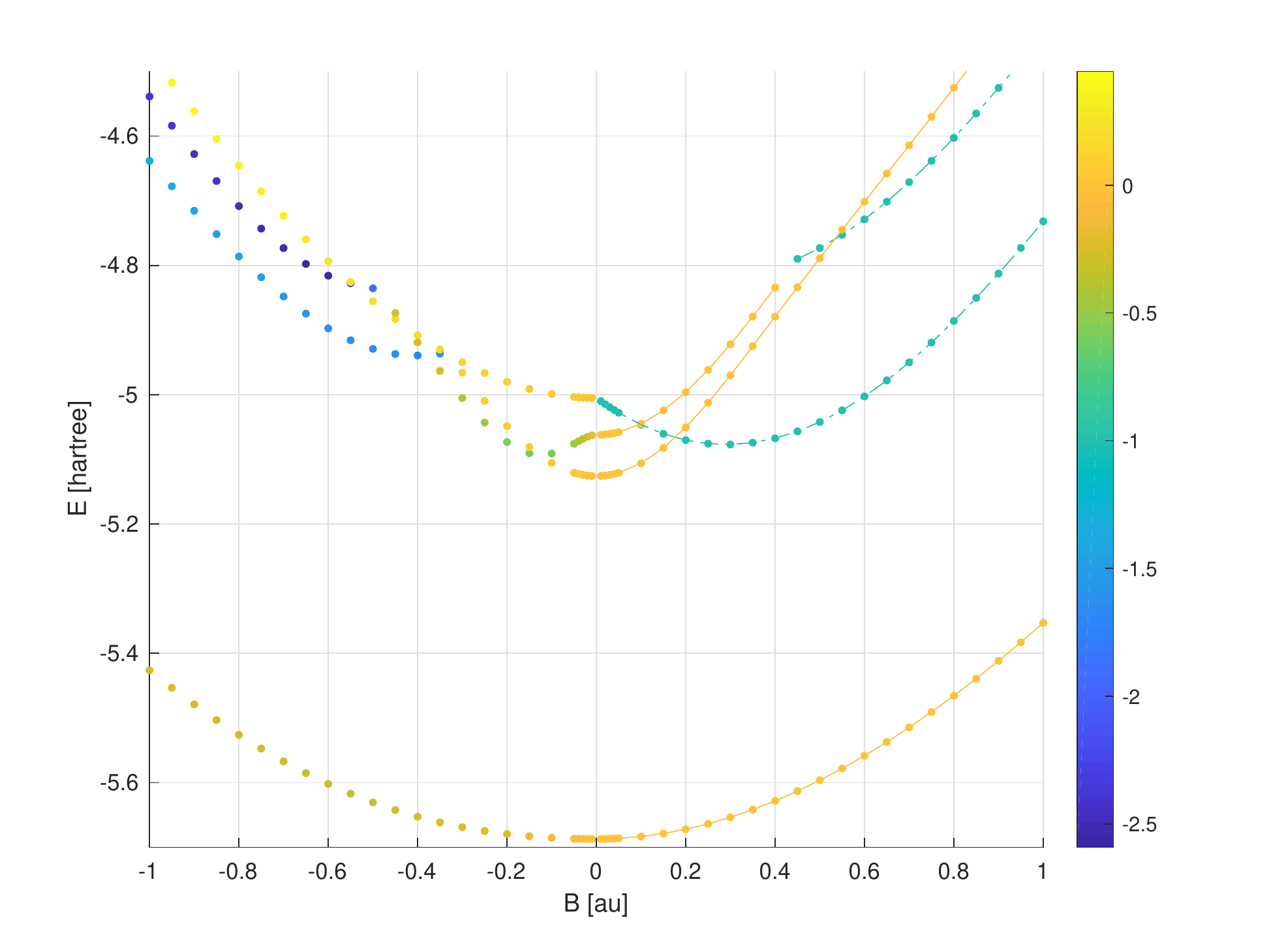}
  \caption{Singlet spectrum as a function of magnetic field for the
    He$_2$ molecule, with fixed bond distance $R = 2a_0$. Perpendicular (parallel) magnetic fields have been mapped to the
    negative (positive) half of the horizontal axis.}
  \label{figSingletEvsB}
\end{figure}

\subsection{States of He$_2$ at a fixed bond distance $R = 2a_0$}

\begin{figure}[h]
  \includegraphics[width=\linewidth]{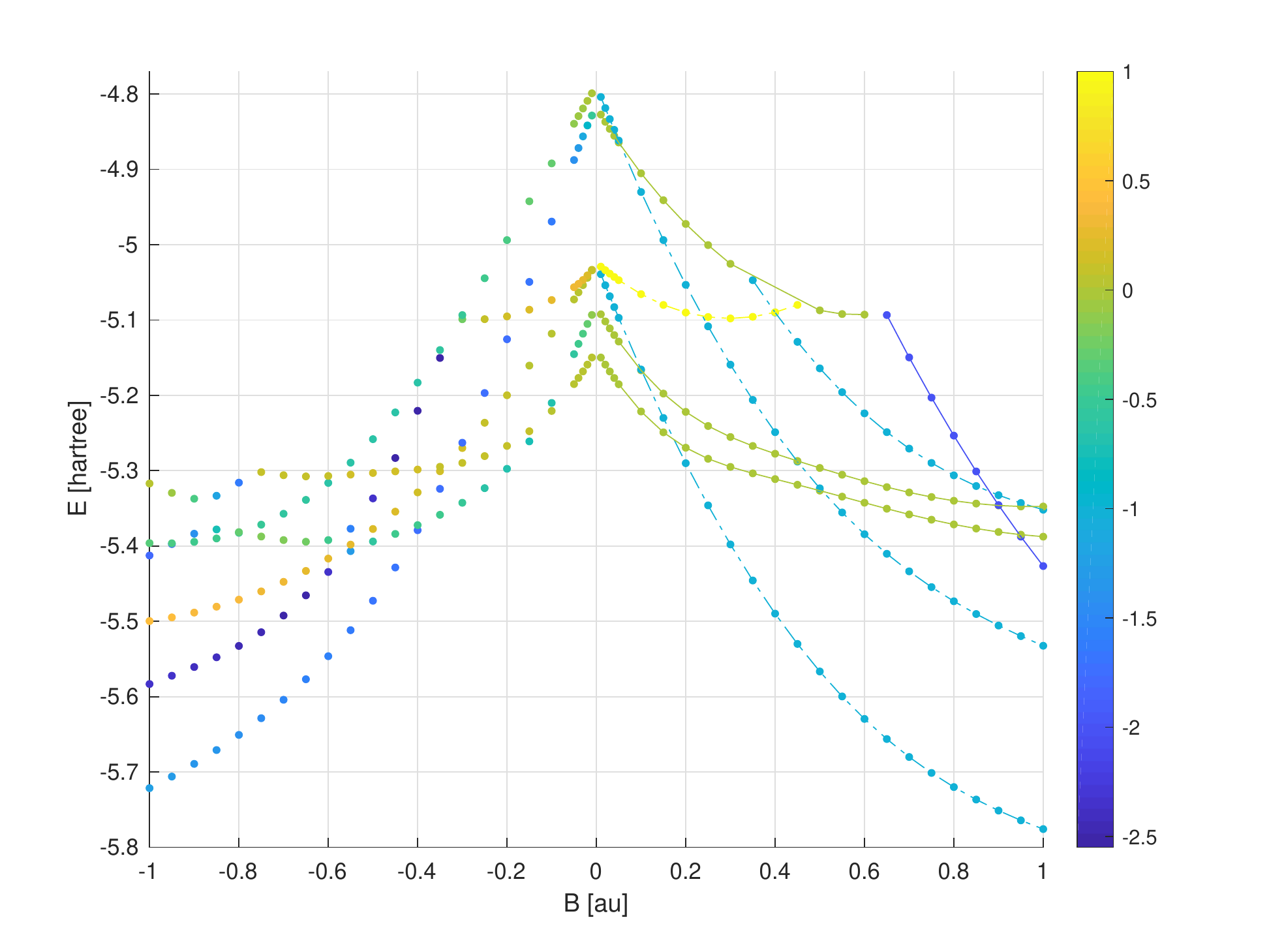}
  \caption{Triplet spectrum as a function of magnetic field for the
    He$_2$ molecule, with fixed bond distance $R = 2a_0$. Perpendicular (parallel) magnetic fields have been mapped to the
    negative (positive) half of the horizontal axis.}
  \label{figTripletEvsB}
\end{figure}

As several interesting minima in the dissociation curves appear at a He--He
bond distance of about $R= 2a_0$ (see below), it is instructive to consider the field dependence of the electronic spectrum at this fixed bond length.
In the following, we consider the singlet and triplet spectra of He$_2$ separately. The energies of singlet and triplet states are plotted in
Fig.~\ref{figSingletEvsB} and Fig.~\ref{figTripletEvsB}, respectively, with the energies in the parallel and perpendicular field orientations
plotted along the positive and negative axes, respectively. 

\subsubsection{Singlet states of He$_2$ at $R = 2a_0$}

In a parallel field, the four lowest singlet states at $R = 2a_0$ are
\smash{$^{1}\Sigma_\text g(\IIorbdouble{\sigma}{}{1s} \IIorbdouble{\sigma}{*}{1s})$}, 
\smash{$^{1}\Sigma_\text u(\IIorbdouble{\sigma}{}{1s} \IIorbNoSpin{\sigma}{*}{1s} \IIorbNoSpin{\sigma}{}{2s})$}, 
and \smash{${^1}\Sigma_\text g(\IIorbdouble{\sigma}{}{1s} \IIorbNoSpin{\sigma}{*}{1s} \IIorbNoSpin{\sigma}{*}{2s}$)} with $\Lambda_{\mathbf{B}}=0$ and
\smash{$^{1}\Pi_\text g(\IIorbdouble{\sigma}{}{1s} \IIorbNoSpin{\sigma}{*}{1s} \IIorbNoSpin{\pi}{}{$-1$})$} 
with $\Lambda_{\mathbf{B}}=-1$, whose energies are
plotted against the field strength along the positive axis in Fig.~\ref{figSingletEvsB}.
While the three sigma states are destabilized diamagnetically in the field, the pi  state is stabilized and becomes
the second singlet state at $B = 0.18B_0$.  In fields stronger than about
$0.6B_0$, the third singlet is \smash{$^{1}\Pi_\text u(\IIorbdouble{\sigma}{}{1s} \IIorbNoSpin{\sigma}{*}{1s} \IIorbNoSpin{\pi}{*}{$-1$})$},
having crossed the two highest sigma states. At this field strength, however, the ground state is no longer a singlet but a triplet, as discussed below. 

In the perpendicular field orientation, where the molecular point group is
C$_\text{2h}$ rather than C$_{\infty \text h}$, 
the loss of cylindrical spatial symmetry manifests itself in more avoided crossings as seen in
Fig.\,\ref{figSingletEvsB}, where the energies of the lowest electronic states are plotted against
the field strength along the negative axis.

In a weak perpendicular magnetic field,  
the ground state is $^1 \text{A}_{\text{g}}(1\orbdouble{a}{g} 1\IITorbdouble{b}{*}{u})$, while the lowest excited states are
\smash{$^1 \text{B}_{\text{u}}(1\orbdouble{a}{g} 1\IITorbNoSpin{b}{*}{u} 2\orbNoSpin{a}{g})$}, 
\smash{$^1 \text{A}_{\text{g}}(1\orbdouble{a}{g} 1\IITorbNoSpin{b}{*}{u} 2\IITorbNoSpin{b}{*}{u})$}, 
and \smash{$^1 \text{B}_{\text{g}}(1\orbdouble{a}{g} 1\IITorbNoSpin{b}{*}{u} 1\orbNoSpin{a}{u})$}. 
These states originate from the same field-free states as do the lowest states in parallel field orientation except that
the third excited state correlates with 
\smash{$^{1}\Pi_\text g(\IIorbdouble{\sigma}{}{1s} \IIorbNoSpin{\sigma}{*}{1s} \IIorbNoSpin{\pi}{}{$\shortparallel$})$},
which contains
a singly occupied \smash{$\pi_\shortparallel = (\pi_+ + \pi_-)/\sqrt{2}$} orbital of a$_\text u$ symmetry 
rather than a singly occupied $\pi_{-1}$ orbital of $\text a_\text u + \text b_\text u$ symmetry 
in the C$_{2 \text h}$ point group.
Hence,
\smash{$^1 \text{B}_{\text{g}}(1\orbdouble{a}{g} 1\IITorbNoSpin{b}{*}{u} 1\orbNoSpin{a}{u})$} is diamagnetic rather than paramagnetic.

Here and in the following, $\pi_\shortparallel$ denotes the $\pi$ component of symmetry a$_{\text{u}}$ parallel to the magnetic field, whereas
$\pi_\perp$ denotes the $\pi$ component of symmetry b$_{\text{u}}$ perpendicular to the field and bond axes. We likewise
use the notation $\pi^\ast_\shortparallel$ for the $\pi$ component of symmetry b$_{\text{g}}$ parallel to the magnetic field, whereas
$\pi^\ast_\perp$ denotes the component of symmetry a$_{\text{g}}$ perpendicular to the field and bond axes. 

In a perpendicular magnetic field of about $B=0.15B_0$,  
\smash{$^1 \text{A}_{\text{g}}(1\orbdouble{a}{g} 1\IITorbNoSpin{b}{*}{u} 2\IITorbNoSpin{b}{*}{u})$} crosses
\smash{$^1 \text{B}_{\text{u}}(1\orbdouble{a}{g} 1\IITorbNoSpin{b}{*}{u} 2\orbNoSpin{a}{g})$} to become the
lowest excited state, stabilized by the antibonding 2s orbital in the magnetic field by the paramagnetic bonding mechanism.
At a field strength of about $B = 0.35B_0$, the state
\smash{$^1 \text{B}_{\text{u}}(1\orbdouble{a}{g} 1\IITorbNoSpin{b}{*}{u} 3\IITorbNoSpin{a}{*}{g})$}, which originates
from the high-lying zero-field state 
${^1}\Pi(\IIorbdouble{\sigma}{}{1s} \IIorbNoSpin{\sigma}{*}{1s} \IIorbNoSpin{\pi}{*}{$\perp$})$
with a singly occupied $\pi^\ast_\perp$ orbital of a$_\text g$ symmetry, 
goes through a narrowly avoided crossing with the second lowest excited state
\smash{$^1 \text{B}_{\text{u}}(1\orbdouble{a}{g} 1\IITorbNoSpin{b}{*}{u} 2\orbNoSpin{a}{g})$}. Around this avoided
crossing, the 2a$_\text g$ orbital changes character from $\sigma_{2\text s}$ to 
\smash{$\pi^\ast_\perp$}, pushing the second excited state
\smash{$^1 \text{B}_{\text{u}}(1\orbdouble{a}{g} 1\IITorbNoSpin{b}{*}{u} 2\orbNoSpin{a}{g})$} 
further down to recross 
\smash{$^1 \text{A}_{\text{g}}(1\orbdouble{a}{g} 1\IITorbNoSpin{b}{*}{u} 2\IITorbNoSpin{b}{*}{u})$},
becoming again the first excited state, but with a HOMO of $\pi_\perp^\ast$ rather than
$\sigma_\text{2s}$ character. 

To summarize, the HOMO of the first excited state in the perpendicular
field orientation is $\sigma_\text{2s}$ from zero field to $0.15B_0$, then becomes
$\sigma_\text{2s}^\ast$ followed by $\pi_\perp^\ast$ at $0.35B_0$. This progression may be
understood in terms of paramagnetic stabilization of the orbitals, noting that the three orbitals have zero, one, 
and two nodal planes, respectively, parallel to the magnetic field vector;
see Table~\ref{tabsorb}.

\begin{table*}
\caption{\label{tabDissocMinB02} The lowest minima on dissociation curves for He$_2$ in a magnetic field $B=0.2B_0$. The quantity $R_{\mathrm{grid}}$ is the bond distance for which the electron configuration, while other quantities are interpolated between grid points on the dissociation curve. All quantities are in atomic units.}
\begin{center}
\begin{tabular}{crcccrcc|lr}
\hline\hline
  spin & $\theta$ & $n$ & $E_{\mathrm{min}}$ & $R_{\mathrm{eq}}$ & $\Lambda_{\mathbf{B}}$ & $E_\infty$ & $E_{\mathrm{dis}}$  & state& $R_{\text{grid}}$\\ \hline 
singlet & $0^{\circ}$ & $0$ & $-5.786650$ & $5.733$ & $0.00$ & $-5.786619$   & $0.000031$ & $^{1}\Sigma_\text g(0.98\,\sigma_{1 \text s}^2 \sigma_{1 \text s}^{*2})$ & 5.800 \\
        &             & $1$ & $-5.070304$ & $1.972$ & $-1.00$ & $-4.934970$  & $0.135334$ & $^{1}\Pi_\text g(0.95\,\sigma_{1 \text s}^2 \sigma_{1 \text s}^{*} \pi_{-1})$ & 2.000 \\
        &             & $2$ & $-5.051661$ & $1.917$ & $0.00$ & $-4.960160$   & $0.091501$ & $^{1}\Sigma_\text u(0.92\,\sigma_{1 \text s}^2 \sigma_{1 \text s}^\ast \sigma_{2s})$ & 1.900 \\
        &             & $3$ & $-4.996800$ & $2.079$ & $0.00$ & $-4.960157$   & $0.036643$ & $^{1}\Sigma_\text g(0.93\,\sigma_{1 \text s}^2 \sigma_{1 \text s}^{*} \sigma_{2 \text s}^{*})$ &2.100 \\
        & $90^{\circ}$ & $0$ & $-5.786667$ & $5.271$ & $-0.00$ & $-5.786618$ & $0.000049$ & $^1 \text{A}_{\text{g}}(0.98\; 1\orbdouble{a}{g} 1\IITorbdouble{b}{*}{u})$ & 5.200 \\
        &              & $1$ & $-5.073291$ & $2.021$ & $-0.57$ & $-4.959465$ & $0.113826$ & $^1 \text{A}_{\text{g}}(0.88\; 1\orbdouble{a}{g} 1\IITorbNoSpin{b}{*}{u} 2\IITorbNoSpin{b}{*}{u})$ & 2.000 \\
        &              & $2$ & $-5.050287$ & $1.896$ & $0.07$ & $-4.959458$  & $0.090829$ & $^1 \text{B}_{\text{u}}(0.92\; 1\orbdouble{a}{g} 1\IITorbNoSpin{b}{*}{u} 2\orbNoSpin{a}{g})$  & 1.900 \\
        &              & $3$ & $-4.980790$ & $1.946$ & $0.07$ & unknown & unknown & $^1 \text{B}_{\text{g}}(0.95\; 1\orbdouble{a}{g} 1\IITorbNoSpin{b}{*}{u} 1\orbNoSpin{a}{u})$ & 1.900 \\
\hline
triplet & $0^{\circ}$ & $0$ & $-5.290267$ & $1.966$ & $-1.00$ & $-5.182528$  & $0.107739$ &  $^{3}\Pi_\text g(0.95\,\orbdouble{$\sigma$}{1s} \IIorbdown{\sigma}{*}{1s} \orbdown{$\pi$}{$-1$})$ & 2.000 \\
         &             & $1$ & $-5.270141$ & $1.937$ & $0.00$ & $-5.206973$   & $0.063168$ & $^{3}\Sigma_\text u(0.89\,\sigma_{1\text{s}}^2 \IIorbdown{\sigma}{*}{1s} \orbdown{$\sigma$}{2\text s})$ & 1.900 \\
         &             & $2$ & $-5.222877$ & $2.103$ & $0.00$ & $-5.206972$   & $0.015905$ & $^{3}\Sigma_\text g(0.88\,\sigma_{1\text{s}}^2 \IIorbdown{\sigma}{*}{1s} \IIorbdown{\sigma}{*}{2s})$& 2.100 \\
        & $90^{\circ}$ & $0$ & $-5.297553$ & $2.006$ & $-0.68$ & $-5.206769$ & $0.090784$ & $^{3} \text{A}_{\text{g}}(0.84\; 1\orbdouble{a}{g} 1\IITorbdown{b}{*}{u} 2\IITorbdown{b}{*}{u})$  & 2.000 \\
        &              & $1$ & $-5.268197$ & $1.918$ & $0.07$ & $-5.206769$  & $0.061428$ & $^3 \text{B}_{\text{u}}(0.89\; 1\orbdouble{a}{g} 1\IITorbdown{b}{*}{u} 2\orbdown{a}{g})$  & 1.900 \\
        &              & $2$ & $-5.200512$ & $1.941$ & $0.07$ & $-5.094688$  & $0.105824$ & $^3 \text{B}_{\text{g}}(0.95\; 1\orbdouble{a}{g} 1\IITorbdown{b}{*}{u} 1\orbdown{a}{u})$ & 1.900 \\
\hline
quintet & $0^{\circ}$ & $0$ & $-4.652235$ & $4.141$ & $-1.00$ & $-4.604572$ & $0.047663$ & $^{5}\Pi_\text g(0.68\,\orbdown{$\sigma$}{1s} \IIorbdown{\sigma}{*}{1s} \orbdown{$\sigma$}{2s} \orbdown{$\pi$}{$-1$})$ & 4.600 \\
        &             & $1$ & $-4.627960$ & $6.052$ & $0.00$ & $-4.626512$ & $0.001448$  & $^{5}\Sigma_\text g(0.73\,\orbdown{$\sigma$}{1s}\IIorbdown{\sigma}{*}{1s} \orbdown{$\sigma$}{2s} \IIorbdown{\sigma}{*}{2s}) $ & 6.200 \\
        & $90^{\circ}$ & $0$ & $-4.686660$ & $4.560$ & $-0.67$ & $-4.627365$ & $0.059295$ & $^{5}\text{A}_{\text{g}} (0.44\; 1\orbdown{a}{g} 1\IITorbdown{b}{*}{u} 2\IITorbdown{b}{*}{u} 3\orbdown{a}{g},  0.32\; 1\orbdown{a}{g} 1\IITorbdown{b}{*}{u} 2\IITorbdown{b}{*}{u} 2\orbdown{a}{g})$ & 4.600 \\
        &              & $1$ & $-4.632662$ & $5.074$ & $-0.59$ & $-4.603487$ & $0.029175$ & $^{5} \text{B}_{\text{u}}(0.82\; 1\orbdown{a}{g} 1\IITorbdown{b}{*}{u} 2\orbdown{a}{g} 3\orbdown{a}{g})$ & 5.000 \\
        &              & $2$ & $-4.609599$ & $5.648$ & $-0.90$ & $-4.602014$ & $0.007585$ & $^{5} \text{A}_{\text{g}}(0.43\; 1\orbdown{a}{g} 1\IITorbdown{b}{*}{u} 2\IITorbdown{b}{*}{u} 3\orbdown{a}{g}, 0.26\; 1\orbdown{a}{g} 1\IITorbdown{b}{*}{u} 2\IITorbdown{b}{*}{u} 2\orbdown{a}{g})$ & 5.800 \\
\hline\hline
\end{tabular}
\end{center}
\end{table*}

\subsubsection{Triplet states of He$_2$ at $R = 2a_0$}

The behaviour of the triplet states in parallel and perpendicular fields are plotted along the positive and negative axes 
of Figure~\ref{figTripletEvsB}, respectively.
The triplet states behave in a similar way to the corresponding open-shell singlets except that the spin Zeeman interaction splits
the triplet states into three $m_\text s$ components, the 
$m_\text{s} = - 1$ and $m_\text{s} = + 1$ components tilted downwards and upwards, respectively. We here consider
the lowest-energy $m_\text{s} = - 1$ components only.

Because of the spin Zeeman interaction, the ground state is the singlet \smash{$^{1}\Sigma_\text g(\IIorbdouble{\sigma}{}{1s} \IIorbdouble{\sigma}{*}{1s})$} only up to a field strength of about
$0.55B_0$ in the parallel field orientation and about $0.65B_0$ in the perpendicular orientation, where the triplet states
\smash{$^{3}\Pi_\text g(\IIorbdouble{\sigma}{}{1s} \IIorbdown{\sigma}{*}{1s} \IIorbdown{\pi}{}{$-1$})$} and 
$^{3} \text{A}_{\text{u}}(1\orbdouble{a}{g} 1\IITorbdown{b}{*}{u} 1\IITorbdown{b}{*}{g})$, respectively, become the ground states, the latter 
originating from 
\smash{$^{3}\Pi_\text u(\IIorbdouble{\sigma}{}{1s} \IIorbdown{\sigma}{*}{1s} \IIorbdown{\pi}{*}{$\perp$})$}. 

In a weak parallel field, with $B \lesssim 0.1B_0$, the two lowest triplet states are 
\smash{$^{3}\Sigma_\text u(\IIorbdouble{\sigma}{}{1s} \IIorbdown{\sigma}{*}{1s} \IIorbdown{\sigma}{}{2s})$} and 
\smash{$^{3}\Sigma_\text g(\IIorbdouble{\sigma}{}{1s} \IIorbdown{\sigma}{*}{1s} \IIorbdown{\sigma}{*}{2s})$}, 
with $\Lambda_{\mathbf{B}} = 0$. 
The next two states are \smash{$^{3}\Pi_\text g(\IIorbdouble{\sigma}{}{1s} \IIorbdown{\sigma}{*}{1s} \IIorbdown{\pi}{}{$-1$})$} and 
\smash{$^{3}\Pi_\text g(\IIorbdouble{\sigma}{}{1s} \IIorbdown{\sigma}{*}{1s} \IIorbdown{\pi}{}{+1})$}, 
which diverge with increasing field strength due to the orbital Zeeman interaction with opposite signs of $\Lambda_{\mathbf{B}} = \pm 1$. 
In fields stronger than about $0.5B_0$,
the lowest states are completely reordered by the spin and orbital Zeeman interactions. 
The lowest triplet is now \smash{$^{3}\Pi_\text g(\IIorbdouble{\sigma}{}{1s} \IIorbdown{\sigma}{*}{1s} \IIorbdown{\pi}{}{$-1$})$}, which is also the electronic ground state of the system,
while the first excited state is \smash{$^{3}\Pi_\text u(\IIorbdouble{\sigma}{}{1s} \IIorbdown{\sigma}{*}{1s} \IIorbdown{\pi}{*}{$-1$},\, \IIorbdouble{\sigma}{*}{1s} \IIorbdown{\sigma}{}{1s} \IIorbdown{\pi}{}{$-1$})$}. 
At one-atomic unit field strength 
$B_0$, the second excited state is \smash{${}^3 \Delta_{\text{u}}(\IIorbdouble{\sigma}{}{1s} \IIorbdown{\sigma}{*}{1s} \IIorbdown{\delta}{}{$-2$})$}.

In a weak perpendicular field, the three lowest electronic triplet states are predominantly
\smash{$^{3} \text{B}_{\text{u}}(1\orbdouble{a}{g} 1\IITorbdown{b}{*}{u} 2\orbdown{a}{g})$},
$^{3} \text{A}_{\text{g}}(1\orbdouble{a}{g} 1\IITorbdown{b}{*}{u} 2\IITorbdown{b}{*}{u})$, 
and $^{3} \text{B}_{\text{g}}(1\orbdouble{a}{g} 1\IITorbdown{b}{*}{u} 1\orbdown{a}{u})$, originating from the field-free states
\smash{$^{3}\Sigma_\text u^+(\IIorbdouble{\sigma}{}{1s} \IIorbdown{\sigma}{*}{1s} \IIorbdown{\sigma}{}{2s})$},
\smash{$^{3}\Sigma_\text g^+(\IIorbdouble{\sigma}{}{1s} \IIorbdown{\sigma}{*}{1s} \IIorbdown{\sigma}{*}{2s})$}, and
\smash{$^{3}\Pi_\text g(\IIorbdouble{\sigma}{}{1s} \IIorbdown{\sigma}{*}{1s} \IIorbdown{\pi}{}{$\perp$})$}. 
At $0.2B_0$, the lowest two triplet states have crossed and the lowest state is now
$^{3} \text{A}_{\text{g}}(1\orbdouble{a}{g} 1\IITorbdown{b}{*}{u} 2\IITorbdown{b}{*}{u})$; see Table~\ref{tabDissocMinB02}.
In the strongest field plotted in Fig.~\ref{figTripletEvsB}, the ground state is 
$^{3} \text{B}_{\text{u}}(1\orbdouble{a}{g} 1\IITorbdown{b}{*}{u} 1\IITorbdown{a}{*}{g})$, 
originating from the field-free state
\smash{$^{3}\Pi_\text u(\IIorbdouble{\sigma}{}{1s} \IIorbdown{\sigma}{*}{1s} \IIorbNoSpin{\pi}{*}{$\perp$})$}. 
As the field increases from zero, this highly-excited state drops below all other triplet states, including 
the $^{3} \text{B}_\text g$ state that originates from
{$^{3}\Pi_\text g(\IIorbdouble{\sigma}{}{1s} \IIorbdown{\sigma}{*}{1s} \IIorbNoSpin{\pi}{}{$\shortparallel$})$}. 
In the process, the $^{3} \text{B}_\text u$ state
acquires a substantial negative AQAM value from the occupied antibonding 
$\IIorbdown{\sigma}{*}{1s}$ and $\IIorbdown{\pi}{*}{$\perp$}$ orbitals. Decreasing from
$0.50$ at field strength $0.01B_0$ to $=-1.42$ at $0.05B_0$, it reaches a minimum value of
$\Lambda_{\mathbf{B}}=-1.75$ at $0.25B_0$, after which it increases again to $-1.24$ at field strength $B_0$. 
We note that the evolution of the lowest triplet state in the perpendicular orientation parallels that of 
the lowest singlet state, the HOMO changing character first
from $\sigma_\text{2s}$ to $\sigma_\text{2s}^\ast$ and then 
from $\sigma_\text{2s}^\ast$ and to $\pi_\perp^\ast$.

\begin{table*}
\caption{\label{tabDissocMinB1} The lowest minima on dissociation
  curves for He$_2$ in a magnetic field $B=B_0$. The quantity
  $R_{\mathrm{grid}}$ is the bond distance for which the electron
  configuration, while other quantities are interpolated between grid
  points on the dissociation curve. All quantities are in atomic
  units.}
\begin{center}
\begin{tabular}{crcccrcc|lr}
\hline\hline
  spin & $\theta$ & $n$ & $E_{\mathrm{min}}$ & $R_{\mathrm{eq}}$ & $\Lambda_{\mathbf{B}}$ & $E_\infty$ & $E_\text{dis}$ & state & $R_{\text{grid}}$  \\ \hline 
singlet & $0^{\circ}$ & $0$ & $-5.454327$ & $4.747$ & $0.00$ & $-5.453984$ & $0.000343$ & $^{1}\Sigma_\text g(0.99\,\sigma_{1\text s}^2 \sigma_{1\text s}^{*2})$ & 4.700 \\
        &             & $1$ & $-4.739260$ & $1.800$ & $-1.00$ & $-4.555238$ & $0.184022$ & $^{1}\Pi_\text g(0.96\,\sigma_{1\text s}^2 \sigma_{1\text s}^{*} \pi_{-1})$ & 1.800 \\
        &             & $2$ & $-4.554578$ & $4.420$ & $-1.00$ & $-4.554309$ & $0.000269$ & $^{1}\Pi_\text u(0.64\,\sigma_{1\text s}^{*2} \sigma_{1\text s} \pi_{-1})$ & 4.600 \\
        &             & $3$ & $-4.423187$ & $1.893$ & $-2.00$ & unknown  & unknown  &  $^{1}\Delta_\text g(0.96\,\sigma_{1\text s}^{*2} \sigma_{1\text s} \delta_{-2})$& 1.900 \\
        & $90^{\circ}$ & $0$ & $-5.455252$ & $3.012$ & $-0.00$ & $-5.453983$ & $0.001269$ & $^{1} \text{A}_{\text{g}}(0.99\; 1\orbdouble{a}{g} 1\IITorbdouble{b}{*}{u})$ & 3.000 \\
        &              & $1$ & $-4.638450$ & $2.009$ & $-1.27$ & $-4.555000$ & $0.083450$ & $^{1} \text{B}_{\text{u}}(0.95\; 1\orbdouble{a}{g} 1\IITorbNoSpin{b}{*}{u} 2\IITorbNoSpin{a}{*}{g})$ & 2.000 \\
        &              & $2$ & $-4.570580$ & $3.136$ & $-1.07$ & $-4.554542$ & $0.016038$ & $^{1} \text{A}_{\text{g}}(0.76\; 1\IITorbdouble{b}{*}{u} 1\orbNoSpin{a}{g} 2\IITorbNoSpin{a}{*}{g})$ & 3.200 \\
        &              & $3$ & $-4.503213$ & $1.617$ & $0.24$ & unknown  & unknown & $^{1} \text{B}_{\text{g}}(0.95\; 1\orbdouble{a}{g} 1\orbNoSpin{a}{u} 1\orbNoSpin{b}{u})$ & 1.600 \\
\hline
triplet & $0^{\circ}$ & $0$ & $-5.782268$ & $1.805$ & $-1.00$ & $-5.644685$ & $0.137583$ &$^{3}\Pi_\text g(0.95\,\sigma_{1\text s}^2 \sigma_{1\text s}^{*} \pi_{-1})$ & 1.800 \\
        &             & $2$ & $-5.428475$ & $1.889$ & $-2.00$ & unknown  & unknown  &$^{3}\Delta_\text g(0.96\,\sigma_{1\text s}^2 \sigma_{1\text s}^{*} \delta_{-2})$ & 1.900 \\
        &             & $2$    & $-5.404984$ & $2.613$ & $-1.00$ & $-5.256417$ & $0.148567$ &$^{3}\Pi_\text u(0.61\,\sigma_{1\text s}^2 \sigma_{1\text s}^{*} \pi_{-1}^{*},0.34\,\sigma_{1\text s}^{*2} \sigma_{1\text s} \pi_{-1})$& 2.600 \\
        &             & $2$    & $-5.406632$ & $5.518$ & $0.00$ & $-5.393051$ & $0.013581$ &$^{3}\Sigma_\text u(0.75\,\sigma_{1\text s}^2 \sigma_{1\text s}^{*} \sigma_{2\text s})$ & 5.400 \\
        & $90^{\circ}$ & $0$ & $-5.721430$ & $1.991$ & $-1.25$ & $-5.644682$ & $0.076748$ & $^{3}\text{B}_{\text{u}}(0.94\; 1\orbdouble{a}{g} 1\orbdown{b}{u} 2\orbdown{a}{g})$ & 2.000 \\
        &              & $1$ & $-5.658760$ & $3.245$ & $-1.10$ & $-5.644682$ & $0.014078$ & $^{3}\text{A}_{\text{g}}(0.70\; 1\orbdouble{b}{u} 1\orbdown{a}{g} 2\orbdown{a}{g},  0.25\; 1\orbdouble{a}{g} 1\orbdown{b}{u} 2\orbdown{b}{u})$ & 3.200 \\
        &              & $2$ & $-5.530958$ & $1.611$ & $0.22$ & $-5.393015$ & $0.137943$ &  $^{3}\text{B}_{\text{g}}(0.95\; 1\orbdouble{a}{g} 1\orbdown{b}{u} 1\orbdown{a}{u})$  & 1.600 \\
        &              & $2$ & $-5.460260$ & $2.416$ & $-1.77$ & $-5.256414$ & $0.203846$ & $^{3} \text{A}_{\text{g}}(0.58\; 1\orbdouble{a}{g} 1\orbdown{b}{u} 2\orbdown{b}{u}, 0.35\; 1\orbdouble{b}{u} 1\orbdown{a}{g} 2\orbdown{a}{g})$ & 2.400 \\
\hline
quintet & $0^{\circ}$ & $0$ & $-5.835333$            & $\infty$        & $-2.00$        & $-5.835333$            & 0           & $^{5}\Delta_\text g(0.92\,\sigma_{1 \text s} \sigma_{1 \text s}^\ast \pi_{-1} \pi_{-1}^\ast)$    & 10.000                        \\
        &             & $1$ & $-5.645344$ & $2.396$ & $-3.00$ & $-5.448259$ & $0.197085$ &  $^{5}\Phi_\text u(0.93\,\sigma_{1\text s} \sigma_{1\text s}^{*} \pi_{-1} \delta_{-2})$ & 2.400 \\
        &             & $1$ & $-5.650264$ & $4.085$ & $-1.00$ & $-5.583994$ & $0.066270$ &  $^{5}\Pi_\text g(0.86\,\sigma_{1\text s} \sigma_{1\text s}^{*} \pi_{-1} \sigma_{2\text s})$ & 4.200 \\
        & $90^{\circ}$ & $0$ & $-5.855142$ & $3.620$ & $-2.21$ & $-5.835368$ & $0.019774$ & $^{5}\text{A}_{\text{g}} (0.91\; 1\orbdown{a}{g} 1\orbdown{b}{u} 2\orbdown{a}{g} 2\orbdown{b}{u})$ & 3.800 \\
        &              & $1$ & $-5.677227$ & $2.660$ & $-0.44$ & $-5.583805$ & $0.093422$ & $^{5}\text{B}_{\text{g}}(0.89\; 1\orbdown{a}{g} 1\orbdown{b}{u} 2\orbdown{a}{g} 1\orbdown{a}{u})$  & 2.600 \\
        &              & $2$ & $-5.618470$ & $3.092$ & $-0.82$ & $-5.583719$ & $0.034751$ & $^{5}\text{A}_{\text{u}} (0.83\; 1\orbdown{a}{g} 1\orbdown{b}{u} 2\orbdown{a}{g} 1\orbdown{b}{g})$ & 3.000 \\
        &              & $3$ & $-5.577629$ & $3.523$ & $-2.83$ & $-5.447773$ & $0.129856$  & $^{5} \text{B}_{\text{u}}(0.87\; 1\orbdown{a}{g} 1\orbdown{b}{u} 2\orbdown{a}{g} 3\orbdown{a}{g})$ & 3.400 \\
\hline\hline
\end{tabular}
\end{center}
\end{table*}

\subsection{Potential-energy curves of He$_2$}

Next, we explore how the energy spectrum varies with the bond distance $R$ and the field orientation $\theta$. 
For visualization purposes, energy curves for perpendicular (parallel)
orientations will in all cases be plotted with a negative (positive)
bond distance. The lowest minima on these dissociation curves are
summarized in Table~\ref{tabDissocMinB02} and \ref{tabDissocMinB1}, for $B = 0.2B_0$ and $B
= B_0$, respectively.

At small bond distances, the dissociation curves are dominated by the nuclear electrostatic repulsion energy, obscuring the united-atom limit. 
We therefore select a cut-off distance $R_\text c$, marked with vertical grey dash--dot lines in each figure. In the region $R < R_\text c$,
we replace the actual energy $E(R)$ by a shifted energy
\begin{align}
   E^\prime(R) &= E(R)  + \frac{Z^2}{R_\text c} \nonumber - \frac{Z^2}{R} \nonumber \\ &- a (R-R_\text c)^3 - b (R-R_\text c)^2 - c (R-R_\text c), 
\end{align}
more suited to the united-atom limit.
The second and third terms remove the singular nuclear repulsion energy,  while the polynomial in $R-R_\text c$ aligns the energy scale.
The shift is state independent;
it vanishes but introduces nondifferentiable kinks and cusps at the cut-off distance $R=R_\text c$. 

In the united-atom limit, as $R\to 0$, 
the molecular orbital basis set becomes linearly dependent, spanning only an orbital space of half the dimension. 
To avoid spurious results from near linear dependence in this region, we therefore avoid very short bond distances in the dissociation curves.

\begin{table*}
\caption{Orbital expectation values $s$ of reflection in the mid-bond plane Eq.~\eqref{sexp} for
  the lowest eight RHF orbitals in
  He$_2$ in a perpendicular field of strength $B_\perp$. A value of $s=+1$ implies perfect symmetry and is associated with
  bonding properties, while a value of $s=-1$ implies perfect antisymmetry
and is associated with a nodal plane and antibonding properties. Due to
field-induced symmetry breaking, intermediate values are typical, indicating mixing of bonding and antibonding properties. The last row contains
the mean absolute $s$ value of the eight listed RHF orbitals for a given bond distance and field strength.}
\label{tabsantibonding}
\scalebox{0.92}{
\begin{tabular}
{lr@{\hskip 5mm}lr@{\hskip 5mm}lr@{\hskip 5mm}lr@{\hskip 5mm}lr@{\hskip 5mm}
 lr@{\hskip 5mm}lr@{\hskip 5mm}lr@{\hskip 5mm}lr@{\hskip 5mm}lr}
\hline\hline
\multicolumn{8}{l}{$B_\perp = 0.2B_0$}& \multicolumn{12}{l}{$B_\perp = 1.0B_0$} \\ 
\multicolumn{2}{l}{$R = 2.0a_0$}& \multicolumn{2}{l}{$R = 3.8a_0$}& \multicolumn{2}{l}{$R = 5.0a_0$}& \multicolumn{2}{l}{$R = 5.8a_0$} &
\multicolumn{2}{l}{$R = 1.8a_0$}& \multicolumn{2}{l}{$R = 2.0a_0$}& \multicolumn{2}{l}{$R = 2.5a_0$}&
\multicolumn{2}{l}{$R = 3.0a_0$}& \multicolumn{2}{l}{$R = 3.2a_0$}& \multicolumn{2}{l}{$R = 3.8a_0$} 
\\ \hline
1a$_\text g$ &   1.00  & 1a$_\text g$ &   0.98 & 1a$_\text g$ &   0.96 & 1a$_\text g$ &    0.94 &
1a$_\text g$ &    0.96 & 1a$_\text g$ &   0.93 & 1a$_\text g$ &   0.86 & 1a$_\text g$ &    0.77 & 1a$_\text g$ &   0.73 & 1a$_\text g$ &     0.62 \\
1b$_\text u$ &$-$0.97  & 1b$_\text u$ &$-$0.97 & 1b$_\text u$ &$-$0.95 & 1b$_\text u$ & $-$0.94 & 
1b$_\text u$ & $-$0.70 & 1b$_\text u$ &$-$0.71 & 1b$_\text u$ &$-$0.71 & 1b$_\text u$ & $-$0.67 & 1b$_\text u$ &$-$0.65 & 1b$_\text u$ &  $-$0.59 \\
2b$_\text u$ &   0.12  & 2b$_\text u$ &   0.17 & 2b$_\text u$ &   0.23 & 2b$_\text u$ &    0.26 & 
2a$_\text g$ &    0.08 & 2a$_\text g$ &   0.12 & 2a$_\text g$ &   0.25 & 2a$_\text g$ &    0.37 & 2a$_\text g$ &   0.40 & 2a$_\text g$ &     0.42 \\
2a$_\text g$ &   0.99  & 2a$_\text g$ &   0.89 & 2a$_\text g$ &   0.73 & 2a$_\text g$ &    0.61 & 
2b$_\text u$ &    0.00 & 2b$_\text u$ &   0.00 & 2b$_\text u$ &   0.01 & 2b$_\text u$ &    0.03 & 2b$_\text u$ &   0.03 & 2b$_\text u$ &     0.04 \\
1a$_\text u$ &   0.99  & 3a$_\text g$ &   0.14 & 3a$_\text g$ &   0.28 & 3a$_\text g$ &    0.30 & 
1a$_\text u$ &    0.90 & 1a$_\text u$ &   0.85 & 1a$_\text u$ &   0.70 & 1a$_\text u$ &    0.54 & 1a$_\text u$ &   0.47 & 1a$_\text u$ &     0.31 \\
3a$_\text g$ &$-$0.11  & 1a$_\text u$ &   0.95 & 1a$_\text u$ &   0.88 & 3b$_\text u$ &    0.06 & 
1b$_\text g$ &    0.14 & 1b$_\text g$ &   0.13 & 1b$_\text g$ &   0.12 & 1b$_\text g$ &    0.12 & 1b$_\text g$ &   0.12 & 1b$_\text g$ &     0.10 \\
3b$_\text u$ &   0.78  & 3b$_\text u$ &   0.62 & 3b$_\text u$ &   0.30 & 1a$_\text u$ &    0.82 & 
3a$_\text g$ &    0.83 & 3b$_\text u$ &   0.26 & 3b$_\text u$ &   0.32 & 3a$_\text g$ &    0.12 & 3a$_\text g$ &   0.11 & 3a$_\text g$ &     0.18 \\
4b$_\text u$ &$-$0.61  & 1b$_\text g$ &$-$0.84 & 1b$_\text g$ &$-$0.81 & 1b$_\text g$ & $-$0.78 & 
3b$_\text u$ &    0.23 & 3a$_\text g$ &   0.78 & 3a$_\text g$ &   0.46 & 3b$_\text u$ &    0.34 & 3b$_\text u$ &   0.33 & 3b$_\text u$ &     0.25 \\
\hline  
& 0.70  && 0.70 && 0.64  && 0.59 && 0.48 && 0.47 && 0.43 && 0.37 && 0.36 && 0.31 \\
\hline\hline
\end{tabular}}
\end{table*}

Finally, we remark that the common notions of bonding and antibonding
orbitals, associated with symmetry and antisymmetry with respect to
mirror reflection $\sigma_{\text{midbond}}$ in the midbond plane, become more complicated and
not well defined in the presence of a magnetic
field. While this symmetry remains exact in a parallel field, a nonparallel magnetic fields breaks it.
As a result, orbitals become superpositions $\phi = a \phi_{+} + b
\phi_{-}$ with $|a|^2 + |b|^2=1$ of symmetric and
antisymmetric components. The expectation value of the mirror reflection operator,
\begin{equation}
    s = \bra{\phi} \sigma_{\text{midbond}} \ket{\phi} = |a|^2 - |b|^2, \label{sexp}
\end{equation}
provides a measure of this mixing. In general, a fraction $|a|^2 = \tfrac{1}{2} (1+s)$ of $\phi$ is
symmetric and bonding, whereas a fraction $|b|^2 = \tfrac{1}{2} (1-s)$ is
antisymmetric and antibonding. In Table~\ref{tabsantibonding}, this expectation value
is given for several bond distances $R$ and perpendicular field strengths $B_{\perp}$ for the reference restricted
Hartree--Fock (RHF) orbitals employed in the FCI calculations. Note that
the triplet and quintet states were also calculated using RHF orbitals.

The perpendicular paramagnetic bonding mechanism, which increases the
magnitude of the angular momentum of antibonding orbitals, can also be
viewed as a mixing of bonding and antibonding orbitals. This is seen
in Table~\ref{tabsantibonding}, where, for example, the orbital
$1\IITorbNoSpin{b}{*}{u}$, which corresponds to the antibonding 1s orbital
$\IIorbNoSpin{\sigma}{*}{1s}$ in the parallel orientation, acquires increasingly strong bonding character in
stronger fields and at longer bond distances. From the listed mean absolute s values in Table~\ref{tabsantibonding},
we note that the $s$ values decrease with increasing field strength and increasing bond distance.
Indeed, strong magnetic fields compress the orbitals and change the relevant length scale,
so that the dissociation limit is reached earlier.

\subsubsection{Singlet potential-energy curves at $B=0.2B_0$} 

\begin{figure}[h]
  \includegraphics[width=\linewidth]{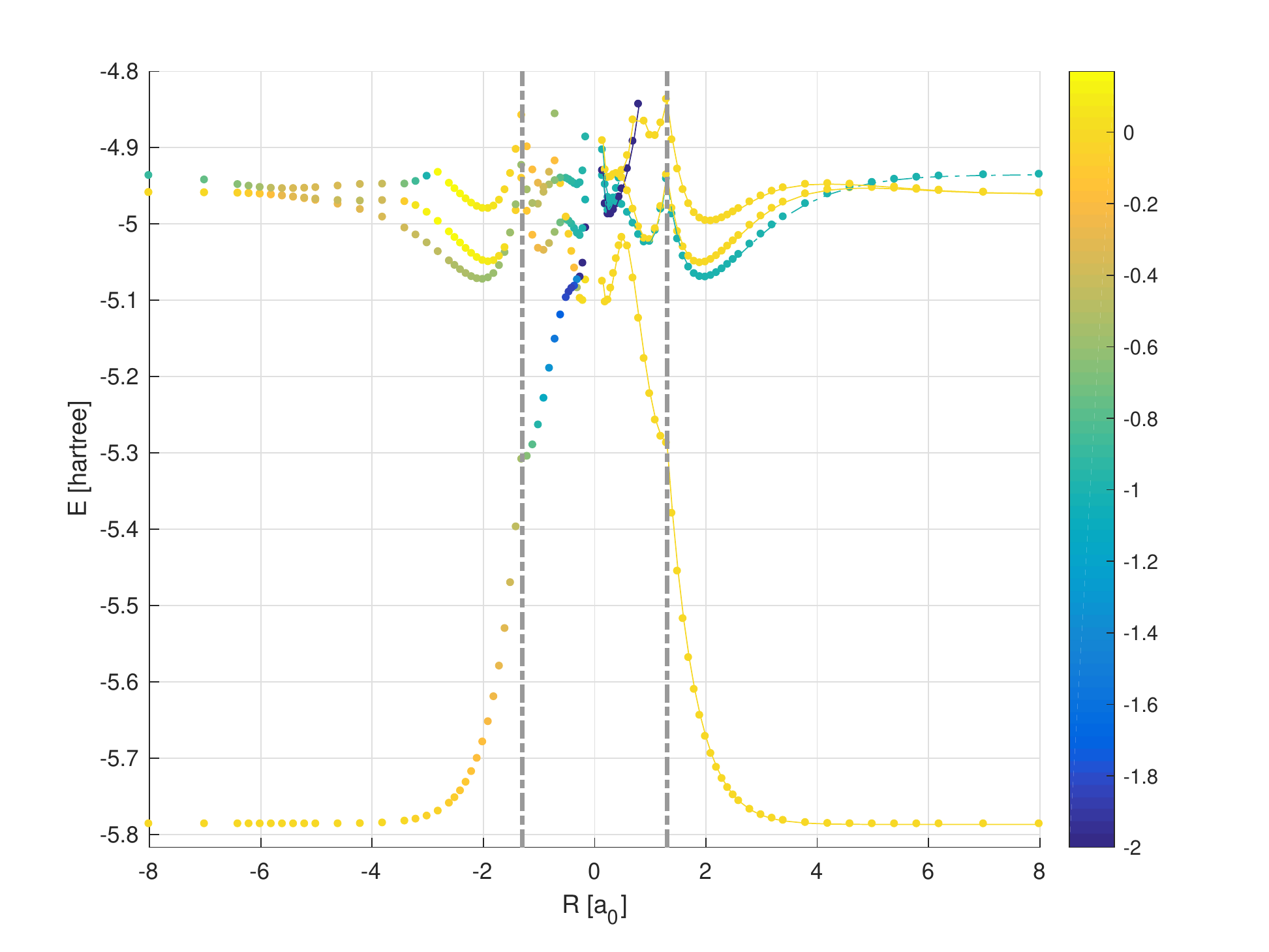}
  \caption{Dissociation curves for singlet states in perpendicular
    (negative half) and parallel (positive half) magnetic field $B =
    0.2B_0$. In the region between grey dashed lines, the curve is
    shifted by the nuclear repulsion energy and an additional
    quadratic fit to align the united atom limit to the same energy
    scale. Plot markers are coloured based on the AQAM value
    $\Lambda_{\mathbf{B}}$; scale indicated on the right.}
  \label{figHe2_B020_singlet_dissoc}
\end{figure}

Singlet dissociation curves at field strength $B=0.2B_0$ are shown in Fig.~\ref{figHe2_B020_singlet_dissoc} with information given in Table\,\ref{tabDissocMinB02}.
In the parallel orientation at $R=2a_0$, the lowest singlet is dominated by the electron configuration $^{1}\Sigma_\text g(\IIorbdouble{\sigma}{}{1s} \IIorbdouble{\sigma}{*}{1s})$. 
The first excited state is $^{1}\Pi_\text g( \IIorbdouble{\sigma}{}{1s} \sigma_{1s}^* \pi_{-1})$  with $\Lambda_{\mathbf{B}}=-1$, while 
second and third states are $^{1}\Pi_\text u(\IIorbdouble{\sigma}{}{1s} \sigma_{1s}^* \sigma_{2s})$ and 
$^{1}\Sigma_\text g(\IIorbdouble{\sigma}{}{1s} \sigma_{1s}^* \sigma_{2s}^*)$ with $\Lambda_{\mathbf{B}}=0$ and the same dissociation limits.
All three excited states are covalently bound with approximately the same equilibrium distance of about $2a_0$, whereas the ground state has a shallow minimum at $5.7a_0$.

The perpendicular orientation gives rise to dissociation curves that
are visually similar. However, the identification of the states
requires care since broken symmetries allow mixing of states that are
distinct in the parallel case. Moreover, viewed as hypersurfaces that
depend on $(R,\theta,B)$, states can be continuously deformed into
each other in a way that is sometimes path dependent due to the
presence of conical intersections. 

At $R=2a_0$ in the perpendicular orientation, the ground state is
$^1 \text{A}_{\text{g}} ( 1\orbdouble{a}{g} 1\IITorbdouble{b}{*}{u} )$  and the lowest three singlet excited states are
$^1 \text{A}_{\text{g}}( 1\orbdouble{a}{g} 1\IITorbNoSpin{b}{*}{u} 2\IITorbNoSpin{b}{*}{u})$, 
$^1 \text{B}_{\text{u}}( 1\orbdouble{a}{g} 1\IITorbNoSpin{b}{*}{u} 2\orbNoSpin{a}{g})$ 
and $^1 \text{B}_{\text{g}}( 1\orbdouble{a}{g} 1\IITorbNoSpin{b}{*}{u} 1\orbNoSpin{a}{u})$, the latter state being replaced by 
$^1\text{B}_{\text{u}}(1\orbdouble{a}{g} 1\IITorbNoSpin{b}{*}{u} 3\orbNoSpin{a}{g})$ at greater  bond distances. 
The ground state has the same parallel and perpendicular dissociation limits but different parallel and perpendicular united-atom limits, tending to
the 1s$^2$2s$^2$ beryllium configuration with $\Lambda_{\mathbf{B}}=0$ in the parallel orientation but to 1s$^2$2p$_{-1}^2$ with $\Lambda_{\mathbf{B}}=-2$ in the perpendicular orientation.
The first excited state in the perpendicular orientation has
$\Lambda_{\mathbf{B}}= -0.6$ arising from the {antibonding orbital
$1\IITorbNoSpin{b}{*}{u}$ and the intermediate orbital
$2\IITorbNoSpin{b}{\circ}{u}$}, slightly less than the
$\Lambda_{\mathbf{B}}= -1.0$ of the first excited state in the parallel orientation (for short bond distances), arising from the singly occupied $\pi_{-1}$ orbital.

\begin{figure}[h]
  \includegraphics[width=\linewidth]{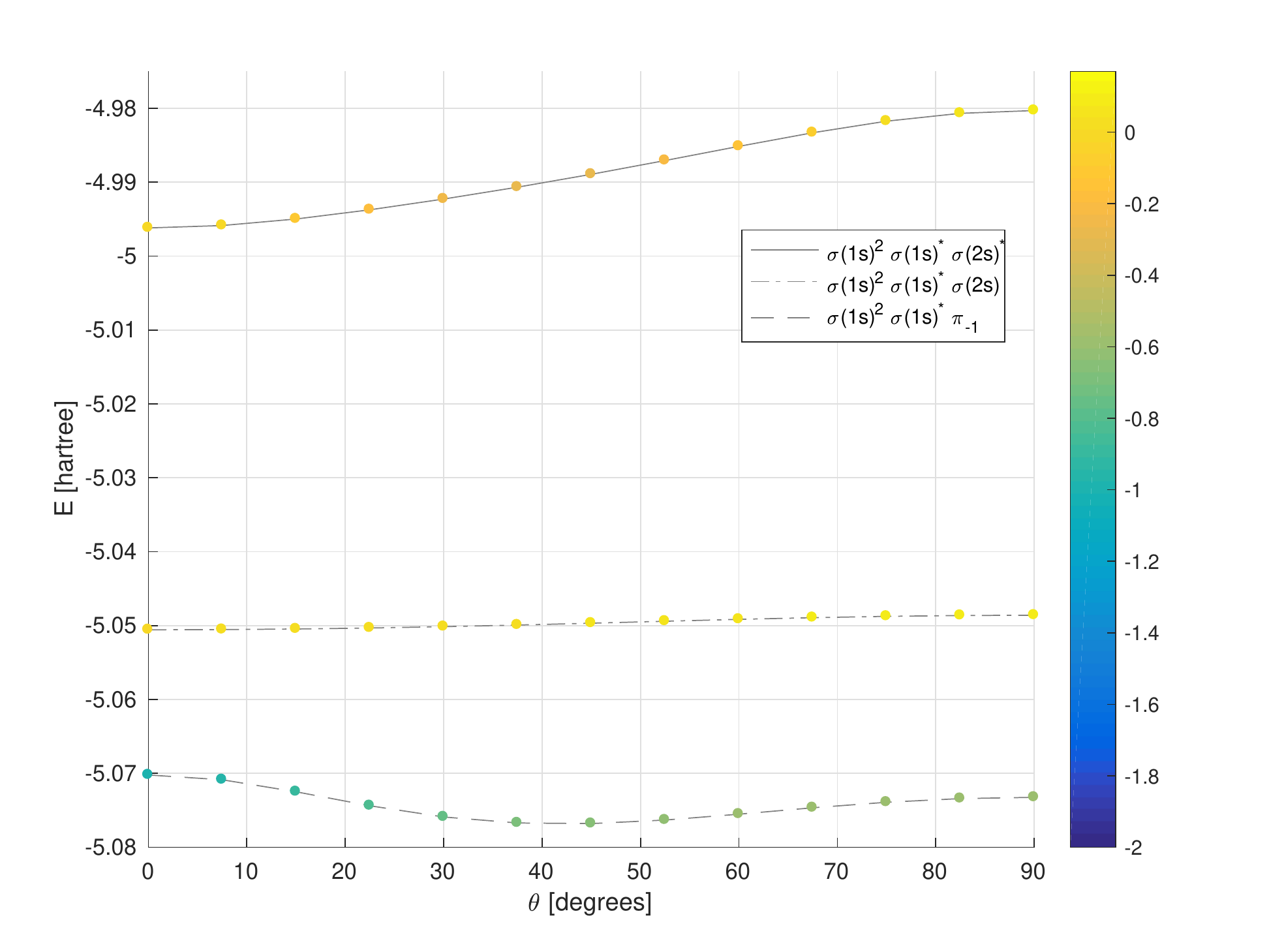}
  \caption{Energies of singlet excited states as a function of angle $\theta$ between the bond
    axis and magnetic field, with magnitudes fixed at $R=2a_0$ and $B = 0.2B_0$, respectively.}
  \label{figHe2_B020_singlet_angle}
\end{figure}

However, as seen in Fig.~\ref{figHe2_B020_singlet_angle}, for a fixed $R=2$~bohr, the first excited states in the parallel configuration smoothly turn into 
the corresponding perpendicular states as the angle $\theta$ is varied. In particular, the antibonding orbital $\sigma_{2\text s}^\ast$ transforms
smoothly into $\pi_{-1}$, both being of a$_\text u$ symmetry in skew orientations.  We note that the first excited state has a minimum at $\theta \approx 40^\circ$, 
which happens since the state a skew angles involves from 
$^{1}\Pi_\text g( \IIorbdouble{\sigma}{}{1s} \sigma_{1s}^* \pi_{-1})$, where
$\sigma^\ast_{1\text s}$ and $\pi_{-1}$ prefer perpendicular and parallel orientations, respectively.

The second and third excited state have $\Lambda_{\mathbf{B}}\approx 0.1$ in the perpendicular orientation. The second excited state also retains the same radial dissociation limit as the corresponding parallel state, while the third excited state acquires a different dissociation limit due to symmetry breaking and orbital mixing. Globally, this indicates conical intersections on the energy surfaces.

In the given basis set, the ground state is bound by about 50 microhartree in the perpendicular configuration and 30 microhartree in the parallel orientation. Hence, at this field strength, perpendicular paramagnetic bonding is negligible in the ground state. In the first excited state, the energy is lowered by 3 millihartree from parallel to perpendicular orientation. However, the dissociation limit is lowered too, leading to a reduction in radial binding energy from 0.14 hartree to 0.11 hartree. By contrast, the second excited state is is essentially unchanged and the third excited state is higher by 16 millihartree in the perpendicular orientation.

\subsubsection{Singlet potential-energy curves at $B=B_0$}

\begin{figure}[h]
  \includegraphics[width=\linewidth]{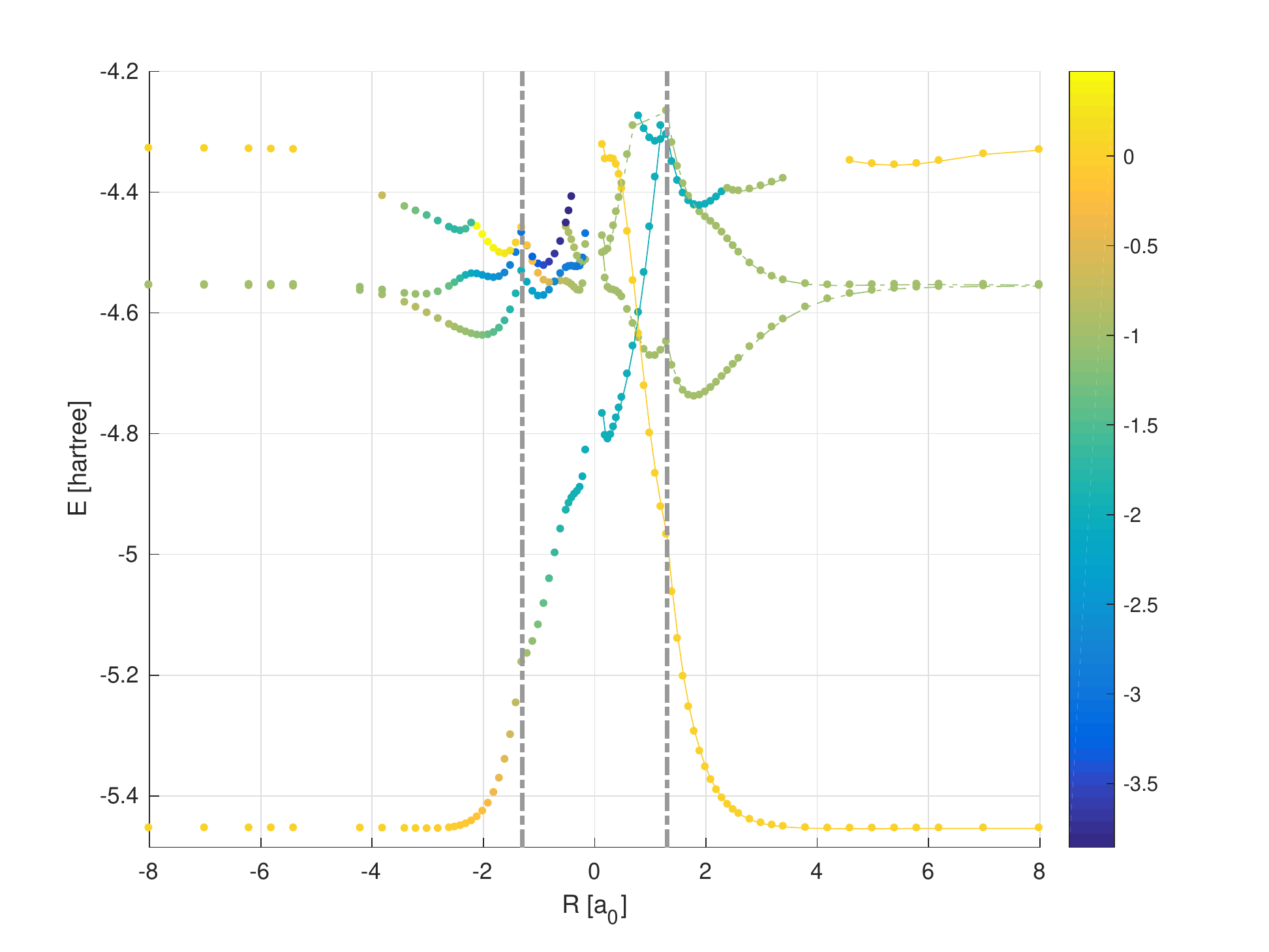}
  \caption{Dissociation curves for singlet states in perpendicular
    (negative half) and parallel (positive half) magnetic field $B = B_0$.}
  \label{figHe2_B100_singlet_dissoc}
\end{figure}

Potential-energy curves for singlet states at $B=B_0$ are shown in Fig.~\ref{figHe2_B100_singlet_dissoc}. 
In the parallel orientation, the weakly bound ground state is dominated by the $^{1}\Sigma_\text g(\IIorbdouble{\sigma}{}{1s} \IIorbdouble{\sigma}{*}{1s})$ configuration at all bond distances, 
just as for field strength $B = 0.2B_0$. Because of the orbital Zeeman effect, the first and second excited states have substantial $\pi$ character with $\Lambda_{\mathbf{B}}=-1$, being
predominantly $^{1}\Pi_\text g(\IIorbdouble{\sigma}{}{1s} \sigma_{1s}^{*} \pi_{-1})$ 
and $^{1}\Pi_\text u(\IIorbdouble{\sigma}{*}{1s} \sigma_{1s} \pi_{-1})$, respectively. The latter state is obtained from the former by promoting one electron from the 
$\sigma_{1\text s}$ bonding orbital to $\sigma_{1\text s}^\ast$ antibonding orbital. The two states therefore dissociate to the same limit but bind in different ways.
The first state is covalently bound with a deep energy minimum at $R=1.80a_0$, while the second is weakly bound with a shallow minimum at $R=4.42a_0$. 
The third excited state has $\delta$ character, being predominantly $^{1}\Delta_\text g(\IIorbdouble{\sigma}{*}{1\text s} \sigma_{1\text s} \delta_{-2})$ with $\Lambda_{\mathbf{B}}=-2$ 
and a minimum at $R=1.89a_0$. We note that the
$^{1}\Sigma_\text u(\IIorbdouble{\sigma}{}{1\text s} \sigma_{1s}^{*} \sigma_{2\text s})$ and $^{1}\Sigma_\text g(\IIorbdouble{\sigma}{}{1\text s} \sigma_{1\text s}^{*} \sigma_{2\text s}^{*})$ states,
which were the second and third excited states at $B=0.2B_0$, 
are not stabilized by the orbital Zeeman interaction  and have therefore been pushed high up in the spectrum by its diamagnetic interaction with the magnetic field.

The dissociation curves in the perpendicular field orientation are substantially different from those in the parallel orientation. However, since the dissociation limits are identical in the two orientations, 
the curves become increasingly similar with increasing bond distance.
The ground state 
$^1\text{A}_{\text{g}}(1\orbdouble{a}{g} 1\IITorbdouble{b}{*}{u})$, originating from the field-free state
\smash{$^{1}\Sigma_\text g^+(\IIorbdouble{\sigma}{}{1s} \IIorbdouble{\sigma}{*}{1s})$},
is stabilized by perpendicular paramagnetic bonding by about 1 millihartree, with equilibrium bond distance $R=3.01$~bohr, which is 1.4~bohr shorter than the bond distance in the parallel field orientation.
While the parallel and perpendicular ground states share the same dissociation limit, they tend to different states in the 
united atom limit---the parallel state becomes $^1\Sigma_{\text{g}}(1\text s^2 2\text s^2)$, while the perpendicular state 
becomes $^1\Sigma_{\text{g}}(1\text s^2 2\text p_{-1}^2)$. 

In the perpendicular orientation, the lowest three singlet excited states are
$^1\text{B}_{\text{u}}(1\orbdouble{a}{g} 1\IITorbNoSpin{b}{*}{u} 2\IITorbNoSpin{a}{*}{g})$, 
$^1\text{A}_{\text{g}}(1\orbdouble{a}{g} 1\IITorbNoSpin{b}{*}{u} 2\IITorbNoSpin{b}{*}{u})$, 
and $^1\text{B}_{\text{g}}(1\orbdouble{a}{g} 1\IITorbNoSpin{b}{*}{u} 1\orbNoSpin{a}{u})$ at $R=1.8a_0$;
at $R=3a_0$, the lowest states are 
\smash{$^1\text{B}_{\text{u}}(1\orbdouble{a}{g} 1\IITorbNoSpin{b}{*}{u} 2\IITorbNoSpin{a}{*}{g})$},
\smash{$^1\text{A}_{\text{g}}(1\IITorbdouble{b}{*}{u} 1\orbNoSpin{a}{g} 2\IITorbNoSpin{a}{*}{g})$}, and
\smash{$^1\text{B}_{\text{g}}(1\orbdouble{a}{g} 1\IITorbNoSpin{b}{*}{u} 2\orbNoSpin{b}{u})$}. 
Thus, while the first excited state retains its overall symmetry and orbital occupation at all distances beyond $R = 1.8$~bohr,
the second state retains
the overall symmetry but changes orbital character and the third state undergoes a level crossing with a state of different symmetry. 

It is interesting to compare the first exited states in the two orientations. In the parallel orientation, the first excited state is 
predominantly $^{1}\Pi_\text g(\IIorbdouble{\sigma}{}{1s} \sigma_{1s}^{*} \pi_{-1})$ of bond order one and a half and a deep minimum at $R=1.80a_0$.
In the perpendicular orientation, the
$^1\text{B}_{\text{u}}(1\orbdouble{a}{g} 1\IITorbNoSpin{b}{*}{u} 2\IITorbNoSpin{a}{*}{g})$ is best described as having orbital configuration
\smash{$\IIorbdouble{\sigma}{}{1s} \sigma_{1s}^{*} \pi_{\perp}^\ast$}, with an antibonding HOMO orbital
$\pi_\perp^\ast$ replacing $\pi_{-1}$. Whereas the orbital Zeeman interaction favours $\pi_{-1}$ in the parallel field orientation, it favours
$\pi_\perp^\ast$ in the perpendicular orientation, by the same mechanism that generates 
paramagnetic bonding. The reduced bond order of one in the perpendicular orientation gives a shallower minimum at a longer bond length $2.01a_0$ compared with the parallel orientation.
We note that the first excited state has $\Lambda_{\mathbf{B}} = -1.3$ at the energy minimum, 
indicating that it has acquired some $\delta$ character. In the united-atom limit, the perpendicular state acquires even more $\delta$ character, as shown by the colour coding in Fig.~\ref{figHe2_B100_singlet_dissoc}. 

The second excited state has a double minimum in the perpendicular orientation. The global minimum occurs at $R=3.14a_0$ with $\Lambda_{\mathbf{B}} = -1.1$, indicating some $\delta$ character. 
The orbital occupation in this region of the dissociation curve is $\IIorbdouble{\sigma}{*}{1s} \sigma_{1s} \pi_\perp^\ast$, with a negative bond order and a strong paramagnetic bonding (more than an order of magnitude stronger than
in the ground state) generated by three electrons occupying antibonding orbitals.

The local minimum 
in the second excited state occurs at the shorter distance of $R=1.81a_0$ and has $\Lambda = -2.5$, indicating a substantial increase in $\delta$ character. 
Compared with the second excited state in the parallel orientation, the energy is much lower (by $0.1E_\text h$ at $R=2a_0$).
Hence, the perpendicular paramagnetic bonding effect is orders of magnitude stronger than in the ground state.

The global picture of the singlet energy surfaces is complicated by level crossings at intermediate bond distances. 
Rotation of the first and second excited states at a fixed bond distance of $R=2a_0$ 
leads to a crossing at roughly $45^{\circ}$, even though the resulting perpendicular states share the same dissociation limit. Hence, at this bond distance, the strongly bound parallel state $\IIorbdouble{\sigma}{}{1s} \sigma_{1s}^{*} \pi_{-1}$ is rotated into a state near the higher minimum on the second excited perpendicular dissociation curve. The very weakly bound second excited state, with $\IIorbdouble{\sigma}{*}{1s} \sigma_{1s} \pi_{-1}$ character, in the parallel orientation is consequently rotated into the more strongly bound first excited state in the perpendicular orientation.

\subsubsection{Triplet potential-energy curves at $B=0.2B_0$}

\begin{figure}[h]
  \includegraphics[width=\linewidth]{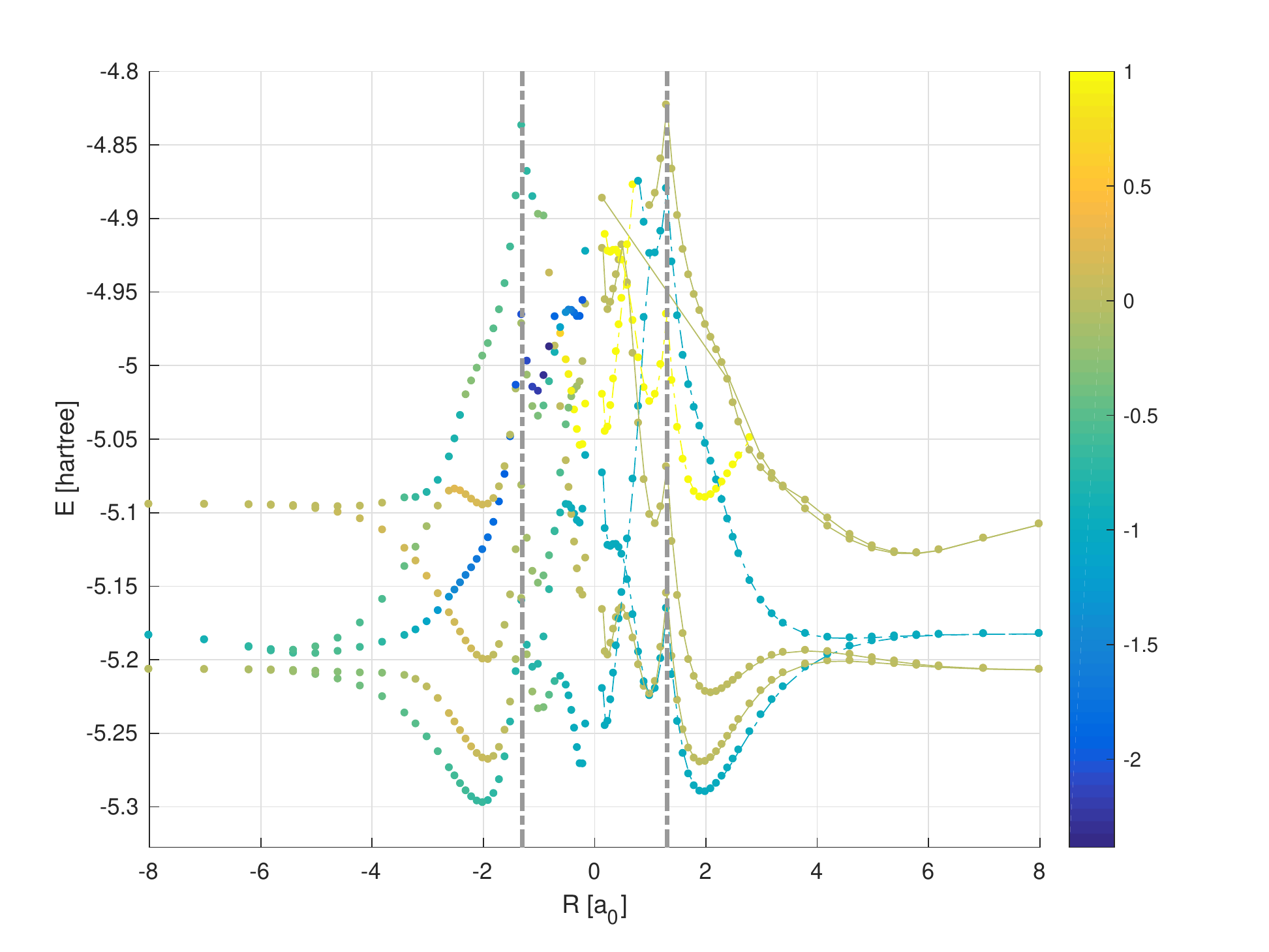}
  \caption{Dissociation curves for triplet states in perpendicular
    (negative half) and parallel (positive half) magnetic field $B =
    0.2B_0$. In the region between grey dashed lines, the curve is
    shifted by the nuclear repulsion energy and an additional
    quadratic fit to align the united atom limit to the same energy
    scale. Plot markers are coloured based on the AQAM value
    $\Lambda_{\mathbf{B}}$; scale indicated on the right.}
  \label{figHe2_B020_triplet_dissoc}
\end{figure}

The lowest triplet states at $B=0.2B_0$ are shown in Fig.~\ref{figHe2_B020_triplet_dissoc}.  
As in the field-free case, the triplet He$_2$ dissociation curves display many features that are analogous to the singlet curves. 
Equilibrium bond distances are roughly 2~bohr. Moreover, there are again conical intersections connecting low-lying states.

\begin{figure}[h]
  \includegraphics[width=\linewidth]{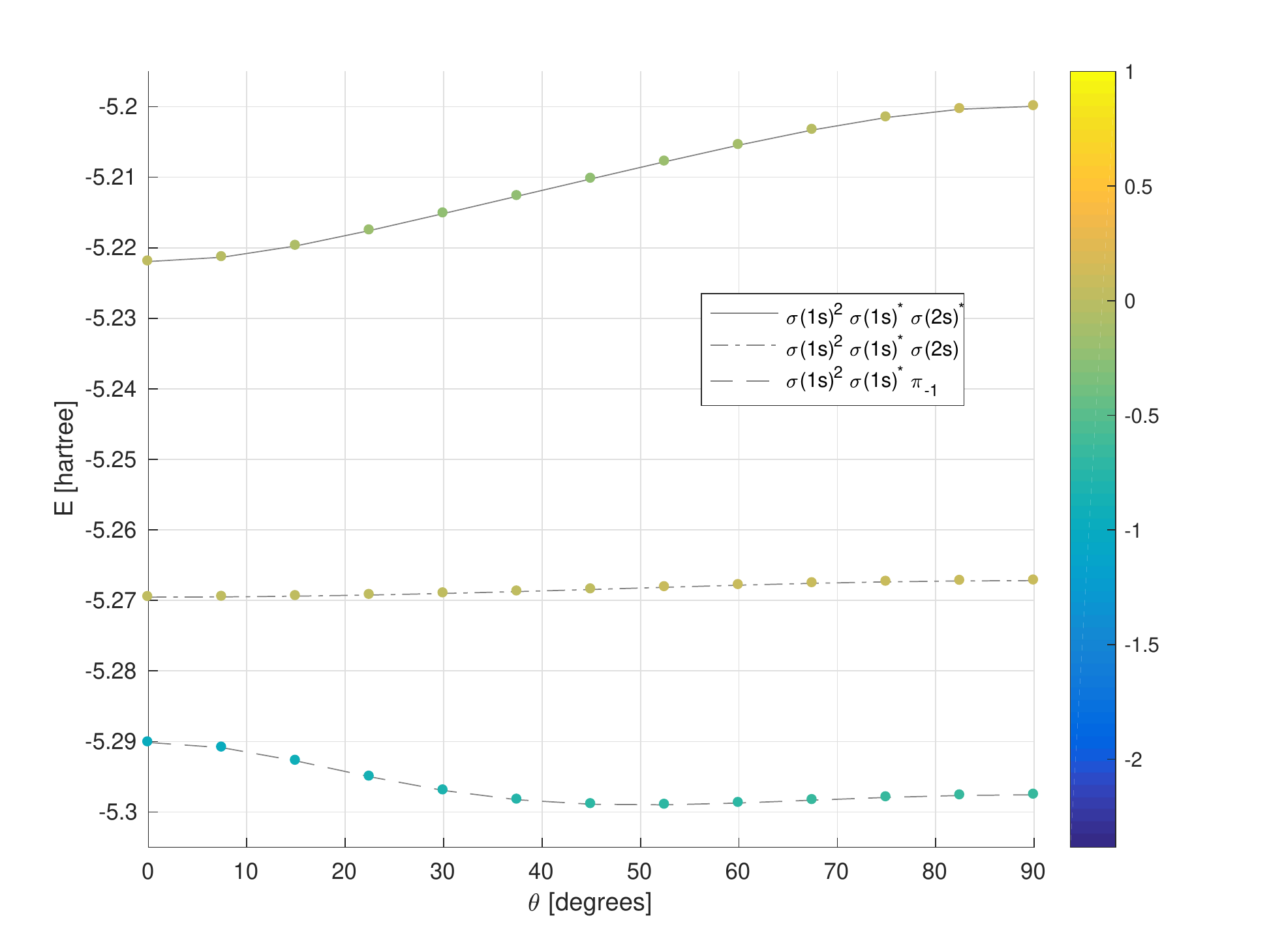}
  \caption{Energies of the lowest triplet states at $R=2a_0$ and $B = 0.2B_0$ plotted against the angle $\theta$ between the bond axis and magnetic axis} 
  \label{figHe2_B020_triplet_angle}
\end{figure}

In the parallel orientation at $R=2a_0$, the lowest triplet is predominantly $^{3}\Pi_\text g(\IIorbdouble{\sigma}{}{1s} \IIorbdown{\sigma}{*}{1s} \IIorbdown{\pi}{}{$-1$})$ with $\Lambda_{\mathbf{B}}=-1$. 
The second and third triplet states have $\Lambda_{\mathbf{B}}=0$ and pure $\sigma$ character, 
with configurations predominantly $\IIorbdouble{\sigma}{}{1s} \IIorbdown{\sigma}{*}{1s} \IIorbdown{\sigma}{}{2s}$ and $\IIorbdouble{\sigma}{}{1s} \IIorbdown{\sigma}{*}{1s} \IIorbdown{\sigma}{*}{2s}$, respectively. 
The fourth triplet state at $R=2a_0$ is related to the first triplet by reversed sign of the angular momentum, having $\Lambda_{\mathbf{B}}=+1$ and an electron configuration dominated 
by $\IIorbdouble{\sigma}{}{1s} \IIorbdown{\sigma}{*}{1s} \IIorbdown{\pi}{}{+1}$. However, at
slightly longer bond distances, the fourth triplet state is instead one with $\Lambda_{\mathbf{B}}=-1$ and 
configuration $\IIorbdouble{\sigma}{}{1s} \IIorbdown{\sigma}{*}{1s} \IIorbdown{\pi}{*}{$-1$}$, which shares the same dissociation limit as the
triplet ground state at $R=2a_0$.

Fixing the bond distance at $R=2a_0$ (close to the equilibrium bond distances of the lowest triplet states in all field orientations)
and plotting the energies as function of the angle $\theta$ between the bond axis and the field vectors, we obtain the curves in Fig.~\ref{figHe2_B020_triplet_angle}. 
In the perpendicular orientation at this bond distance, the lowest triplet is predominantly $^{3} \text{A}_{\text{g}}(1\orbdouble{a}{g} 1\IITorbdown{b}{*}{u} 2\IITorbdown{b}{*}{u})$, where
the $\sigma^\ast_{2\text s}$ HOMO of b$_{\text u}$ symmetry has evolved smoothly from the $\pi_{-1}$ HOMO in the parallel orientation, both being of a$_\text u$ symmetry in skew orientations. 
In the process, the AQAM projection has decreased from $-1$ in the parallel orientation to $-0.7$ in the perpendicular orientation. 
This lowest triplet state has a preferred field orientation of about $50^\circ$, a compromise between the preferred perpendicular
orientation of the antibonding $\sigma$ orbitals and 
preferred parallel orientation of the $\pi$ orbital, in the same way as for singlet states in Fig.~\ref{figHe2_B020_singlet_angle}.
The energy of the lowest triplet is about 7~millihartree lower in the perpendicular orientation than in the parallel orientation, by
paramagnetic stabilization of the antibonding orbitals. 
However, since the paramagnetic stabilization also lowers the dissociation limit, the bond is actually weaker in the perpendicular orientation.

Even though $^{3}\Pi_\text g(\IIorbdouble{\sigma}{}{1s} \IIorbdown{\sigma}{*}{1s} \IIorbdown{\pi}{}{$-1$})$ 
and $^{3} \text{A}_{\text{g}}(1\orbdouble{a}{g} 1\IITorbdown{b}{*}{u} 2\IITorbdown{b}{*}{u})$ are the
lowest parallel and perpendicular triplet states at a bond distance of $2a_0$, smoothly connected to each other by field rotation, they have different radial dissociation limits,
the latter having the same dissociation limit as the parallel states $^{3}\Sigma_\text u(\IIorbdouble{\sigma}{}{1s} \IIorbdown{\sigma}{*}{1s} \IIorbdown{\sigma}{}{2s})$ and 
$^{3}\Sigma_\text g(\IIorbdouble{\sigma}{}{1s} \IIorbdown{\sigma}{*}{1s} \IIorbdown{\sigma}{*}{2s})$. 
The $^{3}\Pi_\text g$ state, on the other hand,  crosses the $^{3}\Sigma_\text u$ and $^{3}\Sigma_\text g$ states around $R = 4a_0$, dissociating
into states of higher energy.
and the minimum at $R=2.1a_0$ is thus a manifestation of the perpendicular paramagnetic bonding mechanism. 

The second triplet state in Fig.~\ref{figHe2_B020_triplet_angle} changes smoothly from
$^{3}\Sigma_\text u(\sigma_{1\text{s}}^2 \IIorbdown{\sigma}{*}{1s} \orbdown{$\sigma$}{2\text s})$ to
$^3 \text{B}_{\text{u}}(1\orbdouble{a}{g} 1\IITorbdown{b}{*}{u} 2\orbdown{a}{g})$ from the parallel to the perpendicular orientation, increasing its energy slightly and its AQAM projection form zero to 0.1.
The third triplet state changes more dramatically (but smoothly) from
$^{3}\Sigma_\text g(\sigma_{1\text{s}}^2 \IIorbdown{\sigma}{*}{1s} \IIorbdown{\sigma}{*}{2s})$ to
$^3 \text{B}_{\text{g}}(1\orbdouble{a}{g} 1\IITorbdown{b}{*}{u} 1\orbdown{a}{u})$ as the HOMO changes from $\sigma^\ast_\text{2s}$ to $\pi_\shortparallel$ character. 
Its energy increases
by about 20 millihartree, while its AQAM projection first decreases to $-0.2$ at $\theta \approx 40^\circ$, after which it increases to 0.1 in the perpendicular orientation.

To summarize, the three lowest triplet states at $R = 2a_0$ differ in their HOMOs, which, in order of increasing energy, are
$\pi_{-1} < \sigma_{2\text s} < \sigma^\ast_{2 \text s}$ in the parallel field orientation (by paramagnetic stabilization of $\pi_{-1}$) 
and $\sigma^\ast_{2\text s} < \sigma_{2 \text s} < \pi_\shortparallel$ in the perpendicular orientation (by paramagnetic stabilization of $\sigma^\ast_{2\text s}$). We note that, 
even though the $\pi$ and $\sigma^\ast_{2\text s}$ orbitals are of
different symmetries in the parallel and perpendicular field orientations,  they are of the same symmetry in skew orientations and may therefore transform smoothly into each other.

\begin{figure}[h]
  \includegraphics[width=\linewidth]{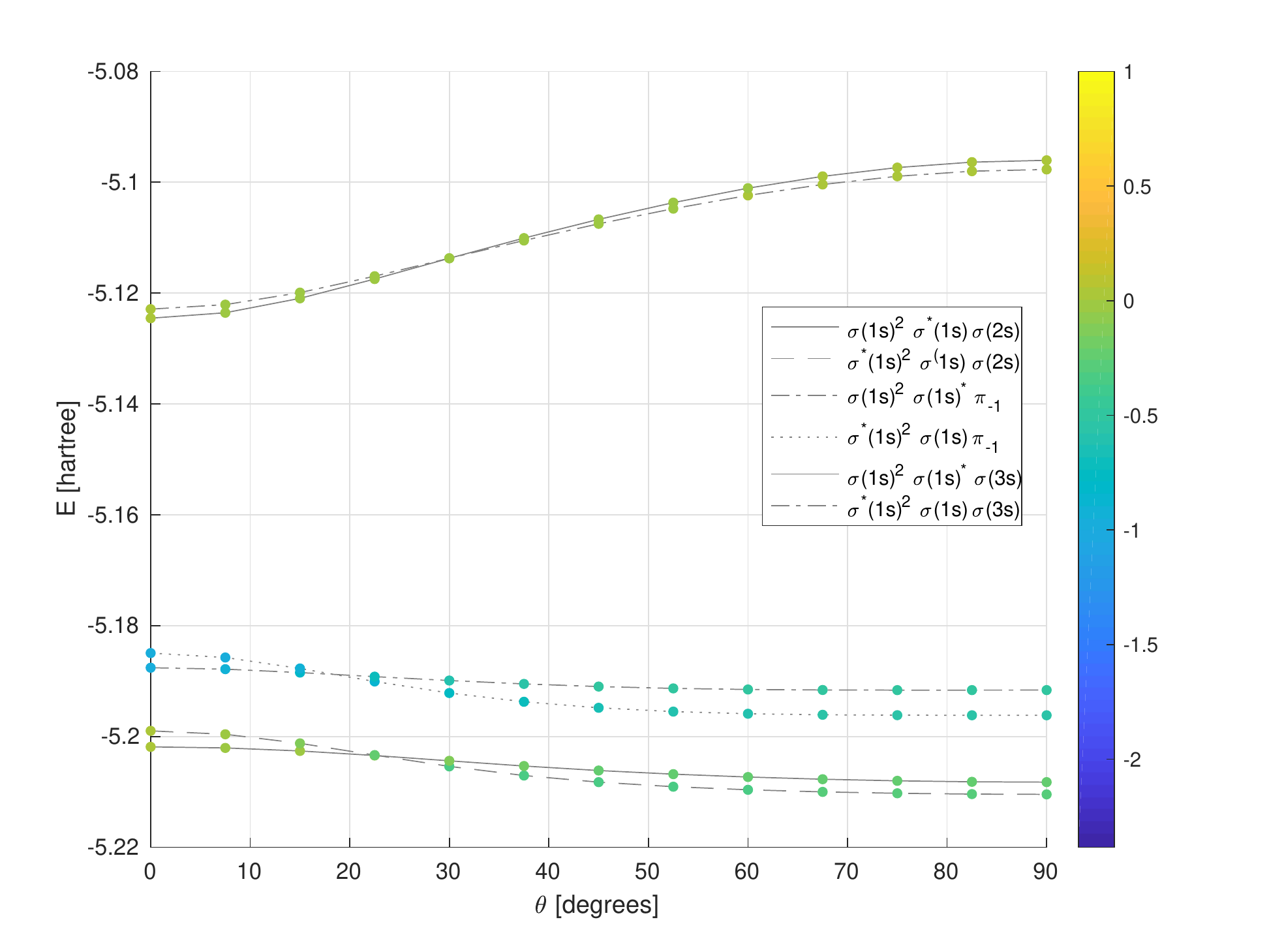}
  \caption{Triplet states as a function of angle between the bond
    axis and magnetic field, with magnitudes fixed at $R=5a_0$ and
    $B = 0.2B_0$, respectively.}
  \label{figHe2_B020_triplet_angle_R5}
\end{figure}

We consider next the electronic states closer to the dissociation limit, at $R = 5a_0$. In Fig.~\ref{figHe2_B020_triplet_dissoc}, there are 
three distinct pairs of states both in the parallel field orientation and in the perpendicular orientation---in Fig.~\ref{figHe2_B020_triplet_angle_R5}, 
we have plotted the energies of the corresponding states against the angle $\theta$ at the fixed bond distance $R=5a_0$. Each pair consists of two close-lying states
with the same dissociation limit but of different symmetries (gerade and ungerade) arising from different occupations of bonding and antibonding orbitals.
Since we are close to the dissociation limit, the electronic states are typically multiconfigurational, with large
contributions from two configurations. We consider the lowest pair of electronic states first.

At $R=5a_0$, the lowest parallel state is predominantly 
$^{3}\Sigma_\text u(\IIorbdouble{\sigma}{}{1s} \IIorbdown{\sigma}{*}{1s} \IIorbdown{\sigma}{}{2s})$ with one occupied antibonding orbital, while the next state
$^{3}\Sigma_\text g(\IIorbdouble{\sigma}{}{1s} \IIorbdown{\sigma}{*}{1s} \IIorbdown{\sigma}{*}{2s},
\IIorbdouble{\sigma}{*}{1s} \IIorbdown{\sigma}{}{1s} \IIorbdown{\sigma}{}{2s})$ has large contributions from two configurations, both with two
occupied antibonding orbitals. 
Although these close-lying states have nearly reached their radial dissociation limits at this bond distance, 
both are lowered in energy as the angle is increased to $90^\circ$, by paramagnetic stabilization of the antibonding orbitals in the nonparallel field.
With two occupied antibonding orbitals, the energy lowering is larger for the $^{3}\Sigma_\text g$ state,
which becomes the lowest state at $22^\circ$. For this state, the AQAM projection changes from zero in the parallel field
orientation to $-0.3$ in the perpendicular orientation; for
the $^{3}\Sigma_\text u$ state, the AQAM projection changes less. In the perpendicular field orientation, the symmetries of the states are
$^{3} \text{A}_{\text{g}}$ for the lower-energy state and $^3 \text{B}_{\text{u}}$ for the higher state.

The next pair of states are 
$^3\Pi_\text g(\IIorbdouble{\sigma}{}{1s} \IIorbdown{\sigma}{*}{1s} \IIorbdown{\pi}{}{$-1$},\IIorbdouble{\sigma}{*}{1s} \IIorbdown{\sigma}{}{1s} \IIorbdown{\pi}{*}{$-1$})$ 
and $^{3}\Pi_\text u(\IIorbdouble{\sigma}{*}{1s} \IIorbdown{\sigma}{}{1s} \IIorbdown{\pi}{}{$-1$}
,\IIorbdouble{\sigma}{}{1s} \IIorbdown{\sigma}{*}{1s} \IIorbdown{\pi}{*}{$-1$})$ 
in the parallel orientation, both with more weight on the configuration containing the bonding orbital $\pi_{-1}$. The 
$^{3}\Pi_\text u$ state is slightly higher in energy, have a doubly occupied $\sigma^\ast_{1\text s}$ orbital in the dominant configuration.
As a result of paramagnetic stabilization, the gerade and ungerade states cross at about $\theta = 30^\circ$. 
Furthermore, with increasing $\theta$, 
the configurations containing the antibonding orbital $\pi_\perp^\ast$ (originating from $\pi_{-1}^\ast$)
become more important than the configurations containing $\pi_\perp$. In the perpendicular orientation, the dominant configurations are
$^{3} \text{B}_{\text{u}}(1\orbdouble{a}{g} 1\IITorbdown{b}{*}{u} 3\IITorbdown{a}{*}{g})$ and
\smash{$^3\text{A}_{\text{g}}(1\IITorbdouble{b}{*}{u} 1\orbNoSpin{a}{g} 3\IITorbNoSpin{a}{*}{g})$} 
where $3\orbNoSpin{a}{g}$ is the $\pi^\ast_\perp$
orbital. Both states 
have minima at roughly $R=5a_0$, which are manifestations of perpendicular paramagnetic bonding.

Notably, there are indications of perpendicular paramagnetic bonding also in states of higher angular momentum. 
Tracing the third triplet state at $R=5a_0$ in the perpendicular 
orientation to shorter bond distances, we find that it develops an AQAM value of $-1.5$ at $R\approx 2a_0$, implying that it has acquired some $\delta$-orbital character, although the orbital Zeeman effect due to the larger magnitude of the angular momentum is not enough to offset other effects, in particular the electrostatic repulsion, at these bond lengths.

\subsubsection{Triplet potential-energy curves at $B=B_0$}

At $B=B_0$, the two lowest parallel electronic states, both with $\Lambda_{\mathbf{B}}=-1$, have the same dissociation limit and are energetically well separated from the other states. 
The lower state is dominated by a single electron configuration $^{3}\Pi_\text g(\IIorbdouble{\sigma}{}{1s} \IIorbdown{\sigma}{*}{1s} \IIorbdown{\pi}{}{$-1$}$),
while the higher $^{3}\Pi_\text u$ state is more mixed, with weights 66\% on $\IIorbdouble{\sigma}{}{1s} \IIorbdown{\sigma}{*}{1s} \IIorbdown{\pi}{*}{$-1$}$
and 29\% on $\IIorbdouble{\sigma}{*}{1s} \IIorbdown{\sigma}{}{1s} \IIorbdown{\pi}{}{$-1$}$ at $R=2a_0$.

\begin{figure}[h]
  \includegraphics[width=\linewidth]{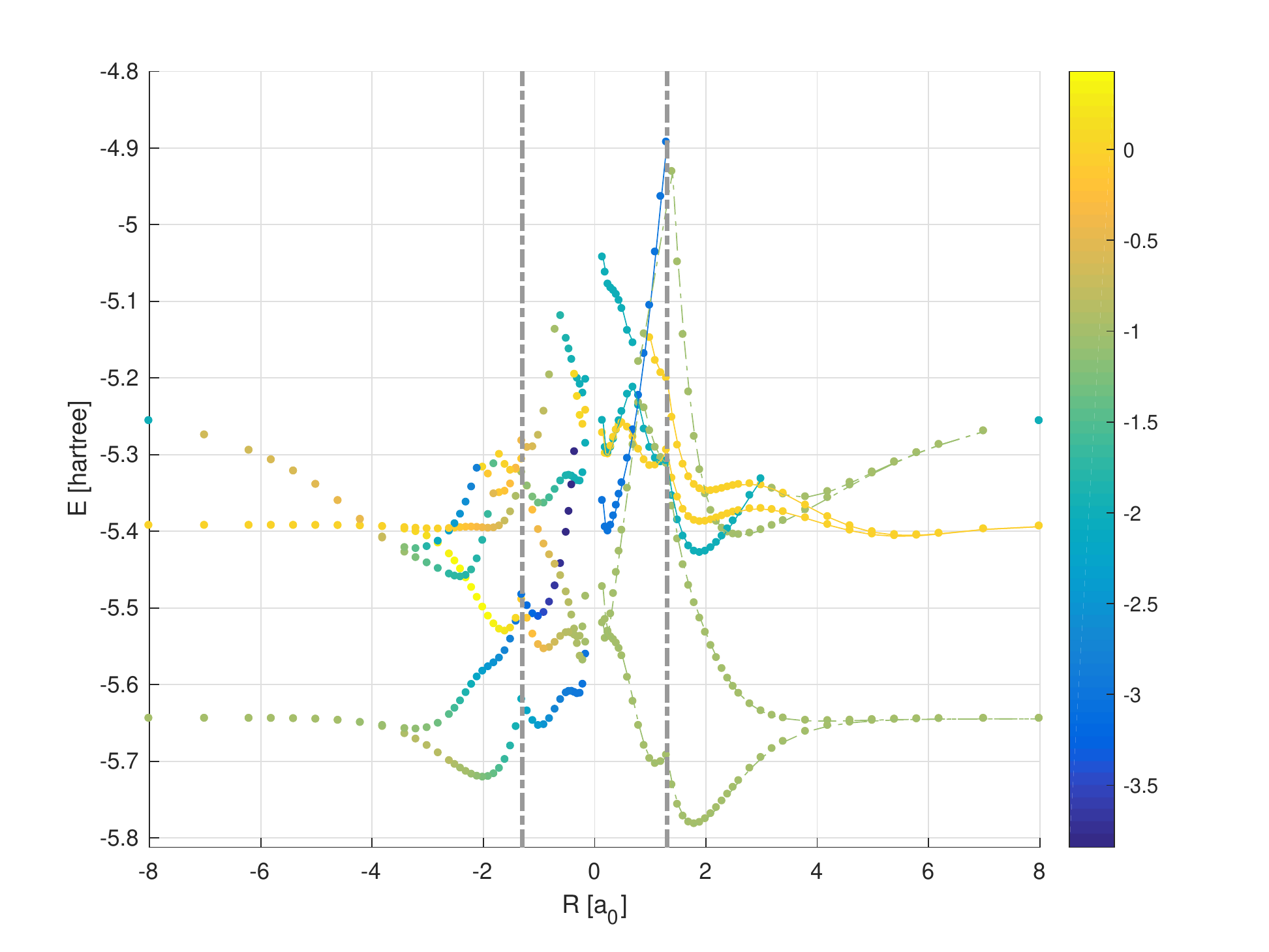}
  \caption{Dissociation curves for triplet states in perpendicular
    (negative half) and parallel (positive half) magnetic field $B = B_0$.}
  \label{figHe2_B100_triplet_dissoc}
\end{figure}

\begin{figure}[h]
  \includegraphics[width=\linewidth]{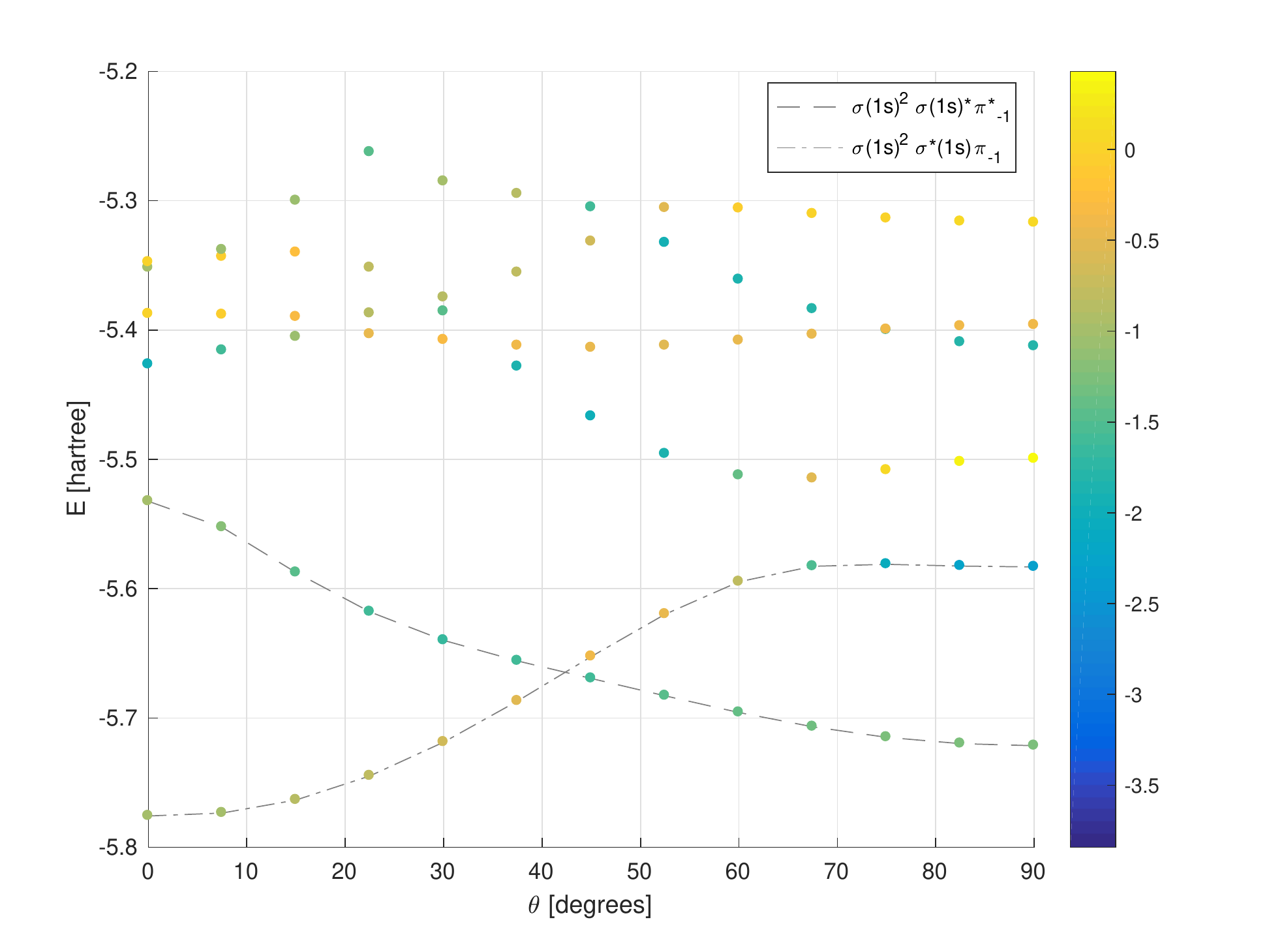}
  \caption{Triplet states as a function of angle between the bond
    axis and magnetic field, with magnitudes fixed at $R=2a_0$ and
    $B = B_0$, respectively.}
  \label{figHe2_B100_triplet_angle}
\end{figure}

The spectrum above these states is more complicated, 
with states closer together and crossings in the interval $2a_0 \leq R \leq 4a_0$.
At $R=2a_0$, the third electronic state is $^3\Delta_\text u(\IIorbdouble{\sigma}{}{1s} \IIorbdown{\sigma}{*}{1s} \IIorbdown{\delta}{}{$-2$})$ with $\Lambda_{\mathbf{B}}=-2$,
while the fourth state is $^3\Sigma_\text u(\IIorbdouble{\sigma}{}{1s} \IIorbdown{\sigma}{*}{1s} \IIorbdown{\sigma}{}{2s})$ with $\Lambda_{\mathbf{B}} = 0$. The fifth
state is again a mixed $^3\Pi_\text u$ state with $\Lambda_{\mathbf{B}}=-1$; it has the same dominant configurations as the second state but with 
weights 65\% on $\IIorbdouble{\sigma}{*}{1s} \IIorbdown{\sigma}{}{1s} \IIorbdown{\pi}{}{$-1$}$ and 29\% on $\sigma_{1s}^{2} \IIorbdown{\sigma}{*}{1s} \IIorbdown{\pi}{*}{$-1$}$.
The sixth state is $^3\Sigma_\text g(\IIorbdouble{\sigma}{}{1s} \IIorbdown{\sigma}{*}{1s} \IIorbdown{\sigma}{*}{2s})$ with $\Lambda_{\mathbf{B}} = 0$.

Tracking the lowest two states from the parallel orientation through a $90^{\circ}$ rotation is straightforward. 
As seen in Fig.~\ref{figHe2_B100_triplet_angle}, at a fixed bond distance of $R=2a_0$, the states cross at about $40^{\circ}$. 
The bound parallel state $^3\Pi_\text g(\IIorbdouble{\sigma}{}{1s} \IIorbdown{\sigma}{*}{1s} \IIorbdown{\pi}{}{$-1$})$ is 
deformed into a nearly unbound, dissociative state $^3\text{A}_{\text{g}}( 1\orbdouble{a}{g} 1\IITorbdown{b}{*}{u} 2\IITorbdown{b}{\circ}{u})$
on the perpendicular side as the bonding $\pi_{-1}$ orbital transforms into the antibonding $\sigma^\ast_{2\text s}$ orbital.
There is, however, a minimum at the larger bond distance of $R=3.2a_0$, with $\Lambda_{\mathbf{B}}=-1.1$ and a depth of 14 millihartree, for this state,
in part generated by paramagnetic bonding. At the same time, the unbound parallel state dominated by
$^3\Pi_\text u(\IIorbdouble{\sigma}{}{1s} \IIorbdown{\sigma}{*}{1s} \IIorbdown{\pi}{*}{$-1$})$ transforms into the 
$^3\text{B}_{\text{u}}( 1\orbdouble{a}{g} 1\IITorbdown{b}{*}{u} 2\IITorbdown{a}{\circ}{g})$ 
with $\Lambda_{\mathbf{B}}=-1.25$, which is bound by perpendicular paramagnetic bonding. 
Compared with the parallel orientation, the energy difference is almost 0.2~hartree---a manifestation of very strong 
perpendicular paramagnetic bonding. However, this state is not the global minimum over all triplet states and geometries, which instead occurs in the parallel orientation.

\subsubsection{Quintet potential-energy curves at $B=0.2B_0$}

\begin{figure}[h]
  \includegraphics[width=\linewidth]{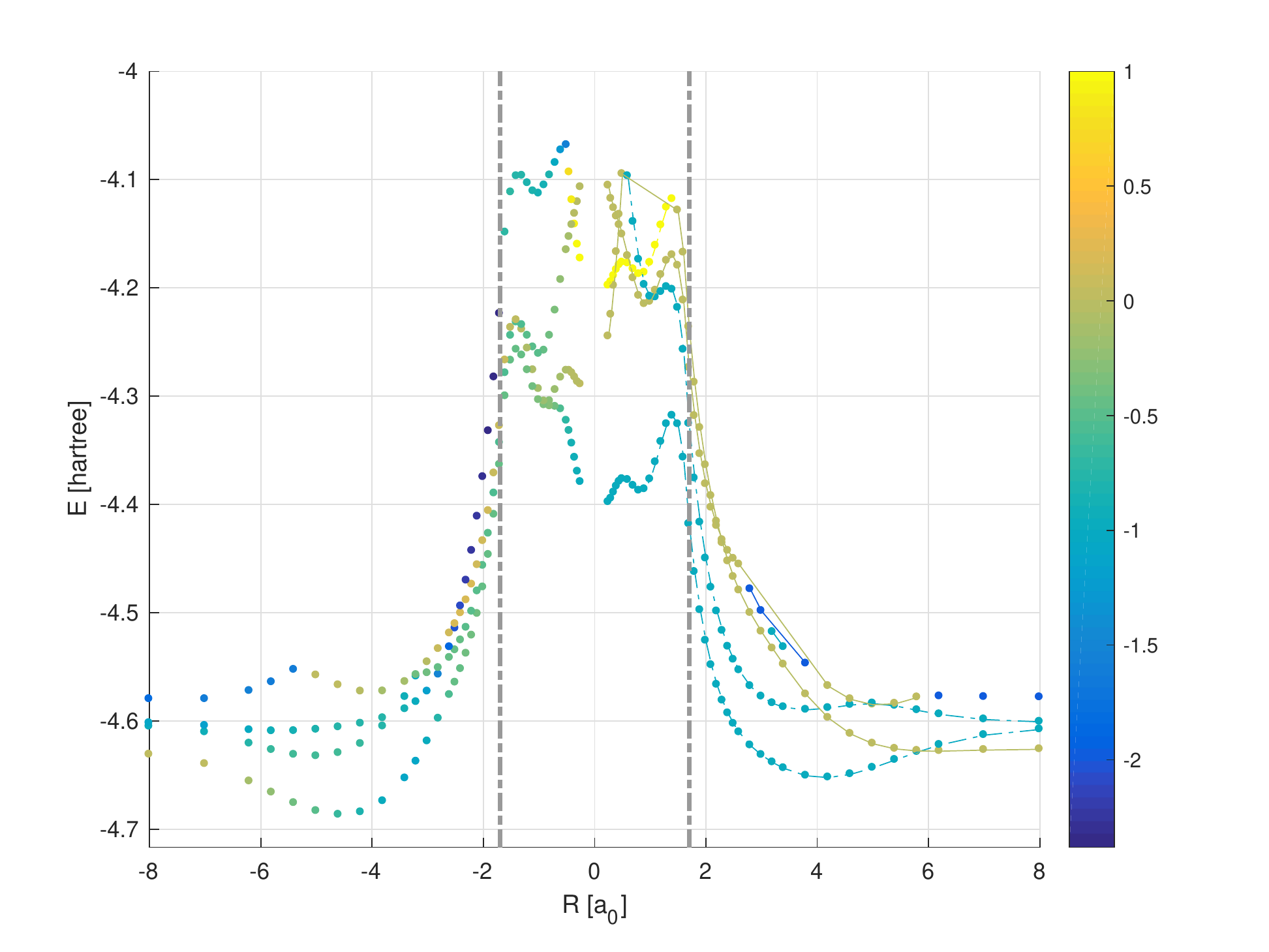}
  \caption{Dissociation curves for quintet states in perpendicular
    (negative half) and parallel (positive half) magnetic field $B = 0.2B_0$.}
  \label{figHe2_B020_quintet_dissoc}
\end{figure}

Dissociation curves for quintet states subject to a field $B=0.2B_0$ are shown in Fig.~\ref{figHe2_B020_quintet_dissoc}. 
In the parallel orientation, the lowest-lying parallel states at $R=4.2a_0$ alternate between $\Lambda_{\mathbf{B}} = -1$ and $\Lambda_{\mathbf{B}} = 0$. 
The lowest quintet state is predominantly $^5\Pi_\text g(\IIorbdown{\sigma}{}{1s} \IIorbdown{\sigma}{*}{1s} \IIorbdown{\sigma}{}{2s} \IIorbdown{\pi}{}{$-1$})$ and covalently bound,
while the second quintet is multiconfigurational
$^5\Sigma_\text g(0.53\,\IIorbdown{\sigma}{}{1s} \IIorbdown{\sigma}{*}{1s} \IIorbdown{\sigma}{}{2s} \IIorbdown{\sigma}{*}{2s}, 0.21\,\IIorbdown{\sigma}{}{1s} \IIorbdown{\sigma}{*}{1s} \IIorbdown{\sigma}{*}{2s} \IIorbdown{\sigma}{}{2p})$
and noncovalently bound.
The third and fourth parallel quintets at $R=4.2a_0$ 
are predominantly $^5\Pi_\text u(\IIorbdown{\sigma}{}{1s} \IIorbdown{\sigma}{*}{1s} \IIorbdown{\sigma}{*}{2s} \IIorbdown{\pi}{}{$-1$})$ and
$^5\Sigma_\text u(\IIorbdown{\sigma}{}{1s} \IIorbdown{\sigma}{*}{1s} \IIorbdown{\sigma}{}{2s} \IIorbdown{\sigma}{}{2p})$, respectively.

\begin{figure}[h]
  \includegraphics[width=\linewidth]{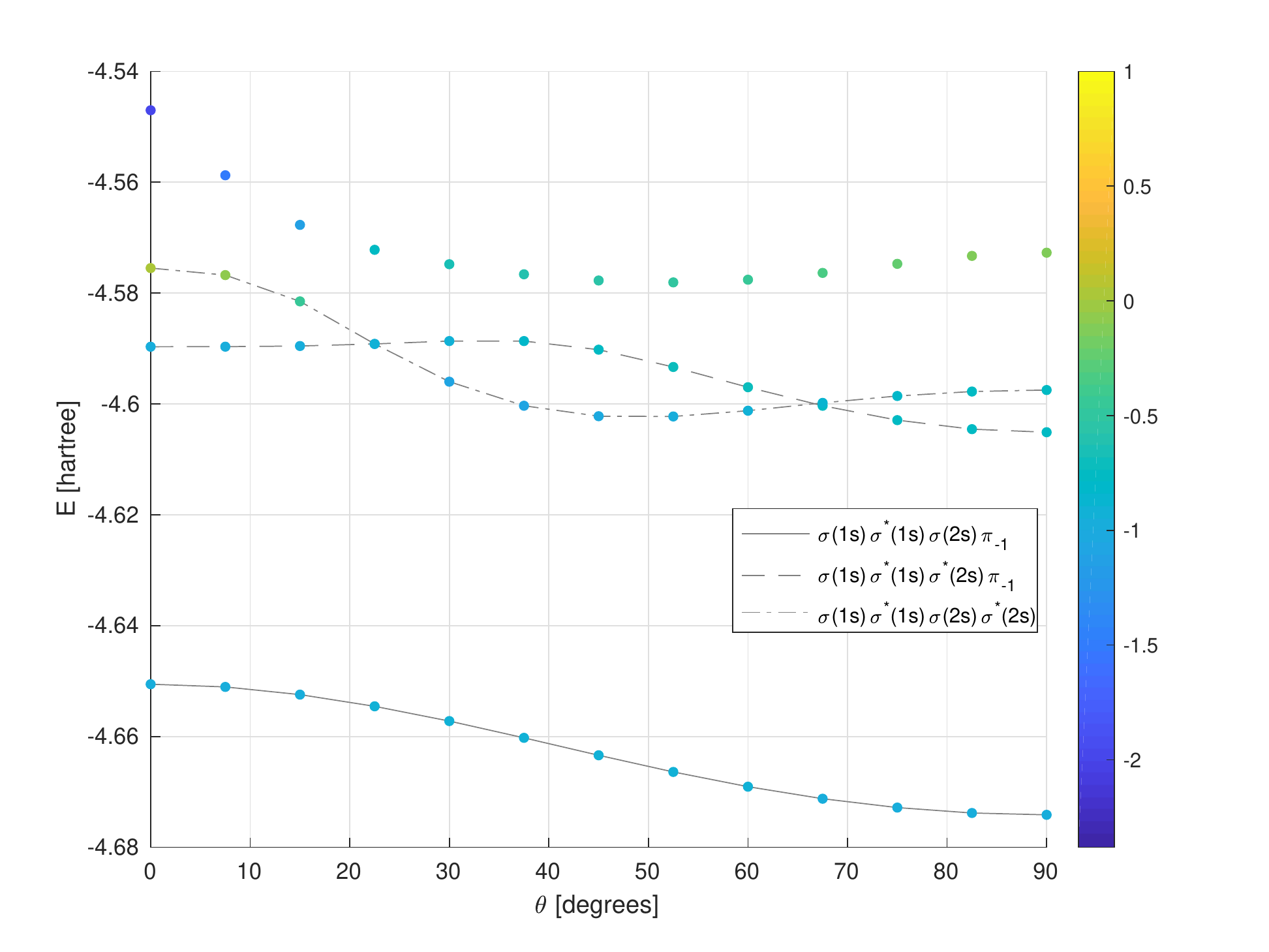}
  \caption{Energies of the lowest quintet states as a function of angle between the bond
    axis and magnetic field, with magnitudes fixed at $R=3.8a_0$ and
    $B = 0.2B_0$, respectively.}
  \label{figHe2_B020_quintet_angle}
\end{figure}

When rotated from parallel ($\theta=0^{\circ}$) to perpendicular ($\theta=90^{\circ}$) orientation at the slightly shorter bond length of $R=3.8a_0$, the lowest quintet state does not undergo any level crossing; see 
Fig.~\ref{figHe2_B020_quintet_angle}. From the parallel to the perpendicular orientation, 
the binding HOMO $\pi_{-1}$ transforms into the antibonding $\sigma_{2
  \text s}^\ast$ orbital of symmetry $\orbNoSpin{b}{u}$, while the antibonding $\pi^\ast_\perp$ of $\orbNoSpin{a}{g}$ symmetry is stabilized paramagnetically. 
The resulting lowest perpendicular state becomes 
$^5 \text{A}_{\text{g}}(0.44\, 1\orbdown{a}{g} 1\IITorbdown{b}{*}{u} 2\IITorbdown{b}{\circ}{u} 3\IITorbdown{a}{\circ}{g}, 0.27 \, 1\orbdown{a}{g} 1\IITorbdown{b}{*}{u} 2\IITorbdown{b}{\circ}{u} 2\orbdown{a}{g})$ where the dominant
configuration has more occupied antibonding than bonding orbitals.  Nevertheless, because of paramagnetic stabilization of antibonding orbitals, the total energy decreases by more then 20 millihartree,
while the dissociation energy increases from 48 to 59 millihartree as the covalent bond in the parallel orientation is replaced by a paramagnetic bond in the perpendicular orientation.
Considering the relatively small magnitude of the magnetic field in this case, this provides an example of paramagnetic bonding that is orders of magnitude stronger than 
the initially reported bonding in the lowest H$_2$ triplet and He$_2$ singlet states~\cite{LANGE_S337_327}. 

As we go from the parallel to perpendicular field orientation,
the AQAM projection of the lowest state decreases in magnitude, from $-1$ to $-0.8$, providing another example where this quantity does not directly capture the energy stabilization by the orbital Zeeman interaction. 
However, the radial dissociation limits are different in the parallel and perpendicular orientations, with the latter corresponding to two helium atoms in the $1\text{s} 2\text{s}$ triplet state. 
Hence, from this perspective, the AQAM value changes from zero in the perpendicular radial dissociation limit to about $-0.7$ at the minimum, correctly indicating a stabilizing orbital Zeeman effect compared to the dissociation limit.

At $R=3.8a_0$, the second and third quintet states in the parallel orientation are $^5\Pi_\text u$ and $^5\Sigma_\text g$, respectively.
As seen from Fig.~\ref{figHe2_B020_quintet_angle}, these states undergo two level crossings from the parallel to perpendicular orientation, at about 20 and 70 degrees. 
The double crossing arises since $^5\Sigma_\text g$ has an energy minimum at about 45 degrees, while $^5\Pi_\text u$ has a maximum at about 35 degrees. 
We note that $^5\Sigma_\text g$ develops a substantial AQAM projection of $-0.8$ at 90 degrees and is even close to $-1.1$ at intermediate angles of 30--40 degrees (i.e., near the energy minimum). 
Both states have a lower energy at 90 degrees than at 0 degrees, with configurations
$^5\text{B}_{\text{u}}(1\orbdown{a}{g} 1\IITorbdown{b}{*}{u} 2\orbdown{a}{g} 3\IITorbdown{a}{\circ}{g})$ and
$^5 \text{A}_{\text{g}}(0.39\, 1\orbdown{a}{g} 1\IITorbdown{b}{*}{u} 2\IITorbdown{b}{*}{u} 2\orbdown{a}{g}, \, 0.32 \, 1\orbdown{a}{g} 1\IITorbdown{b}{*}{u} 2\IITorbdown{b}{\circ}{u} 3\IITorbdown{a}{\circ}{g})$, respectively.

The fourth perpendicular quintet state at $R = 3.8a_0$ is
$^5 \text{B}_{\text{g}}(0.49\, 1\orbdown{a}{g} 1\IITorbdown{b}{*}{u} 2\orbdown{a}{g} 1\orbdown{a}{u}, 0.33 \, 1\orbdown{a}{g} 1\IITorbdown{b}{*}{u} 3\IITorbdown{a}{\circ}{g} 1\orbdown{a}{u})$. 
However, after a level crossing at $R=5.8a_0$, the fourth state is $^5\text{A}_{\text{g}}(0.55 \, 1\orbdown{a}{g} 1\IITorbdown{b}{*}{u} 3\orbdown{a}{g} 3\orbdown{b}{u},  0.30 \, 1\orbdown{a}{g} 1\IITorbdown{b}{*}{u} 2\orbdown{a}{g} 3\IITorbdown{b}{\circ}{u})$.

\begin{figure}[h]
  \includegraphics[width=\linewidth]{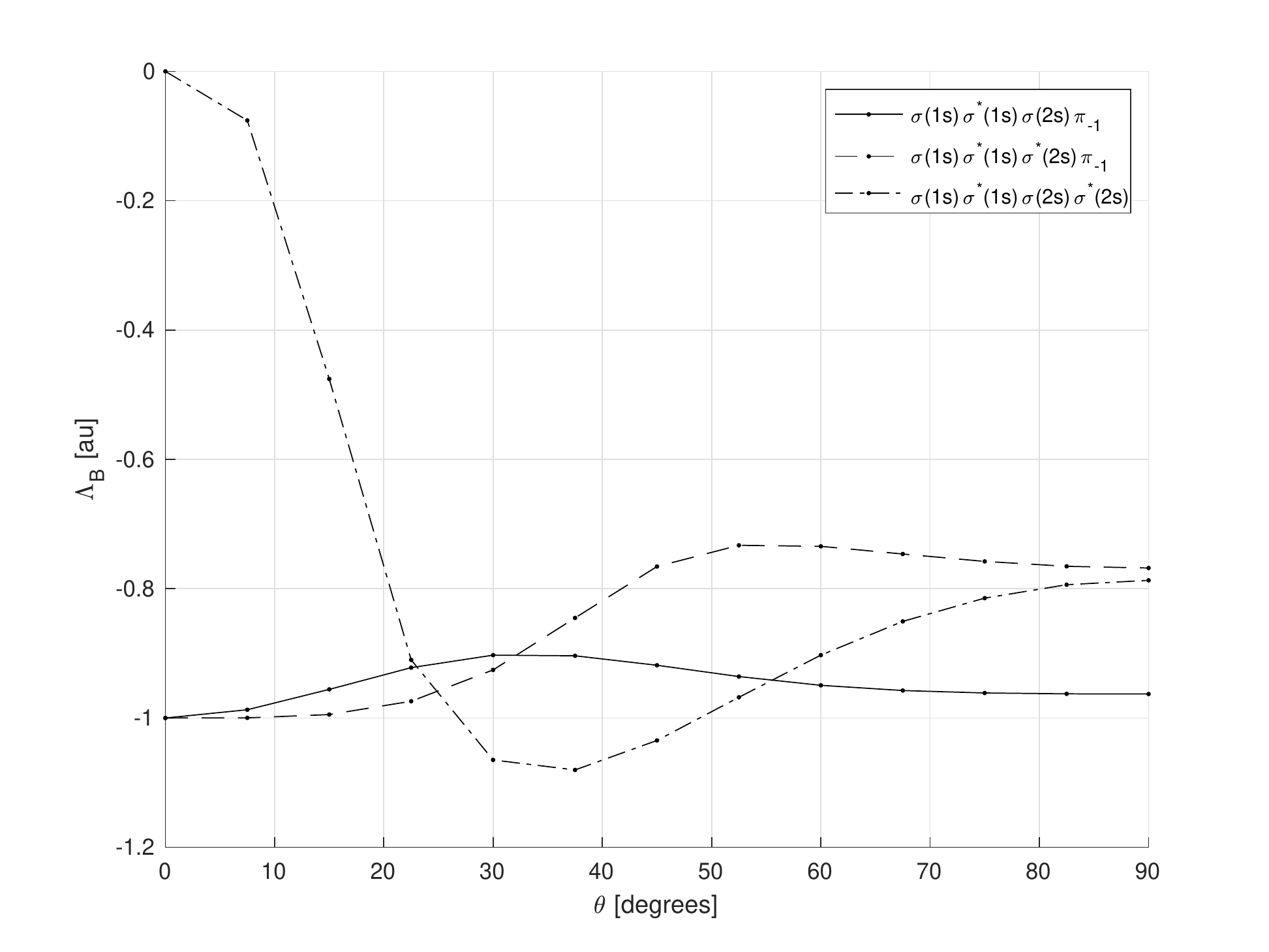}
  \caption{AQAM value for the lowest quintet states as a function of angle between the bond axis and magnetic field, with magnitudes fixed at $R=3.8a_0$ and
    $B = 0.2B_0$, respectively.}
  \label{figHe2_B020_quintet_angle_LB}
\end{figure}

\subsubsection{Quintet potential-energy curves at $B=B_0$}

\begin{figure}[h]
  \includegraphics[width=\linewidth]{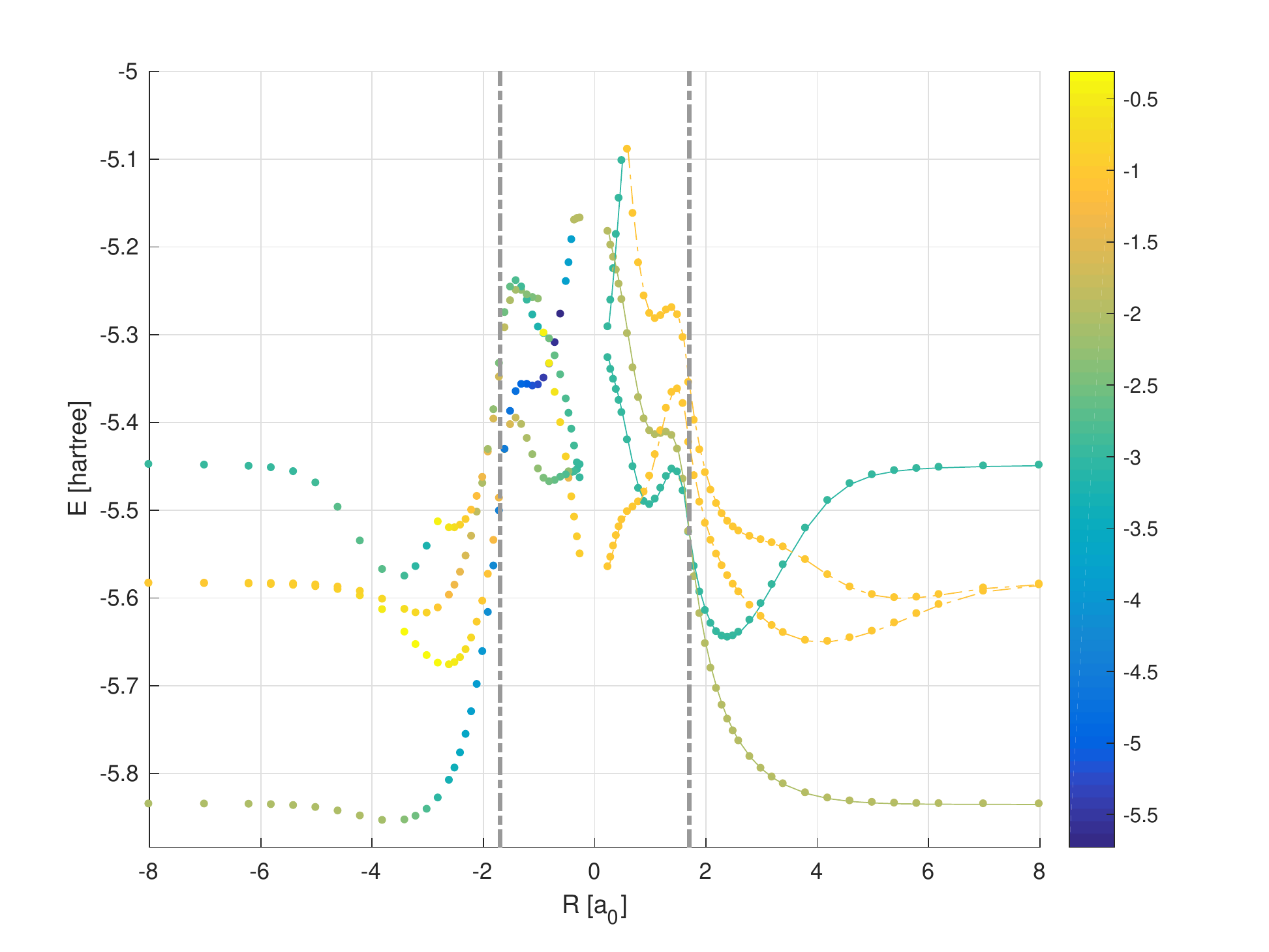}
  \caption{Dissociation curves for quintet states in perpendicular
    (negative half) and parallel (positive half) magnetic field $B = B_0$.}
  \label{figHe2_B100_quintet_dissoc}
\end{figure}

At a field strength of $B=B_0$, the orbital Zeeman interaction  has rearranged the states so that all the states containing only $\sigma$ orbitals are well above those that contain  $\pi$ orbitals. 
The lowest parallel state is dominated by the $^5\Delta_{\text{g}}(\IIorbdown{\sigma}{}{1s} \IIorbdown{\sigma}{*}{1s} \IIorbdown{\pi}{}{$-1$} \IIorbdown{\pi}{*}{$-1$})$ configuration,
which is well below all other states for all bond lengths greater than $2a_0$. 
At a bond distance of $R=4.2a_0$, the second quintet state is dominated by the $^5\Pi_\text g(\IIorbdown{\sigma}{}{1s} \IIorbdown{\sigma}{*}{1s} \IIorbdown{\pi}{}{$-1$} \IIorbdown{\sigma}{}{2s})$  configuration, while
the third quintet state is multiconfigurational
$^5\Pi_\text u(0.69\,\IIorbdown{\sigma}{}{1s} \IIorbdown{\sigma}{*}{1s} \IIorbdown{\pi}{*}{$-1$} \IIorbdown{\sigma}{}{2s},0.22\,\IIorbdown{\sigma}{}{1s} \IIorbdown{\sigma}{*}{1s} \IIorbdown{\pi}{}{$-1$} \IIorbdown{\sigma}{*}{2s})$ 
The fourth quintet state is largely $^5\Phi_\text g(\IIorbdown{\sigma}{}{1s} \IIorbdown{\sigma}{*}{1s} \IIorbdown{\pi}{}{$-1$} \IIorbdown{\delta}{}{$-2$})$. 

In the perpendicular field orientation at field strength $B = B_0$, the lowest quintet states at bond distance $R=3.8a_0$ are 
$^5\text{A}_{\text{g}}(1\orbdown{a}{g} 1\IITorbdown{b}{*}{u} 2\orbdown{a}{g} 2\IITorbdown{b}{\circ}{u})$,
$^5\text{B}_{\text{g}}(1\orbdown{a}{g} 1\IITorbdown{b}{*}{u} 2\orbdown{a}{g} 1\orbdown{a}{u})$, and
$^5\text{A}_{\text{u}}(0.71\, 1\orbdown{a}{g} 1\IITorbdown{b}{*}{u} 2\orbdown{a}{g} 1\IITorbdown{b}{\circ}{g}, 0.21 \, 1\orbdown{a}{g} 1\IITorbdown{b}{*}{u} 2\IITorbdown{b}{\circ}{u} 1\orbdown{a}{u})$.
The fourth quintet state is
$^5\text{B}_{\text{u}}(1\orbdown{a}{g} 1\IITorbdown{b}{*}{u} 2\orbdown{a}{g} 3\IITorbdown{a}{\circ}{g})$; however, at
a shorter bond distance of $R=2.5a_0$, the fourth state has undergone a level crossing and is of the symmetry $^5\text{A}_{\text{g}}(1\orbdown{a}{g} 1\IITorbdown{b}{*}{u} 2\orbdown{a}{g} 3\orbdown{b}{u})$.

\begin{figure}[h]
  \includegraphics[width=\linewidth]{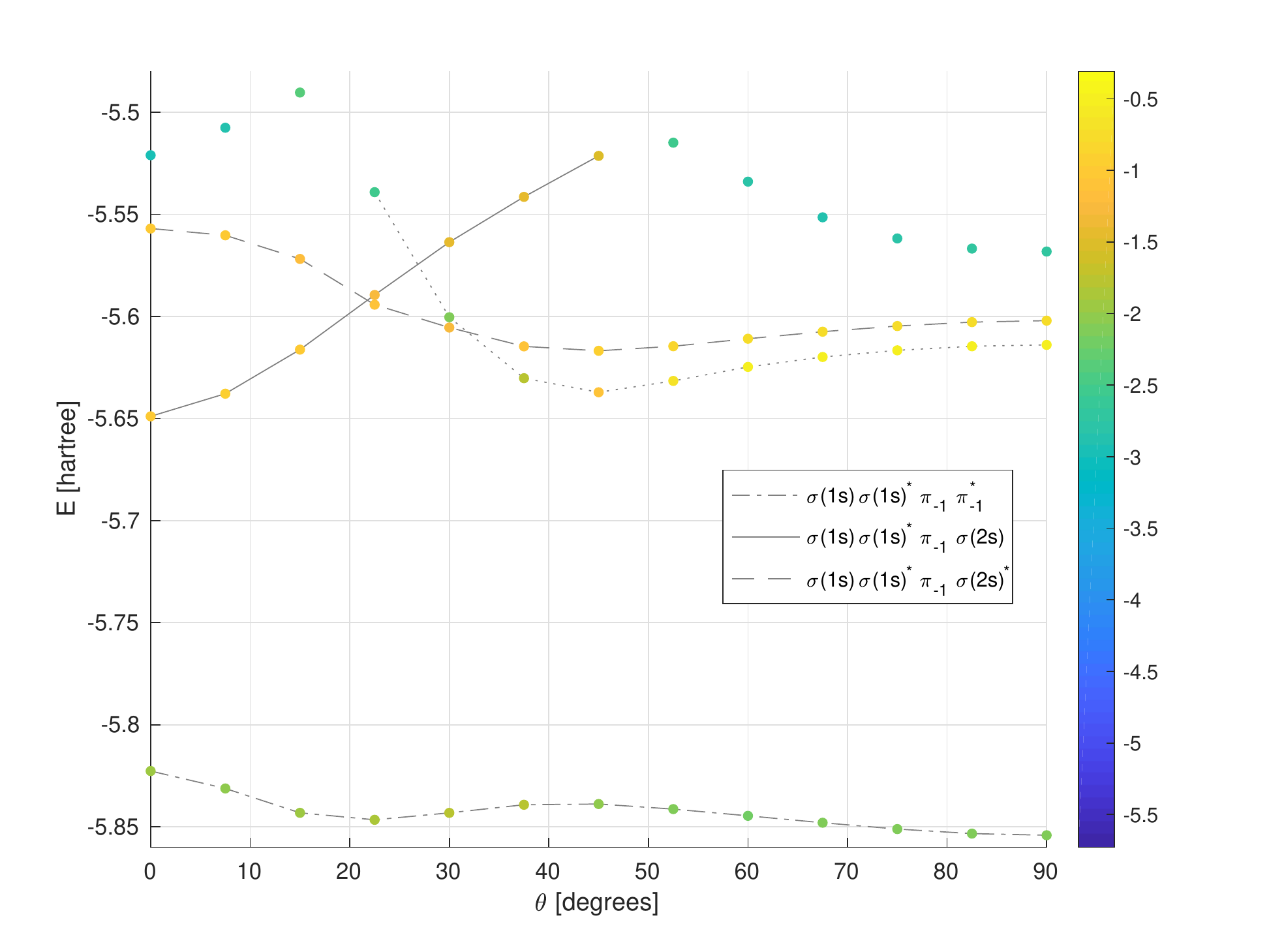}
  \caption{Quintet states as a function of angle between the bond
    axis and magnetic field, with magnitudes fixed at $R=3.8a_0$ and
    $B = B_0$, respectively.}
  \label{figHe2_B100_quintet_angle}
\end{figure}

\begin{figure}[h]
  \includegraphics[width=\linewidth]{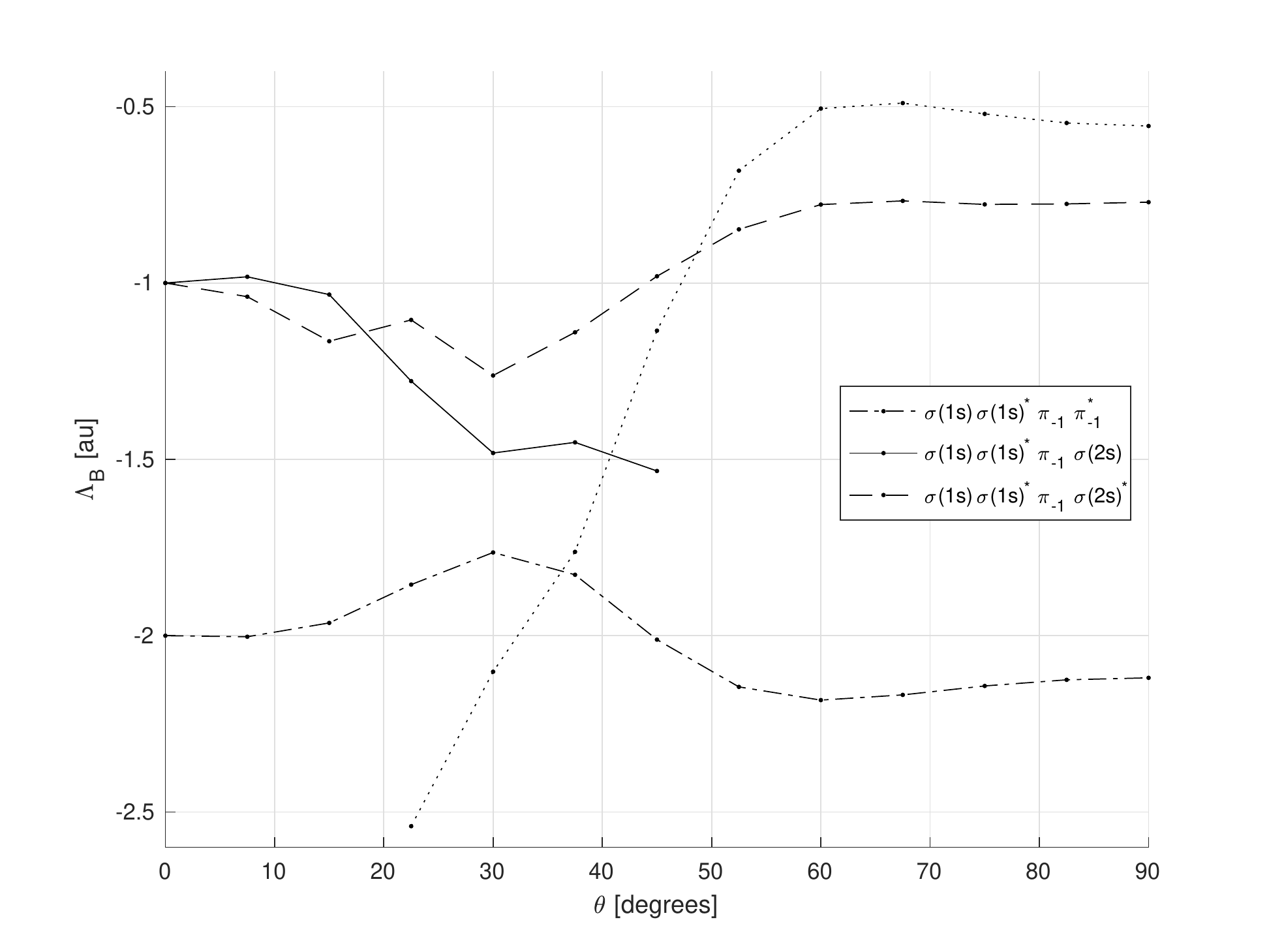}
  \caption{AQAM value for the lowest quintet states as a function of angle between the bond axis and magnetic field, with magnitudes fixed at $R=3.8a_0$ and
    $B = B_0$, respectively.}
  \label{figHe2_B100_quintet_angle_LB}
\end{figure}

Rotation of the parallel states into perpendicular states is comparatively straightforward due to the energy separations between the dissociation curves---see Fig.~\ref{figHe2_B100_quintet_angle} 
for rotation at the bond distance $R=3.8a_0$. The lowest state $^5\Delta_\text g(\IIorbdown{\sigma}{}{1s} \IIorbdown{\sigma}{*}{1s} \IIorbdown{\pi}{}{$-1$} \IIorbdown{\pi}{*}{$-1$})$ 
is further lowered by about 30 millihartree from 0 to 90 degrees. It is paramagnetically bound, 
with a dissociation energy of 20 millihartree and a global minimum located at $\theta = 90^{\circ}$ and a slightly shorter bond distance $R = 3.62a_0$. 

Intriguingly, the rotation curve in Fig.~\ref{figHe2_B100_quintet_angle} has a second local minimum with respect to  $\theta$ at about $25^{\circ}$. 
The corresponding AQAM values in Fig.~\ref{figHe2_B100_quintet_angle_LB} show that this local minimum is not associated with any increase in the magnitude $|\Lambda_{\mathbf{B}}|$. 
The stabilization at $\theta \approx 25^{\circ}$ is therefore of a different origin than the stabilization at $\theta=90^{\circ}$.

\section{Energy surfaces}

\begin{figure}[h]
  \includegraphics[width=\linewidth]{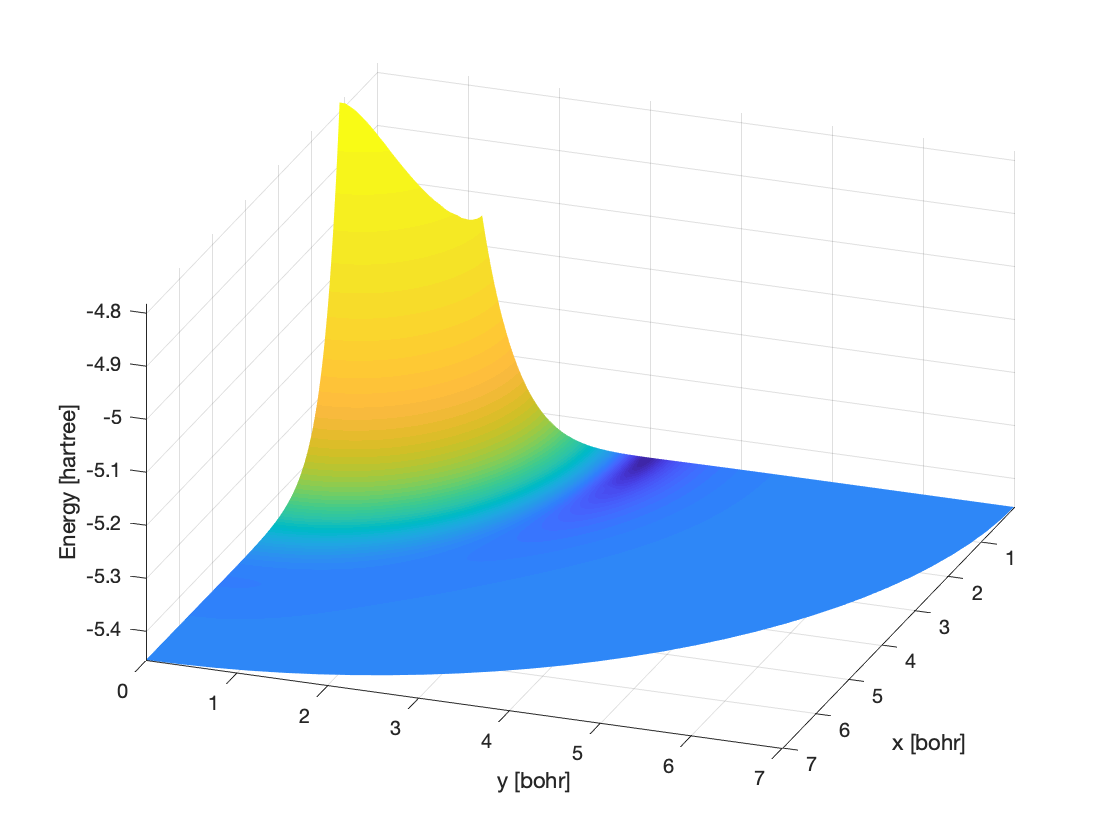}
  \caption{The lowest singlet energy surface in a field of
    $B=B_0$. The axis labels are $x=R\cos(\theta)$ and $y =
    R \sin(\theta)$, so that the left side where $y=0$ (and
    $\theta=0^\circ$) corresponds to the parallel orientation. The
    colour scale is $\ln(\eta + E(R,\theta)-E_{\mathrm{min}})$, with
  $\eta = 10^{-4}$~hartree.}
  \label{figHe2_SingletSurf}
\end{figure}

\begin{figure}[h]
  \includegraphics[width=\linewidth]{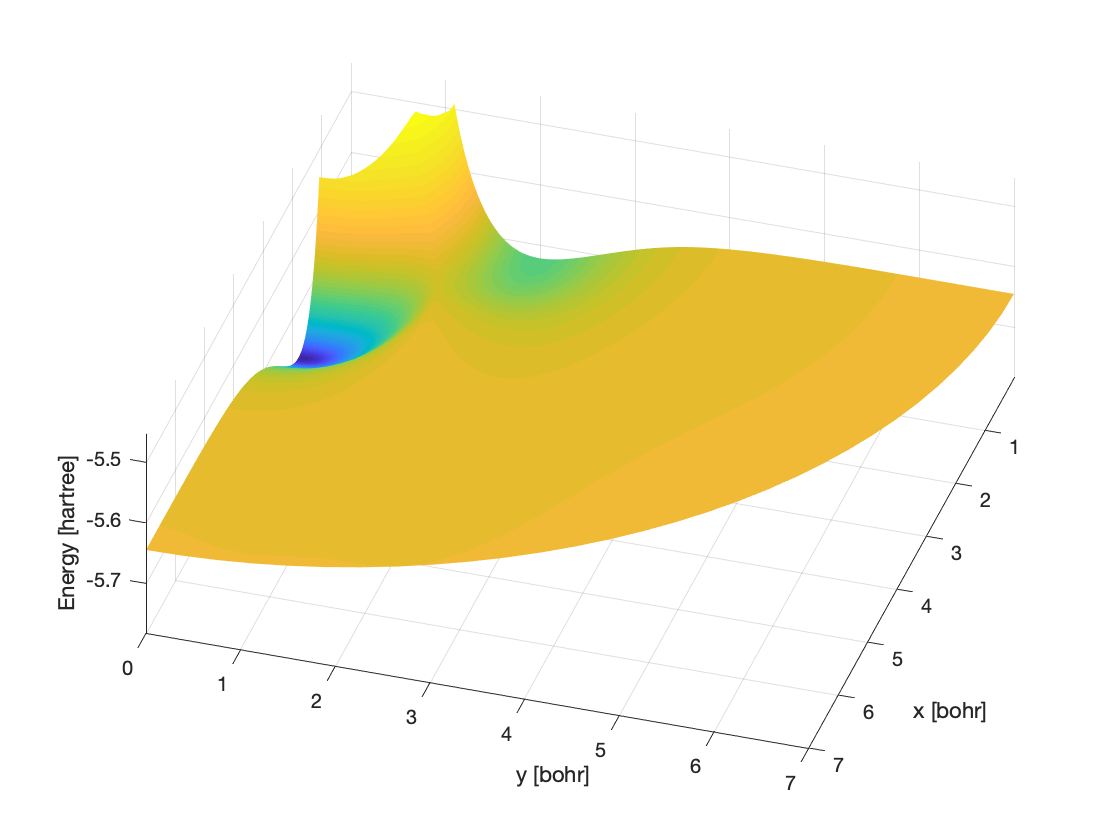}
  \caption{The lowest triplet energy surface in a field of
    $B=B_0$. The axis labels are $x=R\cos(\theta)$ and $y =
    R \sin(\theta)$. The  colour scale is $\ln(\eta + E(R,\theta)-E_{\mathrm{min}})$, with
  $\eta = 5\times 10^{-3}$~hartree.}
  \label{figHe2_TripletSurf}
\end{figure}

\begin{figure}[h]
  \includegraphics[width=\linewidth]{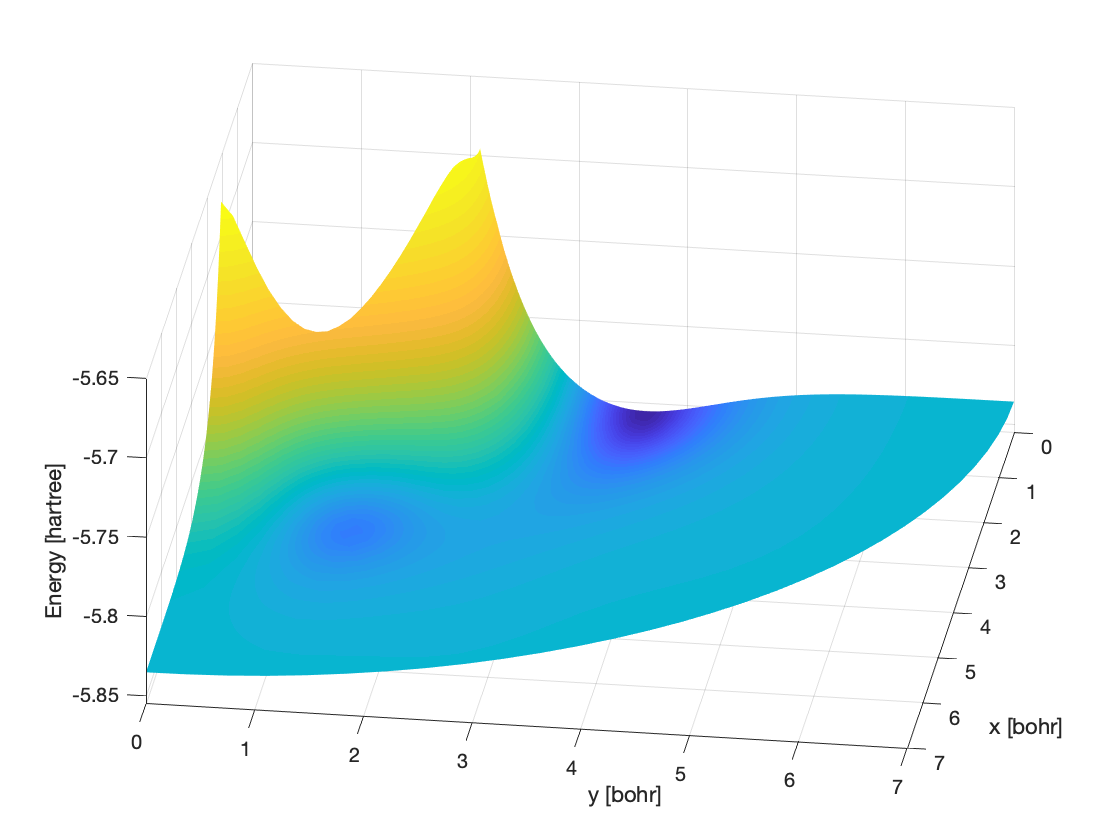}
  \caption{The lowest quintet energy surface in a field of
    $B=B_0$. The axis labels are $x=R\cos(\theta)$ and $y =
    R \sin(\theta)$. The  colour scale is $\ln(\eta + E(R,\theta)-E_{\mathrm{min}})$, with
  $\eta = 5\times 10^{-3}$~hartree.}
  \label{figHe2_QuintetSurf}
\end{figure}

Complete energy surfaces for the lowest singlet, triplet, and quintet
states at $B = B_0$ are shown in Fig.~\ref{figHe2_SingletSurf},
\ref{figHe2_TripletSurf}, and \ref{figHe2_QuintetSurf},
respectively. These surfaces have been computed at the
FCI/Lu-aug-cc-pVTZ level with a correction for basis-set superposition
error (BSSE). The correction is an adapted counterpoise correction, taking
into account the loss of symmetry in a magnetic field and, in
particular, the inequivalence of the parallel and perpendicular
orientations.

The most dramatic feature is seen in the triplet surface in
Fig.~\ref{figHe2_TripletSurf}, which is actually at each $(R,\theta)$
the minimum of two surfaces. One of these crossing states has a
minimum in the perpendicular orientation and the other has a deeper
minimum in the parallel orientation. The level crossing is clearly
seen as a discontinuous ``rift'' that occurs for the shorter bond
distances and angles roughly between $20^\circ$ and $50^\circ$.

On the singlet surface in Fig.~\ref{figHe2_SingletSurf} the
minimum is located at $R=3.01a_0$ and $\theta=90^\circ$ and the
BSSE corrected dissociation energy is $1.264$~millihartree, which differs only negligibly
different from the uncorrected value in Table~\ref{tabDissocMinB1}. On
the triplet surface in Fig.~\ref{figHe2_TripletSurf}, the deepest minimum
occurs at $R=1.80a_0$ and $\theta = 0^\circ$, with a BSSE corrected dissociation
energy of $0.1376$~hartree. The shallower minimum occurs at
$R=1.99a_0$ and has a BSSE corrected dissociation energy of
0.07676~hartree, again negligibly different from the uncorrected
value. Finally, the quintet surface in Fig.~\ref{figHe2_QuintetSurf}
has two minima. The deeper minimum occurs at $R=3.61a_0$,
$\theta=90^\circ$, and has a BSSE corrected dissociation energy of
19.56~millihartree. The shallower minimum is located at $3.5a_0$ and
$\theta=24^\circ$, with a dissociation energy of 11.7~millihartree.

\begin{figure}[h]
  \includegraphics[width=\linewidth]{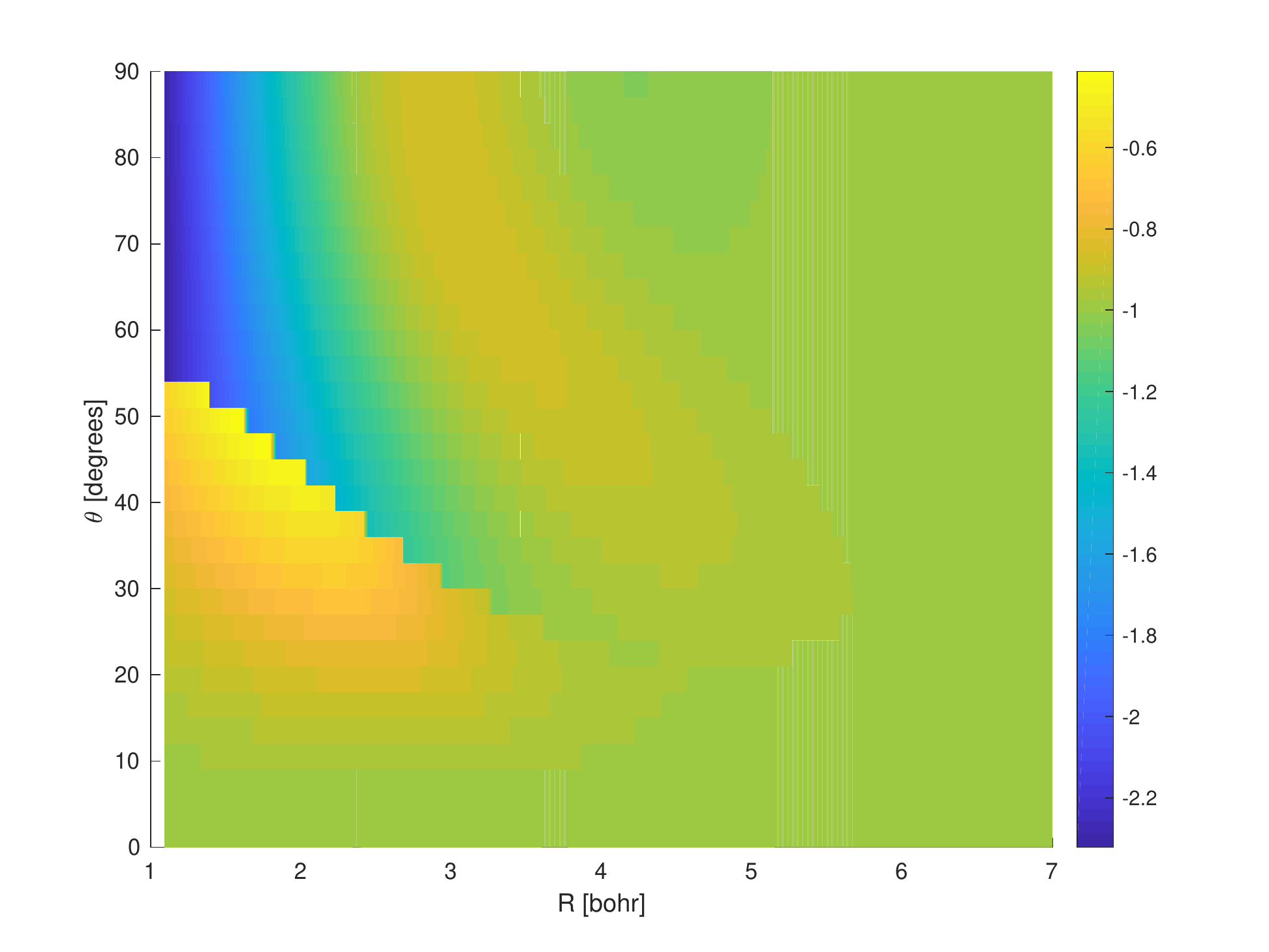}
  \caption{The AQAM projection onto the magnetic field direction for the
    lowest triplet at each value of $(R,\theta)$. The level crossing
    between the two low-lying triplets is clearly manifested in the
    discontinuous ``rift'' that begins at $R\approx 1$~bohr and
    $\theta \approx 50^\circ$.}
  \label{figHe2_TripletAqam}
\end{figure}

\section{Conclusions}
\label{secCONCL}

We have studied the low-lying states of the helium dimer for different spins and magnetic-field strengths.
As expected, the singlet, triplet, and quintet spectra resemble each other to a great degree, since many states have analogues with other total spin. For example, open-shell singlets have direct analogues among triplets. 
In general, all states are subject to a diamagnetic destabilization. However, the spin and orbital Zeeman interactions affect states differently and dramatically reorder the spectra, bringing down states of higher angular momentum. 
Hence, states with $\pi$ and $\delta$ bonding orbitals become increasingly important in strong fields. Moreover, at large field strengths, the spin Zeeman interaction lowers the $m_\text s=-1$ triplets below the singlets. 
For a field strength of $B=B_0$, the globally lowest state is even a paramagnetically bonded quintet state with $D_\text e = 52$\,kJ/mol, oriented perpendicular to the magnetic field. Hence, these field strengths induce an entirely new chemistry of helium atoms.

In general, in addition to the effects of increasing field strength, the orientation with respect to the magnetic field modulates the proportion of $\sigma$, $\pi$ and $\delta$ bonding, which affects the total angular momentum and the orbital Zeeman interaction. 
For nontrivial orientations of the bond axis with respect to the magnetic field, all spatial symmetries except inversion are lost and the canonical angular momentum ceases to be a good quantum number. 
To partially address this complication, we have introduced the almost quantized angular momentum (AQAM) and demonstrated that it is a very useful tool to characterize states in arbitrary orientations. Conical intersections make detailed state classification beyond the
characterization provided by AQAM challenging and poorly defined. In general, energy hypersurfaces become multivalued as functions of the parameters $(R,\theta,B)$. This occurs as an effect of the symmetry breaking, 
which turns true crossings in the parallel orientation into avoided crossings at nontrivial angles. Two states may be continuously deformed into each along some paths in parameter space, but not others. 
In the radial dissociation limit, for instance, the parallel and perpendicular orientations become physically equivalent. 
Nonetheless, at a fixed bond distance, continuously deforming between the parallel and perpendicular orientations can result in a state with a different radial dissociation limit.

Our results show that perpendicular paramagnetic bonding is common in excited electronic
states, although the presence of conical intersections makes the identification somewhat poorly defined and dependent on the which path in parameter space is emphasized.  Moreover, the effect is larger for the more diffuse $\sigma_{2\text s}^*$ compared to the compact $\sigma_{1\text s}^*$ orbital. As a result, the bonding mechanism is also stronger, sometimes by orders of magnitudes, in excited states than the originally described cases (lowest triplet of H$_2$ and lowest singlet of He$_2$). There are some indications of the perpendicular paramagnetic bonding mechanism involving higher angular momentum states (e.g., modulation of $\pi$ into $\delta$ orbitals or $\delta$ into $\phi$ orbitals), although it is difficult to determine the relative contributions from $\sigma^*$ and higher angular momentum orbitals.

\section*{Acknowledgments}

This work was supported by the Research Council of Norway through Grant No.~240674 and CoE Hylleraas Centre for Molecular Sciences Grant No.~262695. This work has also received support from the Norwegian Supercomputing Program (NOTUR) through a grant of computer time (Grant No.~NN4654K).

\newcommand{\noopsort}[1]{} \newcommand{\printfirst}[2]{#1}
  \newcommand{\singleletter}[1]{#1} \newcommand{\switchargs}[2]{#2#1}

\end{document}